\documentclass[twocolumn,tighten,onecolappendix]{aastex61}

\input epsf.tex

\begin{document}
\title{The Gemini/HST Galaxy Cluster Project: Environment Effects on the Stellar Populations in the Lynx Clusters at $z=1.27$}
\author[0000-0003-3002-1446]{Inger J{\o}rgensen}
\affil{Gemini Observatory, 670 N.\ A`ohoku Pl., Hilo, HI 96720, USA}
\author[0000-0001-8698-8280]{Laura C.\ Hunter}
\affil{Gemini Observatory, 670 N.\ A`ohoku Pl., Hilo, HI 96720, USA}
\affil{Department of Astronomy, Indiana University, 727 East 3rd Street, Swain West 318, Bloomington, IN 47405-7105, USA}
\author[0000-0003-4328-9046]{Conor R.\ O'Neill} 
\affil{Gemini Observatory, 670 N.\ A`ohoku Pl., Hilo, HI 96720, USA}
\author{Kristin Chiboucas}
\affil{Gemini Observatory, 670 N.\ A`ohoku Pl., Hilo, HI 96720, USA}
\author{Ryan K.\ Cole}
\affil{Gemini Observatory, 670 N.\ A`ohoku Pl., Hilo, HI 96720, USA}
\affil{Department of Mechanical Engineering, University of Colorado, 427 UCB, 1111 Engineering Dr., Boulder, CO 80309, USA}
\author[0000-0003-3631-7176]{Sune Toft}
\affil{Cosmic Dawn Center (DAWN), Niels Bohr Institute, University of Copenhagen, Lyngbyvej 2, DK-2100 Copenhagen {\O}, Denmark}
\author[0000-0002-2244-0897]{Ricardo P.\ Schiavon}
\affil{Astrophysics Research Institute, Liverpool John Moores University, 146 Brownlow Hill, Liverpool L3 5RF, United Kingdom}

\email{ijorgensen@gemini.edu} 
\email{hunterlc@iu.edu}
\email{conor.oneill.physics@gmail.com}
\email{kchiboucas@gemini.edu}
\email{ryan.cole@colorado.edu}
\email{sune@dark-cosmology.dk}
\email{R.P.Schiavon@ljmu.ac.uk}

\correspondingauthor{Inger J{\o}rgensen}

\submitjournal{Astrophysical Journal}
\date{Accepted for publication June 26, 2019}

\begin{abstract}
Few detailed investigations of stellar populations in passive galaxies beyond $z\approx 1$ are based on
deep spectroscopic observations, due to the difficulty in obtaining such data. 
We present a study of stellar populations, structure, and mass-to-light ratios
of a large sample of bulge-dominated galaxies in the two $z=1.27$ clusters Lynx E and Lynx W, 
based on deep ground-based optical spectroscopy combined with imaging from {\it Hubble Space Telescope}. 
We find that Lynx E has a well-defined core of red passive galaxies, while Lynx W lacks such a core.
If all the sample galaxies evolve similarly in size from $z=1.27$ to the present, 
the data allow only 0.1 dex size-growth at a fixed dynamical mass.
However,
to link the Lynx central galaxies to brightest cluster galaxies similar to those of low 
redshift clusters, the Lynx galaxies would have to grow by at least a factor five, possibly through major merging. 
The mass-to-light ratios and the Balmer absorption lines of the Lynx galaxies 
are consistent with passive evolution of the stellar populations from $z=1.27$ to the present 
and support ages of 1--3 Gyr. 
The galaxies in the outskirts of the clusters contain younger stellar populations than found in the 
cluster cores.
However, when evolved passively to $z\approx 0$ both populations are consistent with
the observed populations in the Coma cluster galaxies. 
The bulge-dominated emission line galaxies in the clusters are dominated by stellar populations
with sub-solar metallicities. Thus, additional 
enrichment of these is required to produce Coma-like stellar populations by $z\approx 0$.
\end{abstract}


\section{Introduction}

Galaxy evolution may be studied by investigating the resulting stellar populations and the 
galaxy structure at $z\approx 0$, commonly referred to as galaxy archaeology, see recent studies
by Cappallari et al.\ (2013a), Conroy et al.\ (2014), McDermid et al.\ (2015), and Parikh et al.\ (2018).  
Alternatively, one may study galaxies over a substantial redshift interval and attempt to establish an evolutionary
timeline consistent with the properties, e.g., van Dokkum \& van der Marel (2007, and references therein), 
Saglia et al.\ (2010), J\o rgensen et al.\ (2017), and Beifiori et al.\ (2017).
In the present paper we focus on the latter technique as it is applied to bulge-dominated galaxies.

Multiple studies have found that since $z=1-2$ the sizes of field galaxies increase
by a factor 3-5 (Toft et al.\ 2009, 2012; Newman et al.\ 2012; Cassata et al.\ 2013; van der Wel et al.\ 2014).
However, it appears that the evolution is accelerated in cluster environments and may happen at earlier epochs
than in the field. Therefore the (remaining) size evolution for cluster galaxies from $z\approx 1.5$ to 
the present is less than found for field galaxies 
(Zirm et al.\ 2012; Papovich et al.\ 2012; Strazzullo et al.\ 2013; J\o rgensen \& Chiboucas 2013; 
J\o rgensen et al.\ 2014; Delaye et al.\ 2014). 
The recent results for three $z=1.3-1.6$ clusters published by Beifiori et al.\ (2017) also
indicate that size evolution for cluster galaxies is smaller than found in the field.
Some of the apparent size growth for passive galaxies may be due to larger more recently 
quenched galaxies entering the samples. Belli et al.\ (2015) find that for galaxies
with masses above $10^{10.7} M_{\sun}$ about half the size growth between $z=2$ and 1.25
originates from such recently quenched galaxies, while the other half represents size growth of already
quenched galaxies.

Stellar population evolution studies beyond $z\approx 1$ have primarily focused on ages through studies of luminosity changes.
Beifiori et al.\ (2017) used new data for 19 galaxies in $z=1.3-1.6$ clusters obtained with
VLT/KMOS to extend the redshift coverage of results regarding the evolution of the mass-to-light (M/L) ratios
of bulge-dominated passive galaxies. 
The authors used their new results together with available literature results covering up to $z=1.3$
(van Dokkum \& Franx 1996; J\o rgensen et al.\ 1999; Kelson et al.\ 2000;
Wuyts et al.\ 2004; Barr et al.\ 2006; J\o rgensen et al.\ 2006;,
Holden et al.\ 2005, 2010; van Dokkum \& van der Marel 2007; Saglia et al.\ 2010;
J\o rgensen \& Chiboucas 2013; J\o rgensen et al.\ 2014) and low redshift reference
data for the Coma cluster (J\o rgensen 1999; J\o rgensen et al.\ 2006)
to further solidify the evidence supporting passive evolution and a formation redshift $z_{\rm form}\approx 2$.
The formation redshift should be understood as the epoch of the last major star formation episode. 
At $z \approx 1$ the massive (Mass $> 10^{11}M_{\sun}$) bulge-dominated galaxies in clusters appear to be 
in place and mostly passively evolving.
Lower mass galaxies may still be added to the red sequence and from then on passively evolve 
(e.g., S\'anchez-Bl\'azquez et al.\ 2009; Choi et al.\ 2014),
but see also Cerulo et al.\ (2016) for results supporting that the red sequence well below $L^{\star}$ is 
fully populated in rich clusters already at $z\approx 1.5$.
Ultimately, the properties of galaxies mapped over a large fraction of the age of the Universe,
may constrain the models for building the galaxies. It is difficult to understand
within the prevailing hierarchical model favored by the $\Lambda$CDM (cold-dark-matter) cosmology, the existence
of such massive passive galaxies with relatively old stellar populations at $z\approx 1$, while
less massive galaxies appear to harbor younger stellar populations, e.g., J\o rgensen et al.\ (2017, and references therein),
see Kauffmann et al.\ (2003) for a discussion of this tension between the observational results and
the hierarchical models of galaxy formation.
However, more recent cosmological simulations like Illustris (Genel et al.\ 2014; Vogelsberger et al.\ 2014; Wellons et al. 2015) 
and UniverseMachine (Behroozi et al.\ 2018) find that massive quiescent galaxies can 
be in place by $z\ga 2$.

Metallicities and abundance ratios for bulge-dominated galaxies up to $z\approx 1$ have been studied from 
spectroscopy both through our project, the Gemini/{\it HST} Galaxy Cluster Project (GCP) 
(J\o rgensen et al.\ 2005, 2017,  J\o rgensen \& Chiboucas 2013),
and the ESO Distant Cluster Survey (S\'anchez-Bl\'azquez et al.\ 2009).
However, few studies have attempted to map the evolution of metallicities and abundance beyond $z\approx 1$,
and samples with deep spectroscopy of $z>1$ (passive) bulge-dominated galaxies are still very small, 
e.g., J\o rgensen et al.\ (2014) and Beifori et al.\ (2017).
Onodera et al.\ (2015) stacked spectra of 24 massive galaxies at $z\approx 1.6$ and established the mean
age, metallicity and abundance ratio $\rm [\alpha /Fe]$ from this composite spectrum. 
The results suggest that the metallicities and abundance ratios of massive galaxies are set already at $z\approx 1.6$,
and the stellar populations may evolve passively from there to agreement with low redshift passive galaxies.
Kriek et al.\ (2016) studied a single massive and passive galaxy at $z=2.1$ and found 
an abundance ratio [Mg/Fe]=0.45, which is significantly higher than in 
the majority of low redshift passive galaxies, and consistent with a very short 
star formation timescale. 

Several authors have used lower resolution {\it Hubble Space Telescope} ({\it HST}) grism spectra to study $z=0.5-2$ galaxies. 
Such recent studies find ages consistent with quiescent galaxies forming a large fraction of their stars
at $z\ga 2$ (e.g., Fumagalli et al.\ 2016; Ferreras et al.\ 2018; Estrada-Carpenter et al.\ 2019).
Metallicities of these galaxies are found to be solar already at $z\approx 2$ 
(Ferreras et al.\ 2018; Estrada-Carpenter et al.\ 2019).

In addition to possibly affecting the speed of the size and mass evolution, the cluster environment
is also known to affect the star formation histories of the galaxies. Of particular importance
for studies of $z>1$ galaxies, it appears that even at $z=1.2-1.5$ centers of very 
massive clusters are devoid of star forming galaxies 
(Gr\"{u}tzbauch et al.\ 2011; Quadri et al.\ 2012; Koyama et al.\ 2013)
possibly because the cluster environment has contributed to the quenching of the star formation.
However, our results for Lynx W (J\o rgensen et al.\ 2014) show that this $z=1.27$ cluster 
appears to be just in the process of quenching star formation. 
In addition, $z\approx 2$ protoclusters still contain actively star forming galaxies, e.g., Wang et al.\ (2016). 
Thus, the main transformation must happen at $z=1-2$ and depends on galaxy mass 
(Quadri et al.\ 2012; Koyama et al.\ 2013) and possibly also on the cluster environment (Tanaka et al.\ 2013).

\begin{figure*}
\begin{center}
\includegraphics[angle=90, scale=0.90]{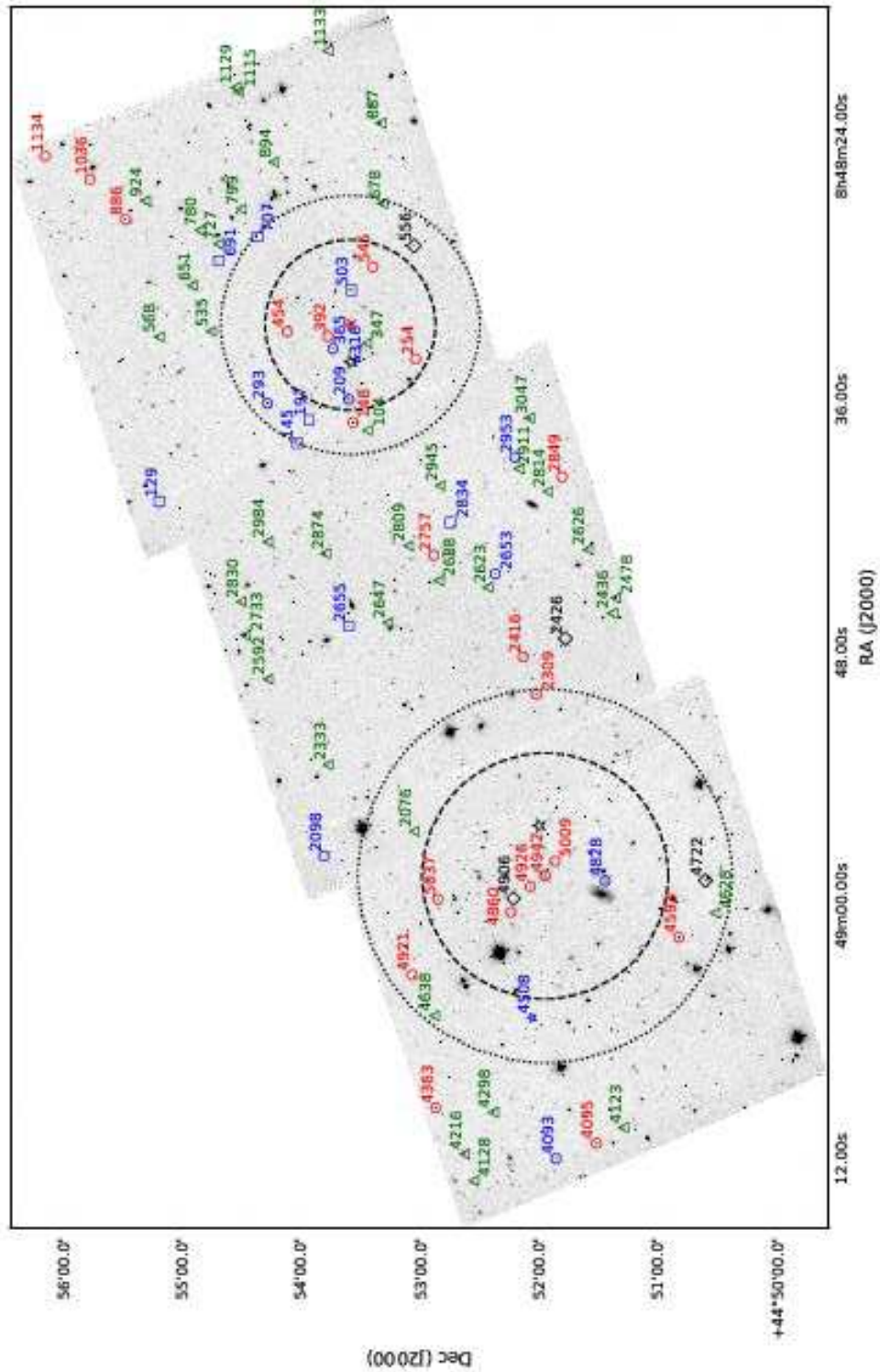}
\end{center}
\caption{
Grayscale image made from the {\it HST}/ACS F850LP observations of the Lynx clusters. 
Our spectroscopic sample for both clusters is marked.
Red circles -- bulge-dominated members with EW[\ion{O}{2}]$\le 5${\AA}. 
Blue circles -- bulge-dominated members with EW[\ion{O}{2}] $> 5${\AA}.
Blue squares -- disk galaxies.
Blue stars -- two AGNs, which are members of the clusters.
Dark green triangles -- non-members. 
Black diamonds -- targets for which the spectra do not allow redshift determination. 
Red star -- the brightest cluster galaxy (BCG) in Lynx W, not part of our spectroscopic sample as it has a triple core, see text.
Dashed and dotted circles -- $R_{500}$ and $R_{200}=1.52 R_{500}$, respectively, centered on the BCGs adopted as cluster centers, see text.
$R_{500}$ is the radius within which the mean over-density of the cluster is 500 times the critical density at the 
cluster redshift.
Black stars -- X-ray centers from Rosati et al.\ (1999).
See Table \ref{tab-phot} in Appendix \ref{SEC-IMAGING} for cross reference between numbering on this figure and that of J\o rgensen et al.\ (2014).
\label{fig-lynxgrey} }
\end{figure*}

In this paper we focus on the structure and stellar populations of the bulge-dominated galaxies in
the two clusters Lynx E and W at $z=1.27$. Lynx E and W are part of the Lynx supercluster. 
The clusters were first spectroscopically confirmed by Rosati et al.\ (1999) and Stanford et al.\ (1997),
while Nakata et al.\ (2005) carried out an extensive photometry survey of the supercluster identifying
not only the two main clusters, but also several smaller groups as part of the supercluster.
Van Dokkum et al.\ (2001) used {\it HST} imaging to investigate
the Lynx W galaxy population in detail. These authors concluded that the cluster contains $\approx 50$\%
bulge-dominated galaxies and that the two most luminous galaxies in the cluster may have been 
in recent interactions, or still in ongoing interactions. One of these is the triple-core
galaxy in the center of the cluster.
In their investigation of a larger part of the Lynx supercluster, Mei et al.\ (2012) focused on 
the morphologies and the color-magnitude relation. 
These authors also concluded that the fraction of bulge-dominated galaxies in the clusters is $\approx 50$\%.
They identify some galaxies as being quenched before being transformed morphologically to bulge-dominated galaxies.
They find that the bulge-dominated galaxies are smaller (at a given mass) than found in lower redshift clusters.
On the other hand, Saracco et al.\ (2014), who also studied sizes and luminosities of Lynx W elliptical galaxies,
concluded that the photometric data were consistent with passive evolution and no additional size 
or mass growth from $z=1.27$ to the present.

The Lynx clusters are part of our $z=1-2$ extension of the GCP.
Using our deep optical spectroscopy together with {\it HST} imaging,
we in this paper investigate to what extent structure evolution is required between $z=1.27$ and the present,
and whether the stellar populations of the galaxies are consistent
with passive evolution from $z=1.27$ to the present. 
In addition, because the two clusters are at identical redshifts but have very different cluster 
masses, we have the opportunity to study cluster environment effects by direct comparison of the galaxy
populations in the two clusters.

The observational data are briefly described in Section \ref{SEC-DATA}. 
We make use of our new low redshift reference sample (J\o rgensen et al.\ 2018b), see Section \ref{SEC-COMPDATA}. 
The Appendices \ref{SEC-IMAGING}, \ref{SEC-SPECTROSCOPY}, \ref{SEC-SPECIM}, and \ref{SEC-LOWZDETAIL}
provide additional detail regarding the data.
In Section \ref{SEC-CLUSTERZ}, we discuss the properties of the two clusters, in particular the masses
compared with masses of other GCP clusters and our low redshift reference clusters.
Section \ref{SEC-METHOD} gives an overview over the methods and models used in the analysis.
We then define the sub-samples of galaxies, Section \ref{SEC-SAMPLE}, and evaluate selection effects for 
the available sample. 
The apparent differences between Lynx E and W are presented in Section \ref{SEC-DIFFERENCES}.
We investigate the possible structure evolution as well as the evolution of the Fundamental Plane
and the M/L ratios in Section \ref{SEC-STRUCTUREFP}.
In Section \ref{SEC-STELLARPOP}, we focus on the stellar populations both as a function of galaxy
velocity dispersion and as a function of cluster environment. In particular, we construct 
composite spectra of samples matching the cluster environments and fit stellar population models to these.
In Section \ref{SEC-DISCUSSION}, we discuss the results in the context of simple models for the 
evolution with redshift.
The conclusions are summarized in Section \ref{SEC-CONCLUSION}.

Throughout this paper we adopt a $\Lambda$CDM cosmology with 
$\rm H_0 = 70\,km\,s^{-1}\,Mpc^{-1}$, $\Omega_{\rm M}=0.3$, and $\Omega_{\rm \Lambda}=0.7$.
Magnitudes are quoted as in the AB system (Oke \& Gunn 1983), except where noted.

\section{Observational data \label{SEC-DATA}}

Sections \ref{SEC-IMAG} and \ref{SEC-SPEC} briefly summarize the observational data for Lynx E and W. 
The reader is referred to Appendices \ref{SEC-IMAGING} and \ref{SEC-SPECTROSCOPY}, as well as
J\o rgensen et al.\ (2014) for details. 
Section \ref{SEC-COMPDATA} summarizes the data for the low redshift reference sample,
see also Appendix \ref{SEC-LOWZDETAIL} and J\o rgensen et al.\ (2018b), as well as 
other GCP data used in the analysis.

\subsection{Imaging of Lynx E and W \label{SEC-IMAG} }

In our analysis we use available {\it HST}/ACS imaging in the filters F775W and F850LP,
obtained in 2004 through HST program ID 9919. 
The three pointings obtained for Lynx E and W cover all galaxies in our spectroscopic sample, 
see Figure \ref{fig-lynxgrey}.
We have reprocessed the {\it HST}/ACS imaging since our analysis of Lynx W in J\o rgensen et al.\ (2014).
The same methods were used as described in Chiboucas et al.\ (2009).
The images were drizzled (Fruchter \& Hook 2002), and then processed with SExtractor (Bertin \& Arnouts 1996). 
For sample selection purposes, we adopted the SExtractor magnitudes {\tt mag\_auto} in F850LP 
as the total magnitudes $z_{tot,850}$,
understanding that {\tt mag\_auto} misses a small fraction of the flux (see, e.g., J\o rgensen et al.\ 2018a for 
detailed simulations of the effect). 
Aperture colors $(i_{775}-z_{850})$ were derived within an aperture with diameter of 0.5 arcsec.
Based on point-spread-function modeling as described in Chiboucas et al.\ (2009), 
we find median FWHM=0.1045 arcsec and 0.1050 arcsec, for the drizzled stacked images in F775W and F850LP, respectively.
Thus, the fixed aperture size may lead to a small systematic error on the colors.
We quantified this from simulations matching the observed properties for $z_{850}=21-24.6$ mag galaxies,
and the PSF and the noise properties of the data. The simulations give an offset in the $(i_{775}-z_{850})$ 
aperture colors of $-0.008$ from the true colors. This difference is not significant for our purpose, as the colors
are used primarily for selection of targets for the spectroscopic observations.
All galaxies in the fields brighter than $z_{tot,850}=24.6$ mag were then fit with the profile
fitting program GALFIT (Peng et al.\ 2002). This limit corresponds to a signal-to-noise ratio of $\approx 25$ 
for the available data, resulting in radii and total magnitudes with formal fitting uncertainties of 10\% or less.
The limit also ensures that our main spectroscopic sample is included in the processing.
We fit both $r^{1/4}$ profiles and S\'{e}rsic (1968) profiles
and derive effective radii, total magnitudes and surface brightnesses.  
This processing was done for the observations in F850LP only. 
Throughout the paper we use circularized effective radii derived from the semi-major and -minor axes as
$r_{\rm e} = (a_{\rm e}\,b_{\rm e})^{1/2}$. 

The photometry is calibrated to the AB system using the synthetic zero points for the filters,
25.654 for F775W, 24.862 for F850LP  (Sirianni et al.\ 2005).
We adopt galactic extinction in the direction of the cluster as provided by 
the NASA/IPAC Extragalactic Database using the calibration from Schlafly \& Finkbeinner (2011). 
The extinction in the two filters are $A_{775}=0.044$ and $A_{850}=0.034$.
The photometric parameters derived using SExtractor and GALFIT are for the spectroscopic
sample listed in Appendix \ref{SEC-IMAGING} Table \ref{tab-phot}.
The re-derived photometry is consistent with our measurements in J\o rgensen et al.\ (2014), see Appendix \ref{SEC-IMAGING}.
For consistency in the current analysis we exclusively use the re-derived photometry.
The full photometric catalog for the three fields will be published in a future paper (I.\ J\o rgensen et al.\ 2019, in prep.).

We calibrate the photometry to rest frame $B$-band (Vega magnitudes)  
using stellar population models from Bruzual \& Charlot (2003). 
We use the improved {\it U}, {\it B}, and {\it V} filter functions
from Ma\'{i}z Appel\'{a}niz (2006) as described in J\o rgensen et al.\ (2018a).
The calibration is given in Appendix \ref{SEC-IMAGING}.

\subsection{Spectroscopy of Lynx E and W \label{SEC-SPEC}}

Spectroscopy of Lynx E and W galaxies was obtained with the Gemini Multi-Object Spectrograph (GMOS-N) on 
Gemini North, see Hook et al.\ (2004) for a description of GMOS-N. 
We used GMOS-N in the multi-object spectroscopic (MOS) mode.
The observations centered on Lynx W are described in J\o rgensen et al.\ (2014). 
Similar observations were obtained for a sample centered on Lynx E in the period UT 2014 Nov 27 to 2015 Jan 12 
under the Gemini programs GN-2014B-Q-22 and GN-2014B-DD-4. 
In both cases we selected galaxies to maximize coverage along the red sequence from the 
brightest cluster galaxy (BCG) to  $z_{\rm tot,850} \approx 24.6$ mag.
Highest priority was given to galaxies within 0.1 mag of the red sequence in $(i_{775}-z_{850})$ versus $z_{\rm tot,850}$.
The red sequence is located at $(i_{775}-z_{850}) \approx 0.9$.
Additional space in the mask was filled with galaxies with $(i_{775}-z_{850})>0.5$ and $z_{\rm tot,850}$ 
in the interval from 21 mag to 25 mag. Some of these turned out to be blue cluster members.
The spectroscopic sample for both Lynx E and W is marked in Figure \ref{fig-lynxgrey}.
The properties of the resulting sample of member galaxies are discussed in Section \ref{SEC-SAMPLE},
where we also detail the final selection of the bulge-dominated galaxies included
in our analysis.

The data processing and determination of spectral parameters follow the methods used in
J\o rgensen et al.\ (2014). Details of the instrumentation and processing are available in Appendix \ref{SEC-SPECTROSCOPY}.
All spectroscopic observations cover the rest frame wavelength range $3600-4150$ {\AA}.
The new data cover fifteen Lynx member galaxies, and have a median signal-to-noise per {\AA} in the rest frame of 24.8.
We use the spectra to derive redshifts, galaxy velocity dispersions and the absorption line indices
CN3883, CaHK, D4000, and H$\zeta _{\rm A}$. 
For galaxies with detectable [\ion{O}{2}] emission the strength of the emission line was measured. 
The measured parameters are available in Table \ref{tab-spec} in Appendix \ref{SEC-SPECTROSCOPY}.
In our analysis we use the redshifts for determining cluster membership and cluster velocity dispersions.
We then focus our analysis on the velocity dispersions, CN3883, H$\zeta _{\rm A}$ and D4000 together 
with the photometric parameters described in the previous section.

\subsection{The Low Redshift Reference Sample and the $z=0.2-0.9$ GCP Clusters \label{SEC-COMPDATA} }

As our low redshift reference sample, we use the bulge-dominated galaxies in the Perseus and Coma clusters 
included in our consistently calibrated spectroscopy from J\o rgensen et al.\ (2018b).
We adopt and calibrate photometry from the Sloan Digital Sky Survey (SDSS) for these galaxies.
The SDSS photometry was calibrated to consistency with our {\it Legacy Data} used in previous GCP papers 
(e.g., J\o rgensen \& Chiboucas 2013, J\o rgensen et al.\ 2014),
and pseudo-S\'{e}rsic radii were derived from the available SDSS size measurements.
Appendix \ref{SEC-LOWZDETAIL} contains additional detail, including comparisons showing
that the  pseudo-S\'{e}rsic radii can be considered equivalent to S\'{e}rsic radii derived
from 2-dimensional profile fits.
In the following, we simply refer to the radii as S\'{e}rsic radii.
The calibrated data will be published in a future paper (I.\ J\o rgensen et al.\ 2019 in prep.).

In the analysis we also use our previous results for the $z=0.2-0.9$ GCP clusters (J\o rgensen \& Chiboucas 2013;
J\o rgensen et al.\ 2017). In particular, we include in the analysis the data for the two highest 
redshift GCP clusters from these papers: RXJ0152.7--1357 at $z=0.83$ and RXJ1226.9+3336 at $z=0.89$.
RXJ0152.7--1357 is a binary cluster most likely in the process of merging (Maughan et al.\ 2003; Jones et al.\ 2004;
Demarco et al.\ 2005; Girardi et al.\ 2005). 
The distance between the centers of the sub-clusters is 0.68 Mpc in the plane of the sky. 
This is significantly smaller than the $\approx 2.5$ Mpc distance in the plane of the sky between the Lynx E and W clusters.  
The masses of each of the components are $M_{500}=1.6\times 10^{14} M_{\sun}$  and $M_{500}=2.0\times 10^{14} M_{\sun}$
for the northern and southern sub-cluster, respectively, similar to the mass of Lynx E. 
Recalibrated values of $M_{500}$ for the RXJ0152.7--1357 sub-clusters are from J\o rgensen et al.\ (2018a).
RXJ1226.9+3336 is a massive cluster, $M_{500}= 4.4 \times 10^{14} M_{\sun}$, with minimal substructure though
it may have experienced a recent merger event (Maughan et al. 2007). 

\begin{figure}
\epsfxsize 8.5cm
\epsfbox{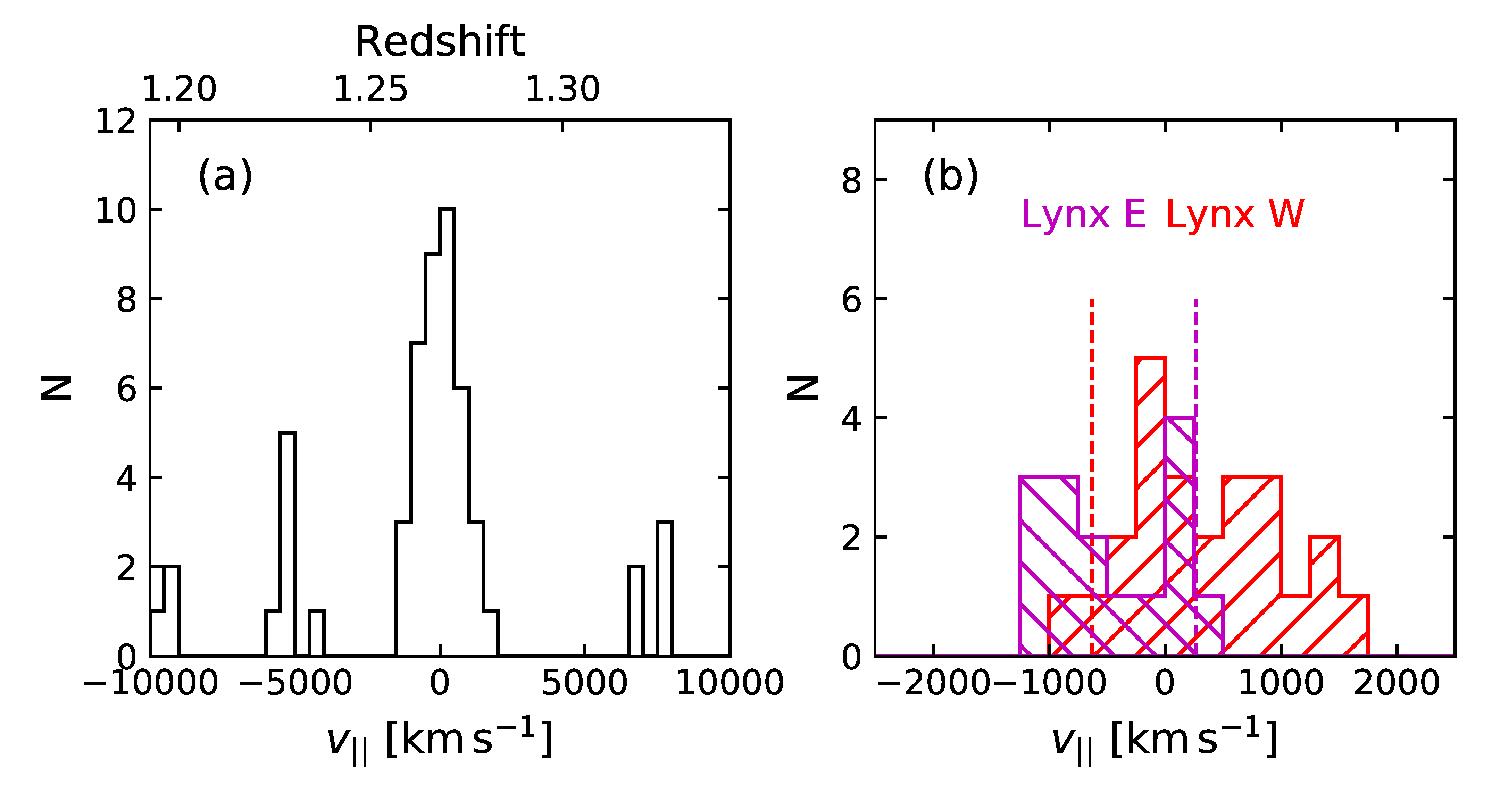}
\caption{Distribution of the radial velocities (in the rest frame) 
relative to the median redshift for all cluster members of Lynx E and W, 
$v_{\|} = c (z - z_{\rm cluster}) / (1+z_{\rm cluster})$.
For this purpose we adopt a common cluster redshift of $z_{\rm cluster}=1.265$.
(a) All spectroscopic data within $\pm 10,000\,{\rm km\,s^{-1}}$ of the common cluster redshift.
(b) Distributions shown with Lynx E and W members color coded.
Magenta histogram -- galaxies considered members of Lynx E; red histogram -- galaxies considered members of Lynx W.
The color-coded vertical lines mark the median radial velocities of the two clusters.
\label{fig-zhist} }
\end{figure}

\begin{deluxetable*}{lcrc crc rrr}
\tablecaption{Cluster Properties\label{tab-redshifts} }
\tablewidth{0pt}
\tabletypesize{\scriptsize}
\tablehead{
\colhead{Cluster} & \multicolumn{3}{c}{BCGs as centers} & \multicolumn{3}{c}{X-ray centers} \\
                  & \colhead{Redshift} & \colhead{$\sigma _{\rm cluster}$} & \colhead{N$_{\rm member}$}
                  & \colhead{Redshift} & \colhead{$\sigma _{\rm cluster}$} & \colhead{N$_{\rm member}$} 
                  & \colhead{$L_{500}$} & \colhead{$M_{500}$} & \colhead{$R_{500}$}  \\
 & & $\rm km~s^{-1}$ & & & $\rm km~s^{-1}$ & & $10^{44} \rm{erg\,s^{-1}}$ & $10^{14}M_{\sun}$ & Mpc \\
\colhead{(1)} & \colhead{(2)} & \colhead{(3)} & \colhead{(4)} & \colhead{(5)} & \colhead{(6)} & \colhead{(7)} & \colhead{(8)} & \colhead{(9)} & \colhead{(10)}
}
\startdata
Lynx E    & $1.2642$ & $568_{-90}^{+68}$ & 15 & $1.2646$ & $580_{-77}^{+65}$ & 16 & 3.73   & 1.65 & 0.518 \\ 
Lynx W    & $1.2706$ & $693_{-87}^{+67}$ & 24 & $1.2706$ & $706_{-79}^{+83}$ & 23 & 0.63   & 0.55 & 0.359 \\
\enddata
\tablecomments{Column 1: Galaxy cluster. Column 2: Cluster redshift, adopting BCGs as cluster centers. 
The uncertainties on the redshifts are 0.001. Column 3: Cluster velocity dispersion, adopting BCGs as cluster centers.
Column 4: Number of member galaxies for which spectroscopy is available, adopting BCGs as cluster centers.
Columns 5--7: Cluster redshift, velocity dispersion, and number of member galaxies, when adopting X-ray centers.
Column 8: X-ray luminosity in the 0.1--2.4 keV band within the radius $R_{500}$. 
Column 9: Cluster mass derived from X-ray data within the radius $R_{500}$.
Column 10: Radius within which the mean over-density of the cluster is 500 times the critical density at the 
cluster redshift.
}
\end{deluxetable*}

\begin{figure}
\epsfxsize 8.5cm
\epsfbox{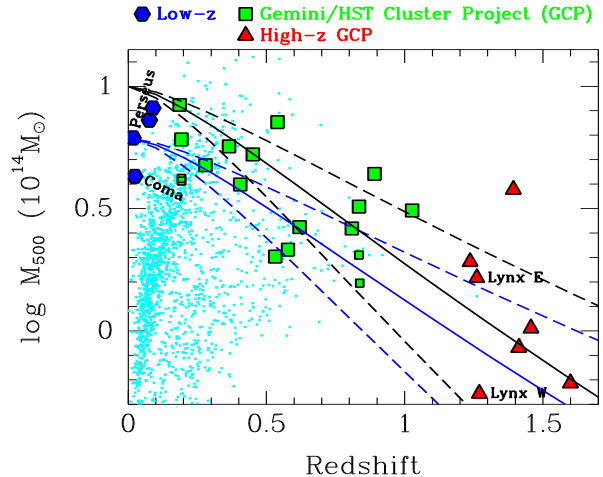}
\caption{The cluster masses, $M_{\rm 500}$, based on X-ray data versus the redshifts of the clusters.
The figure represents an expanded version of similar figures in J\o rgensen et al.\ (2018ab). 
Blue -- the low redshift reference sample, Perseus and Coma used in the present paper are labeled; 
green -- the $z=0.2-1$ GCP cluster sample,
$M_{\rm 500}$ values are from Piffaretti et al.\ (2011) or recalibrated as described in J\o rgensen et al.\ (2018a).
The smaller green points show the data for the components of the binary clusters RXJ0056.2+2622 ($z=0.19$) and 
RXJ0152.7--1357 ($z=0.83$).
Red -- the extension of GCP covering $z=1.2-1.6$ clusters, Lynx E and W are labeled.
The X-ray data for the $z=1.2-1.6$ clusters have been recalibrated to consistency with Piffaretti et al.\
using the offsets derived in J\o rgensen et al.\ (2018a).
Small cyan points -- all clusters from Piffaretti et al.\ shown for reference.
Blue and black lines -- mass development of clusters based on numerical simulations
by van den Bosch (2002).
The black lines terminate at Mass=$10^{15} M_{\sun}$ at $z=0$ roughly
matching the highest mass clusters at $z=0.1-0.2$.
The blue lines terminate at Mass=$10^{14.8} M_{\sun}$ at $z=0$
matching the mass of the Perseus cluster.
The dashed lines represent the typical uncertainty in the
mass development represented by the numerical simulations.
\label{fig-M500} }
\end{figure}

\begin{figure}
\epsfxsize 8.5cm
\epsfbox{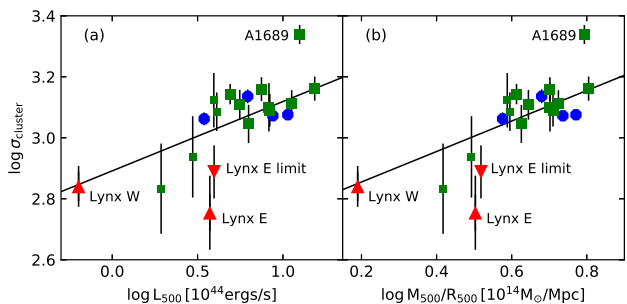}
\caption{ 
Cluster velocity dispersion versus (a) the X-ray luminosity $L_{\rm 500}$ and (b) $M _{\rm 500} \, R _{\rm 500}^{-1}$.
Symbols as in Figure \ref{fig-M500}, with Lynx E and W labeled. 
The point labeled ``Lynx E limit'' shows the cluster velocity dispersion if treating all spectroscopic 
members as members of one cluster, see text. 
For clarity, the point is slightly offset from Lynx E in the X-direction on both panels.
The error bars on the symbols for the Lynx clusters include the systematic uncertainties.
Solid lines show the relations $L \propto \sigma _{\rm cluster}^{4.4}$ (panel a) with the slope adopted from
Mahdavi \& Geller (2001), and 
$M_{\rm 500}\,R_{\rm 500}^{-1} \propto \sigma _{\rm cluster}^{2}$ (panel b).
In both cases, the median zero points are derived from the data for the low redshift reference clusters and the $z \le 1$ GCP clusters, 
excluding Abell 1689 and the binary clusters. 
\label{fig-lsigma_cluster} }
\end{figure}

\section{Cluster Properties\label{SEC-CLUSTERZ}}

In order to determine the cluster redshifts and velocity dispersions, we first examine the
distribution of the galaxies in velocity space. Figure \ref{fig-zhist}a shows the relative radial velocity distribution 
of the spectroscopic sample within $\pm 10,000\, {\rm km\,s^{-1}}$ of the median redshift of the potential cluster members.
The Lynx clusters are well-isolated in velocity space, with a well-defined peak of galaxies within $\pm 2,000 {\rm km\,s^{-1}}$ of 
the median redshift. 
The galaxies present at relative radial velocities of $-4,400\, {\rm km\,s^{-1}}$ and $6,000\, {\rm km\,s^{-1}}$, 
respectively, are at cluster center distances larger than $1.5 R_{500}$ and therefore
unlikely to be gravitationally bound to the clusters. 
Based on this, we limit the possible cluster members in redshift space to $z=1.253-1.283$.

In order to assign the galaxies to either Lynx W or E we need to adopt cluster centers.
We consider (1) the brightest cluster galaxies (BCG), (2) the X-ray centers, or (3) the centroid of 
positions of member galaxies. The BCGs are the triple-core galaxy in Lynx W and ID 4942 in Lynx E.
These and the X-ray centers from Rosati et al.\ (1999) are marked on Figure \ref{fig-lynxgrey}. 
Due to limitations of the MOS mode used for the observations, our spectroscopic sample is too sparse 
in the centers of the clusters to give reliable centroids of the two
clusters. Thus, to derive centroids from member galaxy positions we 
include both our spectroscopic sample and galaxies within three sigma of the red sequence and 
$z_{\rm 850}\le 24.1$ mag. We then proceed to assign each 
possible member to either Lynx W or Lynx E, depending on its closest angular distance to the centers
of the two clusters. 
It turns out that using the centroids as cluster centers results in the same cluster assignments
as if using the BCGs.
If we use X-ray centers instead of the BCGs, ID 2653 is assigned
to Lynx E instead of Lynx W. All other cluster assignments are unchanged. 
We proceed using the BCGs as the cluster centers, while in the following commenting on 
to what extent our results depend on this choice.

We determine the velocity dispersions using the bi-weight method (Beers et al.\ 1990).
The resulting median redshifts and velocity dispersion are listed in Table \ref{tab-redshifts}.
The uncertainties are derived using a boot-strap method as detailed in Beers et al.
Determinations are given for both BCGs as cluster centers and X-ray centers.
Figure \ref{fig-zhist}b highlights
the radial velocity distributions for the two clusters relative to the median redshift of members
of both clusters, 
see Table \ref{tab-spec} in Appendix \ref{SEC-SPECTROSCOPY} for membership information of the individual galaxies
in the spectroscopy sample.
One could argue that it would be better to assign membership based on the closest distance 
to the cluster centers in units of the cluster radii $R_{500}$. 
($R_{500}$ is the radius within which the mean over-density of the cluster is 500 times the critical density at the 
cluster redshift.)
If we take that approach, three
galaxies (IDs 2653, 2655, and 2757) would be assigned to Lynx E, for both choices of cluster centers.
However,
this has no significant effect on the derived cluster redshifts or velocity dispersions.
Specifically, for this alternative membership assignment we find velocity dispersions of $646\, {\rm km\,s^{-1}}$ and 
$675\, {\rm km\,s^{-1}}$ for Lynx E and Lynx W, respectively.
We note that Ettori et al.\ (2009) found a size difference of the clusters of only a factor 1.1, 
while Pascut \& Ponman (2015) found a factor 1.9 difference.  
Because of the considerable uncertainty in the $R_{500}$ measurements from the literature
we choose to proceed with memberships assigned using the closest angular distance to the cluster
centers.
Formally, the uncertainties on the cluster velocity dispersions are $70-90\, {\rm km\,s^{-1}}$, see
Table \ref{tab-redshifts}. The possible uncertainty in the cluster assignments contributes systematic uncertainties
of $30\, {\rm km\,s^{-1}}$ for Lynx W and $80\, {\rm km\,s^{-1}}$ for Lynx E.

In Figure \ref{fig-M500}, we show the cluster masses versus redshifts for the clusters,
together with our GCP cluster sample at $z=0.2-1.0$, the low redshift reference clusters, 
the $z=1.2-1.6$ GCP extension, and for reference the catalog of $z<1$ clusters from Piffaretti et al.\ (2011).
The X-ray data for the GCP clusters have been calibrated to consistency with Piffaretti et al.\ (2011),
see J\o rgensen et al.\ (2018a). The same calibrations are used to bring the
literature data for the $z=1.2-1.6$ GCP clusters, including Lynx E and W, to the same system.
The Lynx X-ray data are from Ettori et al.\ (2004, 2009), Stott et al.\ (2010, Lynx E only), and Pascut \& Ponman (2015).
Table \ref{tab-redshifts} lists the average values of the recalibrated X-ray luminosities, radii and masses 
for Lynx E and W.

\begin{figure}
\epsfxsize 8.5cm
\epsfbox{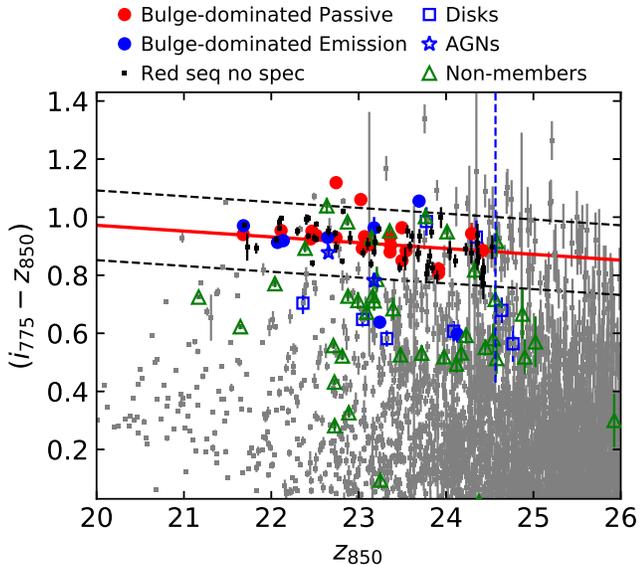}
\caption{
Color-magnitude diagram of galaxies in the Lynx cluster fields. 
Red circles -- bulge-dominated members of the cluster with EW[\ion{O}{2}]$\le 5${\AA}. 
Blue circles -- bulge-dominated members with EW[\ion{O}{2}] $> 5${\AA}.
Small black points -- bulge-dominated galaxies within $3\sigma$ of the red sequence 
and $z_{\rm tot,850} \le 24.6$, but without spectroscopic information.
Blue squares -- disk-dominated members.
Blue stars -- two AGNs, which are members of the clusters.
Green open triangles -- confirmed non-members; 
small black points -- bulge-dominated galaxies within $3\sigma$ of the red sequence,
Small gray points -- all other galaxies in the field.
Red line is the best fit to the 27 member galaxies with $(i_{775}-z_{850})\ge0.8$: 
$(i_{775}-z_{850}) = (0.924 \pm 0.010) - (0.020 \pm 0.010) ( z_{\rm tot,850} - 22.5)$ with rms=0.040. 
Dashed black lines are offset $\pm 3\sigma$ from the best fit lines. 
Vertical blue line shows the magnitude limit for the primary selection of spectroscopic targets.
\label{fig-CM} }
\end{figure}

The mass, $M_{500}$, of Lynx W is significantly lower than that of Lynx E, and of the $z=0.2-1.0$ GCP clusters,
while Lynx E has a mass similar to the lowest mass $z=0.2-1.0$ GCP clusters.
However, due to the expected growth of cluster masses with time (see sample models in Figure \ref{fig-M500}), 
both clusters are viable progenitors for clusters of masses similar to those of Coma and Perseus at $z\approx 0$. 

In Figure \ref{fig-lsigma_cluster}, we show the cluster velocity dispersions versus the
X-ray luminosity and the combination of X-ray masses and radii. The figure includes available data
for the GCP clusters and the four low redshift reference clusters (Perseus, Coma, A2029, and A2142, see J\o rgensen et al.\ 2018b). 
Lynx W falls on the low luminosity and mass extension of the relations for the more massive GCP  
clusters, while Lynx E falls significantly below the relations as its velocity dispersion
is much smaller than expected from its X-ray properties. 
It is not clear whether this discrepancy is due to a misestimate
of the X-ray properties, given the rather low S/N of the X-ray observations, or selection effects
in our spectroscopic sample artificially leading to a non-representative low cluster velocity for the cluster.
We inspected the {\it Chandra} X-ray imaging of the clusters. There are no X-ray point sources very close to the center
of Lynx E, thus no reason to expect that the cluster's X-ray luminosity has been overestimated due to contamination by
point sources.
Jee et al.\ (2006) and Stanford et al.\ (2001) give velocity dispersion for Lynx E and W  of $740^{+113}_{-134}\, {\rm km\,s^{-1}}$
and $650\pm 170\,{\rm km\,s^{-1}}$, respectively. The result from Jee et al.\ is based on weak lensing analysis.
Within the uncertainties these results agree with ours.
If we treat all spectroscopic members as members of one cluster, we find a cluster velocity dispersion of 
$773^{+60}_{-90}\, {\rm km\,s^{-1}}$, which can be considered an upper limit on the cluster velocity dispersion supported by our data.
Figure \ref{fig-lsigma_cluster} shows this as the point labeled ``Lynx E limit''.

\begin{deluxetable*}{ll rl rl}
\tablecaption{Sub-Samples\label{tab-sample} }
\tablewidth{0pc}
\tabletypesize{\scriptsize}
\tablehead{
\colhead{No.} & \colhead{Criteria} & \multicolumn{2}{c}{Lynx E} & \multicolumn{2}{c}{Lynx W} \\
              &                    & \colhead{N} & \colhead{IDs} &  \colhead{N} & \colhead{IDs} 
}
\startdata
1  &  Only redshift \& EW[\ion{O}{2}] from spectra & 
   & & 
  4 & 129 197 2655 2834  \\
2  &  $n_{\rm ser}<1.5$, $\log \sigma$ meas. & 
  & & 4 & 145 503 691 707  \\
3  &  $n_{\rm ser}\ge1.5$, $\log \sigma$ meas., S/N$<$10 & 
   1 & 4921 &
   3 & 254 545 1036  \\
4  &  $n_{\rm ser}\ge1.5$, $\log \sigma$ meas., S/N$\ge$10, EW[\ion{O}{2}]$>$5{\AA}  & 
  3 & 2098 4093 4828 &
  5 & 209 293 365 2653 2953  \\
5  &  $n_{\rm ser}\ge1.5$, $\log \sigma$ meas., S/N$\ge$10, EW[\ion{O}{2}]$\le$5{\AA} & 
  10 & 2309 2416 4095 4363 4593 4860 4926 4942 5009 5037 & 
   7 & 148 392 454 886 1134 2757 2849   \\
6  & $n_{\rm ser}\ge1.5$, $\log \sigma$ meas., S/N$\ge$10, AGN & 
   1 & 4508 &
   1 & 316  \\
\enddata
\tablecomments{Samples 4 and 5 require log Mass $\ge$ 10.3. 
The lower mass $n_{\rm ser}\ge1.5$ galaxy ID 4921 is listed under sample 3.
ID 1036 in sample 3 does not have $\log \sigma$ measured and has no significant [\ion{O}{2}] emission.}
\end{deluxetable*}

To investigate the issue of the cluster masses further, we assess the cluster richness from the number of 
bulge-dominated galaxies on the red sequence within $R_{500}$ for each cluster.
The ratio of the X-ray masses $M_{500}$ is approximately 3:1 with Lynx E being three times as massive as Lynx W.
Figure \ref{fig-CM} shows the color-magnitude (CM) relation for the full field of Lynx E and W covered
by the {\it HST} imaging.
The luminosities of the galaxies in the two clusters
including only galaxies within $R_{500}$, brighter than the limit for our spectroscopic sample, 
and within 3$\sigma$ of the red sequence gives a total luminosity of such galaxies a factor 2.5 
larger for Lynx E than for Lynx W. 
The choice of including only galaxies out to $R_{500}$ is not critical.
If we instead include galaxies to the Lynx E $R_{200} = 1.52 R_{500}$ distance in both clusters, the ratio is two.
Ultimately, a larger spectroscopic sample of cluster members and deeper X-ray observations would be 
needed to provide more secure determinations of the cluster velocity dispersions and the cluster X-ray masses.
For our purpose it is sufficient to know that, assuming that the luminosity ratio traces the mass ratio, 
then Lynx E is at least twice, possibly three times, as massive as Lynx W.
However, see also the analysis by Jee et al.\ (2006), which shows weak lensing masses
of similar sizes for the two clusters, and X-ray mass estimates supporting that Lynx E is 3-4 times
more massive than Lynx W.

\section{The Methods and the Models \label{SEC-METHOD} }

Our analysis focuses on the bulge-dominated galaxies in the two clusters, with the main emphasis on
the possible structure evolution and testing whether passive evolution of the stellar populations
from $z=1.27$ to the present is sufficient to reach properties consistent with our 
low redshift reference sample.
We also address possible differences between the galaxy populations in the two clusters. 

To parameterize the properties of the galaxies we use
(1) the Fundamental Plane (Dressler et al.\ 1987; Djorgovski \& Davis 1987; J\o rgensen et al.\ 1996), 
and the relations between masses, sizes and velocity dispersions; and
(2) the absorption lines as function of galaxy velocity dispersion. 
The zero points of the relations are derived separately for Lynx E and W.
Finally, we investigate high S/N composite spectra of galaxy populations to further 
establish possible differences between stellar populations as a function of cluster environment.

In the analysis involving the photometry and structural parameters, we show results based on 
parameters from the fits with S\'{e}rsic profiles. 
None of our results change significantly, if we had instead used parameters from the fits with $r^{1/4}$ profiles,
and none of the conclusions depend on the choice of profile fitting. 

We adopt the same method for establishing the scaling relations and associated uncertainties on
slopes and zero points as we used in 
J\o rgensen et al.\ (2014, 2017), see also J\o rgensen et al.\ (1996).
The fits are determined by minimizing the sum of the absolute residuals.
Unless otherwise noted, the minimization is done perpendicular to the relations
and the zero points are median zero points. The uncertainties are derived using
a boot-strap method. 

In our analysis we use dynamical masses of the galaxies. 
We adopt ${\rm Mass_{dyn}} = \beta r_e \sigma^2\,G^{-1}$, with $\beta =5$ (Bender et al.\ 1992).
Results from Cappellari et al.\ (2006) show that the approximation gives
a reasonable mass estimate. However, see also discussion in Cappellari et al.\ (2013b) regarding
the possible merits of using a value of $\beta$ dependent on the S\'{e}rsic index $n_{\rm ser}$.
Our results do not depend significantly on whether we use $\beta =5$, or adopt the 
expression for $\beta (n_{\rm ser})$ from Cappellari et al.\ (2006).
Appendix \ref{SEC-BETATEST} contains a comparison of relations and zero point offsets
when using the two different methods of mass determination.

In the analysis we use single stellar population (SSP) models from 
Maraston \& Str\"{o}mb\"{a}ck (2011) for a Salpeter (1955) initial mass function (IMF).
These authors provide spectral energy distributions (SED) for the models. 
As described in J\o rgensen et al.\ (2014), we establish model values
for the indices CN3883 and H$\zeta _A$ from the SEDs.
In addition we use M/L ratios for very similar models from Maraston (2005).
We note that for the rest-frame wavelength interval and ages of interest for our analysis,
the SEDs from Maraston \& Str\"{o}mb\"{a}ck (2011) are almost identical 
those from Vazkedis et al.\ (2010) for solar abundance ratio models. 
The main difference between the two sets of models is the inclusion of contributions from 
thermally pulsating asymptotic giant branch (TP-AGB) stars in the Maraston \& Str\"{o}mb\"{a}ck models,
while these stars are not included in the models from Vazkedis et al. 
However, the TP-AGB stars do not contribute significantly to the flux in the rest-frame wavelength interval
of interest here, even for the $\approx 1$Gyr old stellar populations relevant for our analysis.

J\o rgensen et al.\ (2014) list linear model relations between measurable
parameters and the logarithm of the age and the metallicity [M/H], 
valid for the Maraston \& Str\"{o}mb\"{a}ck (2011) models. 
As in that paper, we use these relations to aid our analysis. 
In particular, we derive the expected changes of the measurable parameters with redshift
under the simple assumption of passive evolution of the stellar populations.
In these models, it is assumed that after an initial period of star formation
the galaxies evolve passively without any additional star formation. 
The models are usually parameterized by a formation
redshift $z_{\rm form}$, which corresponds to the approximate epoch of the last 
major star formation episode.
In particular, if the Lynx galaxies and the low redshift reference sample share a common formation epoch, 
at a lookback time of $t_{\rm form}$, 
we can then derive that from their age difference in log space, $\Delta \log {\rm age}$,
\begin{equation}
t_{\rm form} = \frac {\left ( t_{\rm lookback,Lynx} 10^{\Delta \log {\rm age}} - t_{\rm lookback,low-z} \right )}{( 10^{\Delta \log {\rm age}}-1 ) }
\end{equation}
where $t_{\rm lookback,Lynx}$ and $t_{\rm lookback,low-z}$ are the lookback times for 
the redshifts of the Lynx sample and the low redshift reference sample, respectively.  
With the aid of models, the age difference may be determined from the difference in M/L ratios or line strengths.
Using our adopted cosmology, $t_{\rm form}$ can be converted to the formation redshift $z_{\rm form}$.

Finally, we use the parameterization of structural changes due to mergers established
in Bezanson et al.\ (2009). Specifically, for minor mergers
\begin{equation}
\Delta \log r_e = 2\, \Delta \log {\rm Mass} = -4\, \Delta \log \sigma
\end{equation}
where $r_e$ is the half-light radius, and $\sigma$ is the velocity dispersion.
For major mergers  
\begin{equation}
\Delta \log r_e = \Delta \log {\rm Mass}
\end{equation}
while the velocity dispersion is unchanged.
As done by Bezanson et al.\ (2009), we assume that the mergers do not involve star formation, 
ie.\ they are dry mergers.

In the absence of mergers, we implicitly assume that the galaxies we observe
in the Lynx clusters can be considered progenitors to the galaxies in the low redshift
reference sample. However, the low redshift reference sample contains galaxies with 
stellar populations younger than $\approx 9$ Gyr and are therefore too young to have been passive at $z=1.27$
(J\o rgensen et al.\ 2018b). 
This progenitor bias is discussed in detail by van Dokkum \& Franx (2001).
We also consider progenitor bias for our investigation of the structure evolution
originating from newly quenched galaxies being added to the sample of passive galaxies since $z=1.27$ 
and that such galaxies may be larger than the older passive galaxies, see Belli et al.\ (2015). 
We return to both effects of progenitor bias in the discussion (Section \ref{SEC-DISCUSSION}).

\begin{figure}
\epsfxsize 8.5cm
\epsfbox{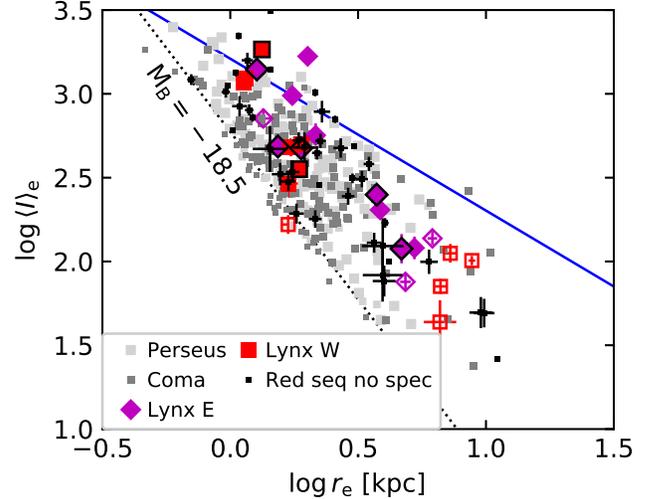}
\caption{
Effective radii versus mean surface brightnesses for the low redshift reference sample and the 
Lynx E and W samples of bulge-dominated galaxies. 
Light gray -- Perseus cluster members; dark gray -- Coma cluster members;  
red boxes -- Lynx W; magenta diamonds -- Lynx E.
Open symbols for Lynx E and W -- galaxies with EW[\ion{O}{2}]$>$5{\AA}.
Symbols with black edges -- Lynx E and W passive galaxies within $R_{500}$ of the cluster centers.
Small black points -- bulge-dominated Lynx E and W galaxies with $z_{850}\le 24.2$, within $3\sigma$
of the red sequence, and without spectroscopic observations.
The data for the Lynx sample have been offset by 
$\Delta \log \langle I \rangle _{\rm e} = -0.75$ to take into account the luminosity offset 
relative to the low redshift reference sample.
After applying this offset, the distribution in the $\log r_{\rm e}$--$\log \langle I \rangle _{\rm e}$ space
of the Lynx spectroscopic sample, as well as the full Lynx sample, resembles that of the 
low redshift reference sample.
\label{fig-lre_lie} }
\end{figure}

\begin{figure}
\begin{center}
\epsfxsize 8.5cm
\epsfbox{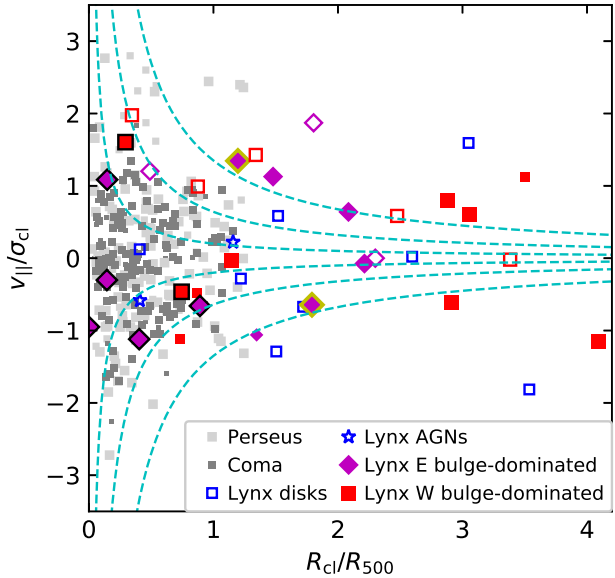}
\end{center}
\caption{
The distribution of the Lynx galaxies in the phase space diagram of line-of-sight velocities relative to
the cluster velocity dispersions, $v_{||}/\sigma _{\rm cl}$, versus the cluster center distances, $R_{\rm cl}$, in units of $R_{500}$.
Large magenta diamonds -- Lynx E bulge-dominated galaxies, galaxies with EW[\ion{O}{2}]$>5$ {\AA} are shown as open points. 
Symbols with yellow edges mark the two post-starburst galaxies ID 2416 and 4593.
Large red squares -- Lynx W  bulge-dominated galaxies, galaxies with EW[\ion{O}{2}]$>5$ {\AA} are shown as open points.
Symbols with black edges show the passive bulge-dominated galaxies within $R_{500}$ of the cluster centers.
Small magenta diamonds -- Lynx E bulge-dominated galaxies for which the spectra have S/N $< 10$.
Small red squares -- Lynx W  bulge-dominated galaxies for which the spectra have S/N $< 10$.
Blue open squares -- disk galaxies including galaxies for which only EW[\ion{O}{2}]$>5$ {\AA} can be measured. 
Blue stars -- AGNs.
The Perseus and Coma cluster members in our low redshift reference sample are shown as light gray and dark gray points, respectively.
Cyan lines show caustics with constant $v_{||}/\sigma _{\rm cl} \times R_{\rm cl} / R_{500}$ values of 0.2, 0.64, and 1.35
(Noble et al.\ 2015), see text.
\label{fig-phasespace} }
\end{figure}

\section{Samples for Analysis \label{SEC-SAMPLE} }

In order to ensure consistency between final sample selection for our Lynx analysis in the present paper and 
the samples used in our analysis of the $z=0.2-0.9$ GCP clusters we proceed as follows.
Following J\o rgensen et al.\ (2014), we divide the cluster members into sub-samples according to available 
spectroscopic parameters, S/N,  S\'{e}rsic index and the strength of the [\ion{O}{2}] emission.
As noted above we assign the galaxies to either Lynx E and Lynx W, depending on their proximity
to the adopted cluster centers.
The samples are listed in Table \ref{tab-sample}.
Our main sample consists of sub-samples (4) and (5), which are the bulge-dominated
galaxies with EW[\ion{O}{2}]$> 5${\AA} and $\le 5${\AA}, respectively, and spectroscopic S/N $\ge 10 {\rm \AA}^{-1}$
in the rest frame of the galaxies. 
The galaxies with EW[\ion{O}{2}]$\le 5${\AA} are considered passive galaxies and in the following referred to as such.
The limit in S/N is equivalent to uncertainties on the velocity dispersions
of $\le 0.125$ dex (see Appendix \ref{SEC-SPECTROSCOPY}). The galaxies have total magnitudes of $z_{850}\le 24.2$. 
Two galaxies, IDs 316 
and 4508, listed as sample (6) have significant emission in the high excitation neon lines 
[\ion{Ne}{5}]3426{\AA} and [\ion{Ne}{3}]3869{\AA}, most likely originating from active galactic nuclei (AGNs) 
(Schmidt et al.\ 1998; Mignoli et al.\ 2013). ID 4508 also coincides with a {\it Chandra} X-ray point-source.
These two galaxies are excluded from the analysis.

\begin{figure*}
\begin{center}
\epsfxsize 17cm
\epsfbox{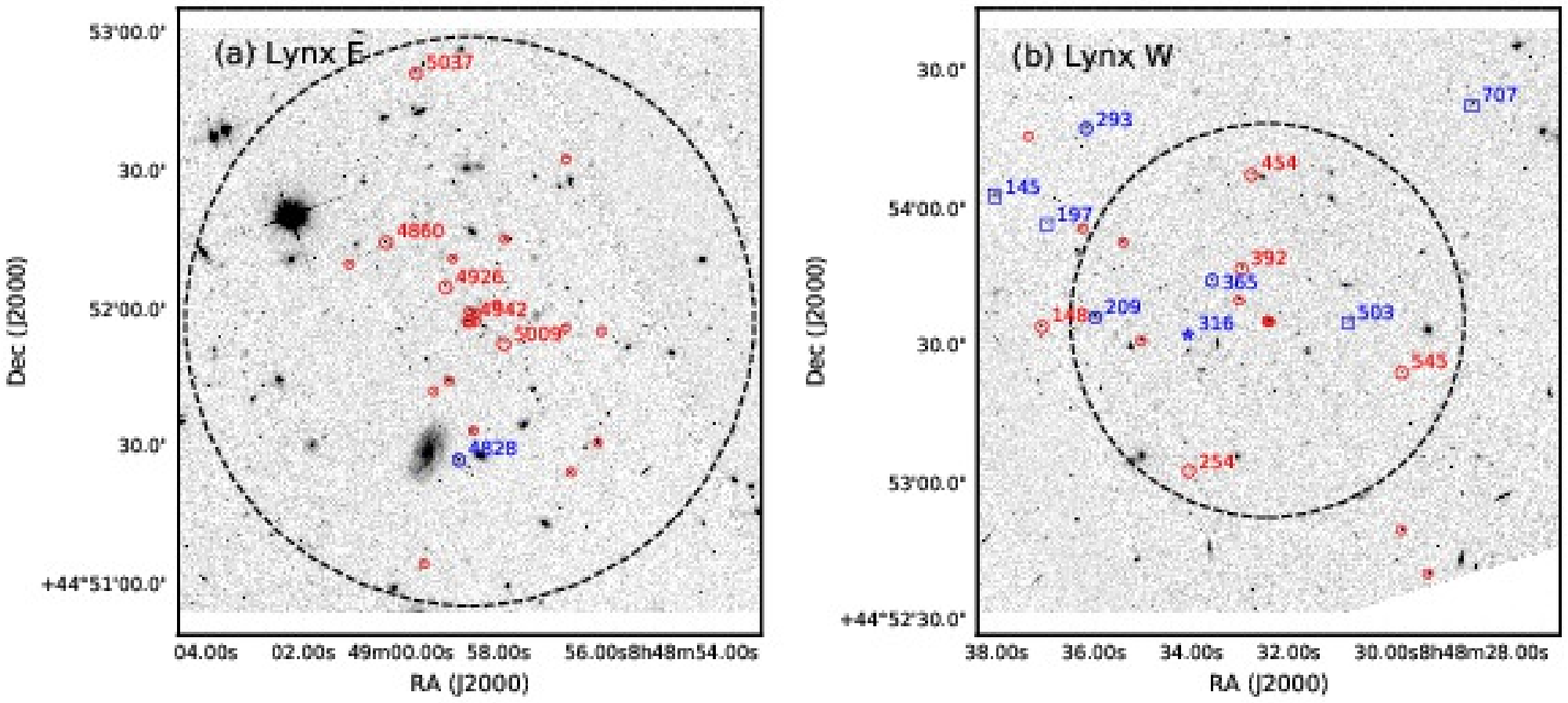}
\end{center}
\caption{
Grayscale images of the Lynx cluster cores made from the {\it HST}/F850LP images.
Red circles labeled with IDs  -- bulge-dominated members with EW[\ion{O}{2}]$\le 5${\AA}. 
Blue circles -- bulge-dominated members with EW[\ion{O}{2}] $> 5${\AA}.
Blue squares -- disk galaxies.
Blue star -- AGN in Lynx W.
Small red circles without IDs -- bulge-dominated galaxies within $3\sigma$ of the red sequence for the clusters 
and brighter than $z_{850}=24.6$ mag but not included in our spectroscopic sample.
Dashed circles mark $R_{500}$ for each of the clusters.
\label{fig-lynxgrayscale_subimages} }
\end{figure*}

Except for allowing galaxies with lower S/N spectra as part of the analysis, 
the selection criteria for sample (5) are the same as used for our analysis of $z=0.2-0.9$ GCP clusters 
(J\o rgensen \& Chiboucas 2013; J\o rgensen et al.\ 2017). 
The bulge-dominated galaxies with EW[\ion{O}{2}] $> 5${\AA} (sample 4) 
are included in the present analysis, as we assume that their fairly weak emission originates from the now decreasing
star formation rather than on-going low level star-formation. We want to investigate if these galaxies therefore
can passively evolve into galaxies similar to the low redshift reference sample galaxies.
In the following, unless explicitly stated otherwise the figures show only the Lynx galaxies from
samples (4) and (5).

For bulge-dominated galaxies within 0.12 mag ($3\sigma$) of the  red sequence, the spectroscopic sample is 41\% complete to $z_{850} = 24.2$, 
the magnitude of the faintest galaxies included in the analysis. 
Comparing the Lynx sample of bulge-dominated galaxies with spectroscopy to the full sample along the red sequence,
we find that a Kolmogorov-Smirnov test gives a $\approx 30$\% probability that the two samples are drawn 
from the same parent distribution in $z_{850}$. 
In addition, if we offset the Lynx sample of bulge-dominated galaxies with spectroscopy
by $\Delta \log L =-0.75$ to account for the average luminosity evolution, the luminosity distribution
of the offset sample and that of the low redshift reference sample are in agreement with being
drawn from the same parent sample.

To evaluate if our sample is biased in sizes or surface brightnesses, Figure \ref{fig-lre_lie} 
shows these two parameters, $\log r_{\rm e}$ and $\log \langle I \rangle _{\rm e}$, versus each other
for the bulge-dominated galaxies in our spectroscopic Lynx samples (4) and (5). 
The figure also includes bulge-dominated galaxies with $z_{850}\le 24.2$ mag and within 
0.12 mag ($3\sigma$) of the red sequence for Lynx, but without spectroscopy.
The Lynx data have been offset by $\Delta \log \langle I \rangle _{\rm e} = -0.75$ to take 
into account the luminosity evolution.
The low redshift reference sample is overlaid.
The distributions of the Lynx samples closely resemble the distribution of the low redshift reference 
sample, without any obvious bias in the coverage of both parameters.
The spectroscopic sample is limited at effective radii $r_{\rm e}\ge 1$ kpc, which is 
similar to limits of literature samples (e.g., Saglia et al.\ 2010; Beifiori et al.\ 2017).

In summary, we find no obvious selection effects in luminosities, effective radii or surface brightnesses,
that may bias our results. 

\section{The Lynx Cluster Differences \label{SEC-DIFFERENCES} }

Examining the grayscale figure of the two clusters, Figure \ref{fig-lynxgrey}, 
it is striking that the core of Lynx E is dominated by passive bulge-dominated galaxies.
The core of Lynx W appears to be roughly a 50:50 mix of passive bulge-dominated galaxies and disk galaxies 
plus emission line bulge-dominated galaxies.
We investigate this closer using phase space diagram of line-of-sight velocities relative to
the cluster velocity dispersions, $v_{||}/\sigma _{\rm cl}$, versus the cluster center distances in units of 
$R_{500}$, see Figure \ref{fig-phasespace}.
On the figure, the cyan dashed lines denote caustics with constant $v_{||}/\sigma _{\rm cl} \times R_{\rm cl}/R_{500}$ 
values of 0.2, 0.64 and, 1.35.
Following Noble et al.\ (2015), galaxies between the caustics of 0.64 and 1.35 are recently accreted, while those outside 1.35 are still infalling.
This is in general agreement with the majority of the low redshift reference cluster galaxies being 
inside the 0.64 caustic and therefore accreted early in the formation of the clusters. 
Both the Lynx W and Lynx E emission line bulge-dominated galaxies are primarily recently accreted or still infalling,
while the passive galaxies inside $R_{500}$ are also inside the 0.64 caustic and thus can be assumed to have been accreted earlier.
The Lynx E sample contains two post starburst galaxies (marked with yellow outlines in Figure \ref{fig-phasespace}). 
These are also outside the 0.64 caustic.
The Lynx E cluster core is significantly richer in bulge-dominated galaxies on the red sequence than seen for the Lynx W core.
To illustrate this we show grayscale figures of the cluster cores with all bulge-dominated galaxies on the red sequence marked, 
see Figure \ref{fig-lynxgrayscale_subimages}.

As we proceed to the quantitative analysis of the data in the following sections, we highlight in the figures
the passive bulge-dominated galaxies located in the cores of the two clusters 
(marked with black outlines in Figure \ref{fig-phasespace}). In Section \ref{SEC-COMPOSITES},
we directly investigate the stellar populations of composite spectra of these galaxies compared to 
the rest of the passive bulge-dominated galaxies in the clusters and 
to the bulge-dominated galaxies with significant [\ion{O}{2}] emission.

\begin{figure}
\epsfxsize 8.5cm
\epsfbox{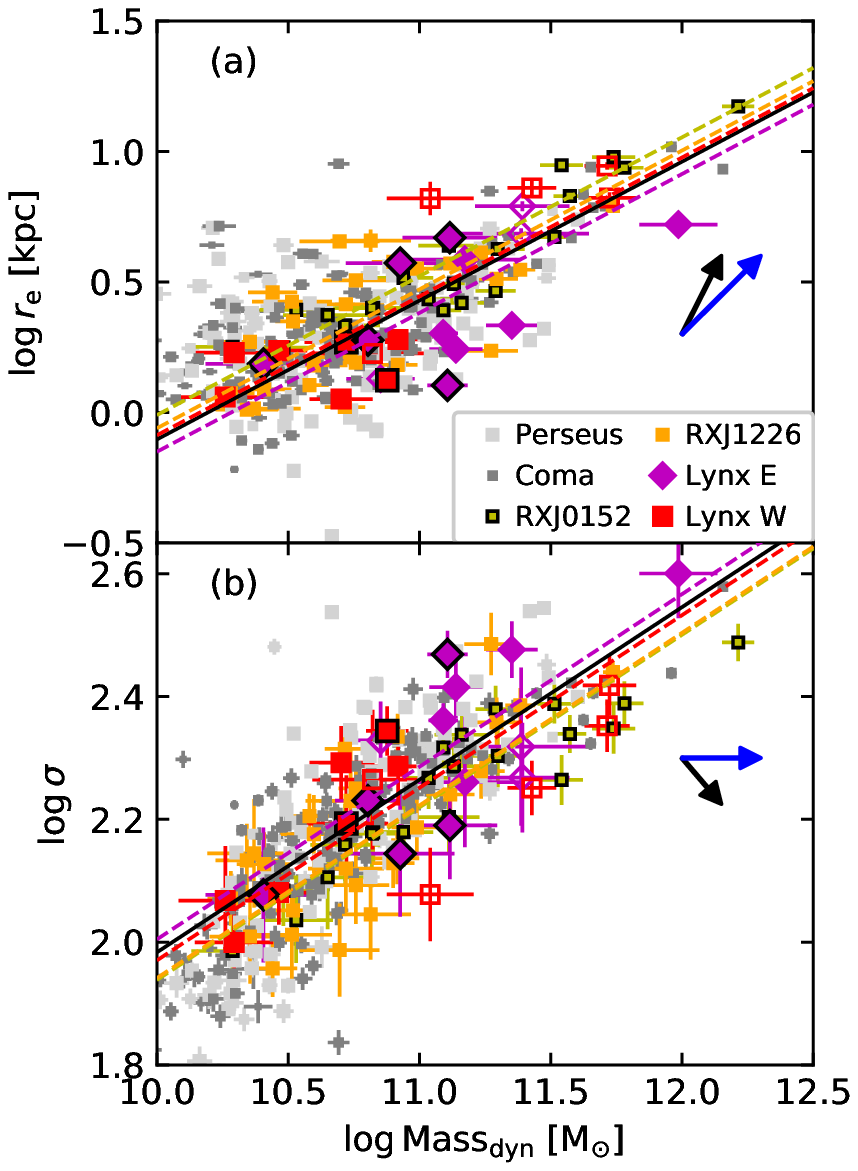}
\caption{
Effective radii and velocity dispersions versus dynamical masses. 
Light gray -- Perseus cluster members; dark gray -- Coma cluster members;  
magenta diamonds -- Lynx E; red boxes -- Lynx W. 
Open symbols for Lynx E and W show galaxies with EW[\ion{O}{2}]$>$5{\AA}.
Symbols with black edges show Lynx E and W passive galaxies within $R_{500}$ of the cluster centers.
Small yellow boxes -- passive bulge-dominated galaxies in RXJ0152.7--1357.
Small orange boxes -- passive bulge-dominated galaxies in RXJ1226.9+3332.
Solid black lines -- best fit relations to the low redshift reference sample (Perseus and Coma). 
Dashed colored lines -- the low redshift relations offset to the median zero points of the higher redshift 
clusters, color coded to match the symbols: black -- low redshift sample; yellow -- RXJ0152.7--1357; 
orange -- RXJ1226.9+3332; magenta -- Lynx E; red -- Lynx W.
Arrows mark minor (black) and major (blue) merger tracks from Bezanson et al.\ (2009) equivalent to a factor two increase 
in size, $\Delta \log r_e = 0.3$. For clarity, the model tracks are shown offset from the data.
\label{fig-lrelmass} }
\end{figure}

\section{Galaxy Structure and the Fundamental Plane \label{SEC-STRUCTUREFP}}

\begin{figure*}
\epsfxsize 17cm
\epsfbox{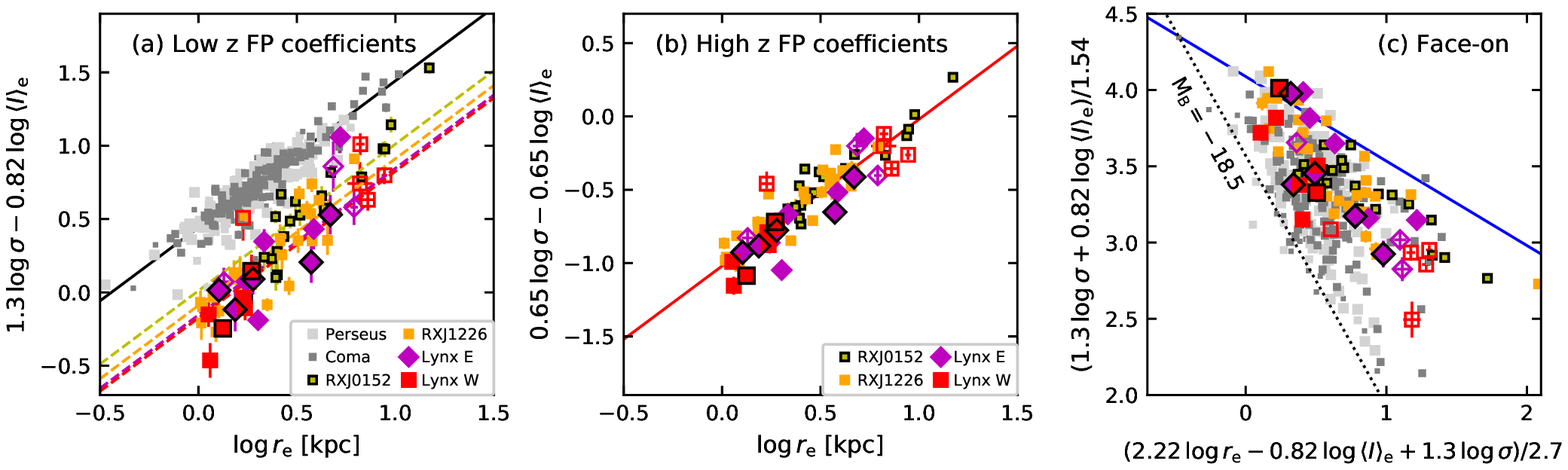}
\caption{
The Fundamental Plane (FP) shown edge-on (panels a and b) and face-on (panel c).  Symbols as in Figure \ref{fig-lrelmass}.
Black solid line -- best fit relation for the Coma cluster sample from J\o rgensen et al.\ (2014), offset
to the zero point of the larger reference sample of Coma and Perseus galaxies. 
Dashed colored lines -- the low redshift relations offset to the median zero points of the higher redshift 
clusters, color coded to match the symbols: black -- low redshift sample; yellow -- RXJ0152.7--1357; 
orange -- RXJ1226.9+3332; magenta -- Lynx E; red -- Lynx W.
The Lynx E and W lines fall very close to each other.
Panel (b) shows the FP edge-on for the high redshift coefficients, see Table \ref{tab-FPfit}.
The best fit is shown as the median zero point for the full sample.
On panel (c), the data for RXJ0152.7--1357 and RXJ1226.9+3332 have been offset in $\log \langle I \rangle _e$ by $-0.6$,
while the Lynx data were offset by $-0.75$, in both cases approximately matching their luminosity evolution.
Blue line -- the exclusion zone from Bender et al.\ (1992).
Black dotted line -- magnitude limit for the low redshift reference sample.
The apparent lack of high redshift galaxies in the lower left of the FP face-on view is due to 
the difference in FP coefficients and that this face-on view is not exactly face-on for the high redshift samples.
\label{fig-FP} }
\end{figure*}

\begin{figure*}
\epsfxsize 14cm
\begin{center}
\epsfbox{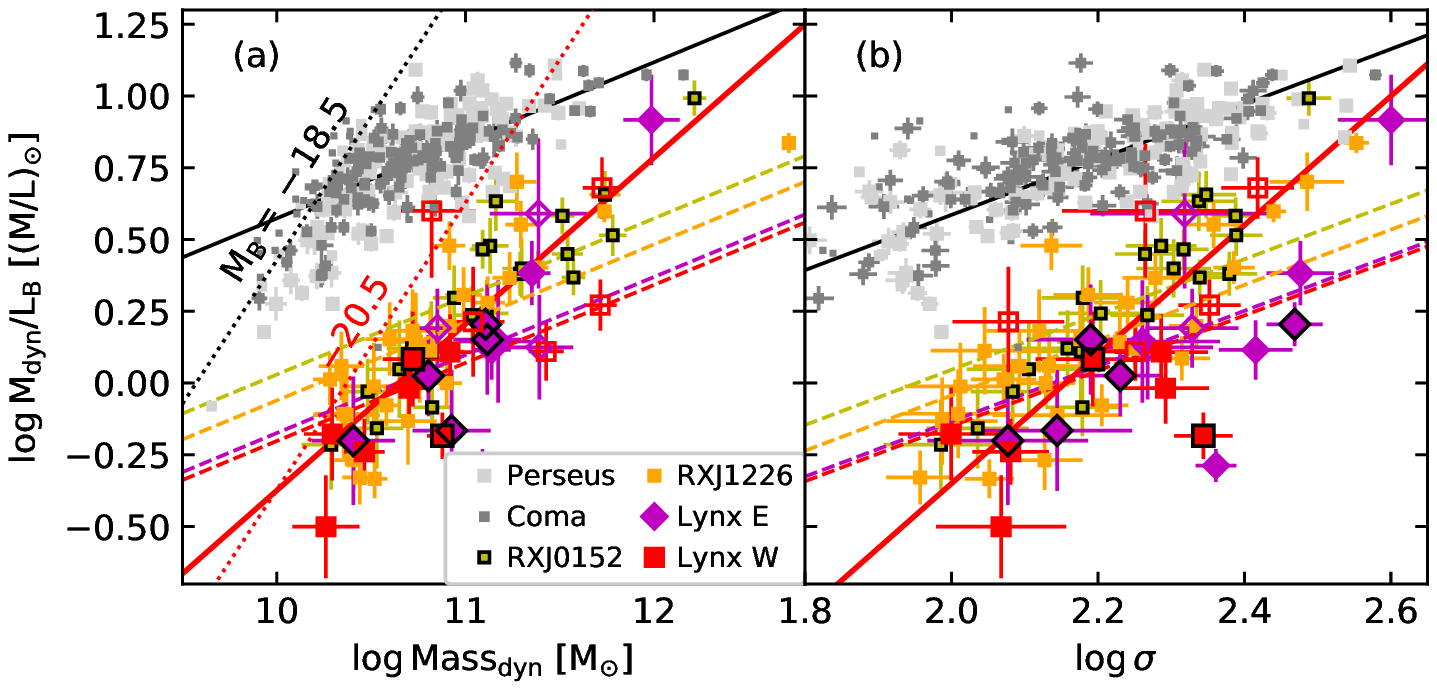}
\end{center}
\caption{ 
The dynamical M/L ratios versus the dynamical masses and versus the velocity dispersions. 
Symbols as in Figure \ref{fig-lrelmass}.
Black lines -- best fit relations to the low redshift reference sample (Perseus and Coma).
Dashed colored lines -- the low redshift relations offset to the median zero points of the higher redshift 
clusters, color coded to match the symbols: black -- low redshift sample; yellow -- RXJ0152.7--1357; 
orange -- RXJ1226.9+3332; magenta -- Lynx E; red -- Lynx W.
Solid red lines -- best fit relations to the higher redshift clusters, shown at the median zero point for the 
samples.
Dotted lines on panel (a) show the magnitude limits of the samples, $M_{\rm B,abs}=-18.5$ for the low 
redshift reference sample and typically $M_{\rm B,abs}=-20.5$ for the higher redshift samples.
\label{fig-MLonly} }
\end{figure*}

All scaling relations relevant for our 
analysis of the possible structure evolution and the evolution
of the FP are summarized in Tables \ref{tab-relations} and \ref{tab-FPfit},
and shown in Figures \ref{fig-lrelmass}, \ref{fig-FP}, and \ref{fig-MLonly}.
For all relations, we list in the tables the zero points for the low redshift reference sample 
and Lynx samples relative to the Coma relations from J\o rgensen \& Chiboucas (2013). 
We also establish relations based on our new larger low redshift reference sample described in Section \ref{SEC-COMPDATA},
and give zero points for the reference sample and the Lynx samples relative to these relations. 
All slopes for the relations based on our new larger low redshift reference sample are consistent within $1\sigma$ 
with the slopes from J\o rgensen \& Chiboucas (2013).
On the figures we show the new relations, except for the FP where we choose to show
the relation edge-on and face-on using the same coefficients as in J\o rgensen \& Chiboucas (2013).
This choice was made to make it easier to compare our results visually with other published results
since many of these use the coefficients from J\o rgensen \& Chiboucas (2013).
The tables list the results for Lynx cluster assignments based on adopting the BCGs as cluster centers.
There are no significant differences in the results if using the X-ray centers for the cluster assignments.
Specifically, in all cases the zero point changes are less than half the uncertainties on the zero points (derived as ${\rm rms}\, N_{\rm gal}^{-1/2}$).

\begin{deluxetable*}{lrrr rrr rrr}
\tablecaption{Structure Relations and the Fundamental Plane\label{tab-relations} }
\tablewidth{0pc}
\tabletypesize{\scriptsize}
\tablehead{
\colhead{Relation} & \multicolumn{3}{c}{Low redshift} & 
  \multicolumn{3}{c}{Lynx E} &\multicolumn{3}{c}{Lynx W}  \\
 & \colhead{$\gamma$} & \colhead{$N_{\rm gal}$} & \colhead{rms} 
 & \colhead{$\gamma$} & \colhead{$N_{\rm gal}$} & \colhead{rms} &  \colhead{$\gamma$} & \colhead{$N_{\rm gal}$} & \colhead{rms}  \\
\colhead{(1)} & \colhead{(2)} & \colhead{(3)} & \colhead{(4)} 
& \colhead{(5)} & \colhead{(6)} & \colhead{(7)} & \colhead{(8)} & \colhead{(9)} & \colhead{(10)} 
}
\startdata
$\log r_e      = (0.57 \pm 0.06) \log {\rm Mass} + \gamma$\tablenotemark{a}  & -5.836 & 228 & 0.21 & -5.881 & 13 & 0.19 & -5.815 & 12 & 0.17  \\  
$\log r_e      = (0.53 \pm 0.04) \log {\rm Mass} + \gamma$\tablenotemark{b}  & -5.423 & 228 & 0.21 & -5.470 & 13 & 0.19 & -5.407 & 12 & 0.17  \\   
$\log \sigma   = (0.26 \pm 0.03) \log {\rm Mass} + \gamma$\tablenotemark{a}  & -0.602 & 228 & 0.10 & -0.578 & 13 & 0.09 & -0.615 & 12 & 0.09  \\ 
$\log \sigma   = (0.28 \pm 0.02) \log {\rm Mass} + \gamma$\tablenotemark{b}  & -0.826 & 228 & 0.10 & -0.805 & 13 & 0.09 & -0.840 & 12 & 0.09  \\ 
$\log \rm{r_e} = (1.30 \pm 0.08) \log \sigma  - (0.82 \pm  0.03) \log \langle I \rangle _e + \gamma $\tablenotemark{a}
                                                                             & -0.429 & 228 & 0.09 &  0.154 & 13 & 0.21 & 0.174 & 12 & 0.21 \\
$\log \rm{r_e} = (1.23 \pm 0.07) \log \sigma  - (0.84 \pm  0.02) \log \langle I \rangle _e + \gamma $\tablenotemark{b}
                                                                             & -0.219 & 228 & 0.09 &  0.374 & 13 & 0.21 & 0.402 & 12 & 0.22 \\
$\log \rm{M/L} = (0.24 \pm 0.03) \log {\rm Mass} + \gamma$\tablenotemark{a}  & -1.818 & 228 & 0.12 & -2.537 & 13 & 0.25 & -2.563 & 12 & 0.26   \\ 
$\log \rm{M/L} = (0.27 \pm 0.02) \log {\rm Mass} + \gamma$\tablenotemark{b}  & -2.162 & 228 & 0.12 & -2.894 & 13 & 0.24 & -2.922 & 12 & 0.25   \\ 
$\log \rm{M/L} = (1.07 \pm 0.12) \log \sigma + \gamma$\tablenotemark{a}  & -1.592 & 228 & 0.11 & -2.302 & 13 & 0.24 & -2.308 & 12 & 0.27   \\ 
$\log \rm{M/L} = (0.96 \pm 0.10) \log \sigma + \gamma$\tablenotemark{b}  & -1.356 & 228 & 0.11 & -2.059 & 13 & 0.24 & -2.076 & 12 & 0.27   \\ 
\enddata
\tablecomments{Column 1: Scaling relation. Column 2: Zero point for the low redshift sample (average of the zero points for Perseus and Coma). 
Column 3: Number of galaxies included from the low redshift sample. 
Column 4: rms in the Y-direction of the scaling relation for the low redshift sample.
Columns 5, 6, and 7: Zero point, number of galaxies, rms in the Y-direction for the Lynx W sample.
Columns 8, 9, and 10: Zero point, number of galaxies, rms in the Y-direction for the Lynx E sample.
}
\tablenotetext{a}{Slopes adopted from J\o rgensen \& Chiboucas (2013), zero points use effective radii (and surface brightnesses) from fits with S\'{e}rsic profiles, $r_{\rm e}=(a_{\rm e}\,b_{\rm e})^{1/2}$.}
\tablenotetext{b}{Slopes from best fit to Coma and Perseus samples. Effective radii (and surface brightnesses) from fits with S\'{e}rsic profiles, $r_{\rm e}=(a_{\rm e}\,b_{\rm e})^{1/2}$.}
\end{deluxetable*}

\begin{deluxetable*}{llrr}
\tablecaption{The Fundamental Plane and Relations for the M/L Ratios \label{tab-FPfit} }
\tablewidth{0pc}
\tabletypesize{\scriptsize}
\tablehead{
\colhead{Cluster} & \colhead{Relation\tablenotemark{a} } & \colhead{$N_{\rm gal}$} & \colhead{rms}
}
\startdata
Coma\tablenotemark{a}           & $\log \rm{r_e} = (1.30 \pm 0.08) \log \sigma  - (0.82 \pm  0.03) \log \langle I \rangle _e -0.443 $ & 105 & 0.08 \\  
Coma, Perseus\tablenotemark{b}  & $\log \rm{r_e} = (1.23 \pm 0.07) \log \sigma  - (0.84 \pm  0.02) \log \langle I \rangle _e -0.219 $ & 228 & 0.09 \\  
Lynx E+W                        & $\log \rm{r_e} = (0.55 \pm 0.39) \log \sigma  - (0.61 \pm  0.11) \log \langle I \rangle _e +1.120 $ &  25 & 0.14 \\  
RXJ0152.7--1357, RXJ1226.9+3332, Lynx E+W\tablenotemark{c}  & $\log \rm{r_e} = (0.65 \pm 0.21) \log \sigma  - (0.65 \pm  0.09) \log \langle I \rangle _e + \{ 1.000,1.000,1.060\} $ &  74 & 0.09,0.11,0.15 \\  
Coma\tablenotemark{a}           & $\log \rm{M/L} = (0.24 \pm 0.03) \log {\rm Mass} -1.754$ & 105 & 0.09 \\ 
Coma, Perseus\tablenotemark{b}  & $\log \rm{M/L} = (0.27 \pm 0.02) \log {\rm Mass} -2.162$ & 228 & 0.12 \\ 
Lynx E+W                        & $\log \rm{M/L} = (0.70 \pm 0.16) \log {\rm Mass} -7.580$ &  25 & 0.23 \\ 
RXJ0152.7--1357, RXJ1226.9+3332, Lynx E+W\tablenotemark{c}  & $\log \rm{M/L} = (0.58 \pm 0.09) \log {\rm Mass} -\{ 6.163,6.136,6.222\}$ &  74 & 0.13,0.17,0.22 \\  
Coma\tablenotemark{a}            & $\log \rm{M/L} = (1.07 \pm 0.12) \log \sigma -1.560$ & 105 & 0.11 \\ 
Coma, Perseus\tablenotemark{b}  & $\log \rm{M/L} = (0.96 \pm 0.10) \log \sigma -1.356$ & 228 & 0.11 \\ 
Lynx E+W                        & $\log \rm{M/L} = (2.38 \pm 0.89) \log \sigma -5.281$ &  25 & 0.30 \\ 
RXJ0152.7--1357, RXJ1226.9+3332, Lynx E+W\tablenotemark{c}  & $\log \rm{M/L} = (2.25 \pm 0.26) \log \sigma -\{ 4.775,4.787,4.989\}$ & 74 & 0.12,0.19,0.29 \\  
\enddata
\tablenotetext{a}{Relations for Coma adopted from J\o rgensen \& Chiboucas (2013).}
\tablenotetext{b}{Velocity dispersions from J\o rgensen et al.\ (2018b), photometry as described in Section \ref{SEC-COMPDATA}.}
\tablenotetext{c}{RXJ0152.7--1357, RXJ1226.9+3332, and Lynx E+W fit with parallel relations. The zero points and rms for the
three samples are listed in the same order as the clusters.}
\end{deluxetable*}

\subsection{Sizes and Velocity Dispersions \label{SEC-SIZE} }

Figure \ref{fig-lrelmass} shows the effective radii and velocity dispersions versus the dynamical masses.
The figure includes our low redshift sample, the Lynx samples (4) and (5), and for reference the samples 
in RXJ0152.7--1357 and RXJ1226.9+3332. 
As described in J\o rgensen \& Chiboucas (2013), we found no evolution in size or velocity dispersion 
for the RXJ0152.7--1357 and RXJ1226.9+3332 samples compared to the Coma cluster galaxies.
The galaxies in Lynx E and W follow the same relations as the low redshift reference sample.
Further, relative to the low redshift reference sample relations (black lines) the Lynx samples have no
significant offsets. Formally we find a difference of  $\Delta \log r_e = -0.02 \pm 0.04$ with the Lynx galaxies 
being marginally smaller than the low redshift galaxies, cf.\ Table \ref{tab-relations}.
Similarly the offset in velocity dispersions is insignificant, $\Delta \log \sigma = 0.003 \pm 0.019$.
Galaxies in Lynx with emission of EW[\ion{O}{2}]$>$5 {\AA} appear to be larger 
and have lower velocity dispersions than the passive galaxies.
Deriving the offsets relative to the low redshift reference sample from only the passive 
galaxies we find median offsets of $\Delta \log r_e = -0.06 \pm 0.05$ and $\Delta \log \sigma = 0.03 \pm 0.04$.
Thus, we adopt 0.1 dex as the upper $1\sigma$ limit on the bulk size evolution between $z=1.27$ and the present.

Figure \ref{fig-lrelmass} also shows models from Bezanson et al.\ (2009).
These will be discussed in Section \ref{SEC-DISCUSSION}.

\subsection{The Fundamental Plane \label{SEC-FP} }

Figure \ref{fig-FP} shows the FP edge-on and face-on for the Lynx samples (4) and (5)
together with our low redshift reference sample and the RXJ0152.7--1357 and RXJ1226.9+3332 samples.
Panels (a) and (c) use the FP coefficients for the Coma cluster sample from J\o rgensen \& Chiboucas (2013), offset
to the zero point of the larger reference sample of Coma and Perseus galaxies. The zero point 
for the larger reference sample is in agreement with that from J\o rgensen \& Chiboucas within 0.014.
The edge-on view in panel (b) uses the FP coefficients fit to the joint sample
of the RXJ0152.7--1357, RXJ1226.9+3332 and Lynx samples, see Table \ref{tab-FPfit}.

In Figure \ref{fig-MLonly}, we show the FP as the M/L ratios versus the dynamical masses and versus
the velocity dispersions. In both cases, we show the best fits to the new larger reference sample (black lines).
The dashed colored lines show the low redshift relation offsets to the median zero points of the higher redshift samples.

Figures \ref{fig-FP}a and \ref{fig-MLonly}a both indicate that the $z=0.8-1.3$ galaxies 
follow different relations than the low redshift reference sample. 
In our previous papers J\o rgensen et al.\ (2006, 2007) and J\o rgensen \& Chiboucas (2013), we presented
simulations and discussions of this difference in the relations between $z=0.8-0.9$ and $z\approx 0$ samples.
We do not repeat these tests here, but simply provide for reference the fit of 
the FP to the three clusters RXJ0152.7--1357, RXJ1226.9+3332 and Lynx as parallel planes, allowing 
differences in zero points. 
Similarly, we fit the M/L-mass relation for the three clusters as parallel relations.
Lynx E and W are required to have the same zero point, as we find no indication
of a zero point difference between these two clusters. 
Figure \ref{fig-FP}b shows the resulting FP edge-on visualizing that the scatter of the high redshift data 
relative to this relation is significantly lower than the scatter relative  to the low redshift FP shown
in Figure \ref{fig-FP}a. 
The observed scatter for the Lynx sample in the direction of $\log r_e$ decreases from 0.21 to 0.15.
Similarly in Figure \ref{fig-MLonly}a, we show the common fit to the high redshift clusters. 
The magnitude limits of the samples are shown in Figure \ref{fig-MLonly}a. While it is possible
that the magnitude limit contributes to the steepness of the high-redshift relation, it is unlikely that even
with a significantly larger and deeper sample the galaxies would simply populate an offset version of the 
low redshift relation. At a minimum, the scatter of the low mass galaxies relative to the relation
would be significantly higher than seen at low redshift. 

The offset of the Lynx sample relative to the low redshift reference sample indicated by the FP and 
the M/L-mass relation, in the absence of structural evolution 
is equivalent to $\Delta \log M/L = -0.75 \pm 0.05 $, cf.\ Table \ref{tab-relations},
with the Lynx galaxies having lower M/L ratios than galaxies in the low redshift reference sample.
The uncertainty is dominated by the scatter in the relation for the Lynx sample.
If due to an age difference, the offset in M/L ratio is equivalent to 
$\Delta \log {\rm age} = \log ({\rm age_{low-z}}/{\rm age_{Lynx}}) = 0.79 \pm 0.05$ using
the relation between M/L and age based on models from Maraston (2005), see J\o rgensen et al.\ (2014, Table 6). 
The zero points for Lynx E and W are not significantly different from each other. 
Assuming a common formation epoch for the Lynx galaxies and the low redshift sample galaxies we then find 
$z_{\rm form} = 1.89 _{-0.10}^{+0.14}$, for the adopted cosmology.
This also means that the ages of stellar populations in the Lynx galaxies are $\approx 1.5$ Gyr.
Adopting the steeper M/L-mass relation results in younger ages for the low mass galaxies, and older ages for the higher mass galaxies.
At masses of $10^{10.5}M_{\sun}$ we find $z_{\rm form} \approx 1.8$ or ages of the stellar populations of 
$\approx 1$ Gyr, while at $10^{11.5}M_{\sun}$ $z_{\rm form} \approx 3.7$ and ${\rm age} \approx 3$ Gyr.

If we allow for the structural evolution of $\Delta \log r_e = 0.1$ ($\approx 1\sigma$ more than the formal offset
for the passive galaxies only) and assume changes in the other parameters 
as predicted by the dry minor merger tracks from Bezanson et al.\ (2009), as described in Section \ref{SEC-METHOD}, 
then the M/L ratios are unaffected while the masses increase by $\Delta \log {\rm Mass} = 0.05$. 
However, with a slope of 0.27 of the M/L-mass relation at $z\approx 0$, the additional required change 
in the M/L ratios would only be $\Delta \log M/L \approx 0.01$, which is insignificant given the uncertainties.
Changes in the M/L ratios due mergers of galaxies with different M/L ratios whether due to metallicity
or age differences are indirectly restricted due to the limits on the mass growth allowed by the data. 
For example, increasing the mass by $\Delta \log {\rm Mass} = 0.05$ by adding 
1 Gyr old stellar populations to a more massive galaxy with 3 Gyr old stellar populations will 
decrease the M/L of the resulting merger by $\Delta \log M/L < 0.1$ relative to the more 
massive galaxy in the encounter. This estimate is based on model M/L values from Maraston (2005). 

Turning our attention to the FP face-on (Figure \ref{fig-FP}c), we examine whether the high redshift 
samples populate the FP similarly as seen for the low redshift reference sample. 
To account for the average luminosity evolution,
the RXJ0152.7--1357 and RXJ1226.9+3332 samples have been offset by $\Delta \log \langle I_e \rangle = -0.6$,
while the Lynx sample was offset by $\Delta \log \langle I_e \rangle = -0.75$.
Once these offsets are applied, the high redshift samples populate the FP similarly as the low redshift reference sample,
with the exception of the area at low X and Y-values in this view.
The lack of galaxies in the area corresponds to the FP for the $z=0.8-1.3$ samples not being seen exactly face-on 
in this view but rather slightly tilted due to the difference in FP coefficients. 
The zone of avoidance for $z\approx 0$ galaxies as established by Bender et al.\ (1992)
is shown in Figure \ref{fig-FP}c as the solid blue line. 
Once offset for the luminosity evolution, the $z=0.8-1.3$ samples respect this same zone of avoidance. 

In summary, the Lynx sample zero point offsets relative to the low redshift reference samples can be interpreted as 
luminosity evolution only, and leads to an average formation redshift of $z_{\rm form} = 1.89 _{-0.10}^{+0.14}$.
The steeper M/L-mass relation (and FP) indicates that the low mass galaxies are younger than the higher mass galaxies,
with the ages spanning from 1 to 3 Gyr.

\begin{figure*}
\epsfxsize 17cm
\epsfbox{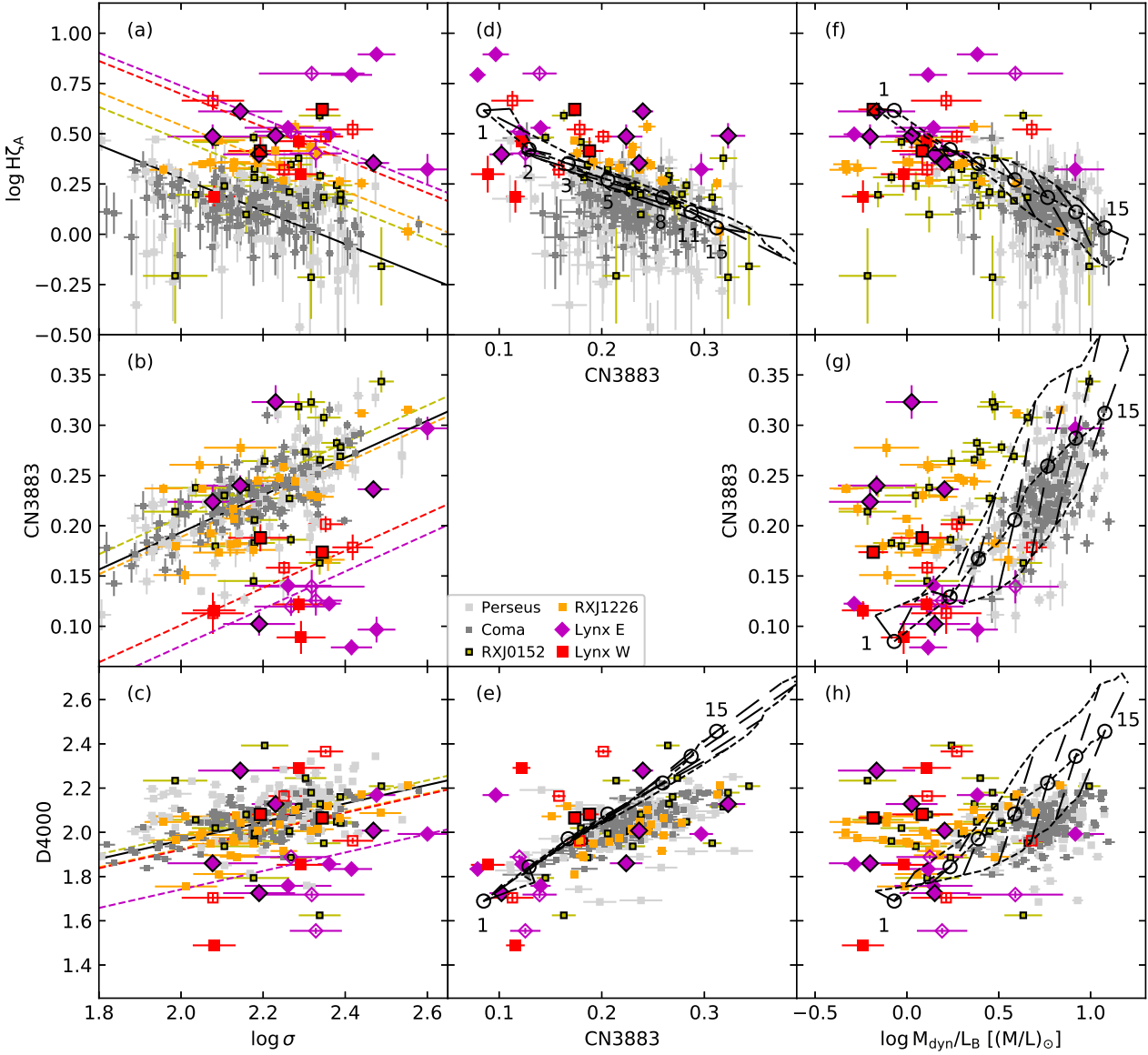}
\caption{ 
Stellar population parameters versus each other.
Symbols as in Figure \ref{fig-lrelmass}.
Only Lynx galaxies with all three indices (H$\zeta _{\rm A}$, CN3883, D4000) measured are included on the figure.
Panels (a)--(c) show absorption line strengths versus velocity dispersions. 
The solid line on each panel shows the best fit relation for the low redshift reference sample.
The dashed lines show the relations offset to the median zero point for each of the other clusters,
color coded match the symbols: black -- low redshift sample; yellow -- RXJ0152.7--1357; 
orange -- RXJ1226.9+3332; magenta -- Lynx E; red -- Lynx W.
Panels (d) and (e) show H$\zeta _{\rm A}$ and D4000 versus CN3883.
Black dashed lines -- model values based on SSP models from Maraston \& Str\"{o}mb\"{a}ck (2011).
The models are degenerate in age and metallicity [M/H]. 
Panels (f)--(h) show the line indices versus the M/L ratios with the stellar population models overlaid.
The black open circles correspond
to models with solar [M/H] and ages of 1, 2, 5, 8, 11, and 15 Gyr as labeled in panel (d) below the points, 
see text for details. In panels (e)-(h) only the 1 and 15 Gyr model points are labeled.
\label{fig-stellarpops} }
\end{figure*}

\section{Stellar Populations \label{SEC-STELLARPOP} }

We can gain a more detailed understanding of the stellar populations of the Lynx galaxies 
by using the line indices, H$\zeta _{\rm A}$, CN3883, and D4000,
together with the velocity dispersions and the M/L ratios. We first establish scaling relations between
the indices and the velocity dispersions and compare the data to SSP models (Section \ref{SEC-INDICES}).
Then in Section \ref{SEC-COMPOSITES} we assemble the spectra into three composite spectra and fit these with SSP models.

\subsection{Line strengths and mass-to-light ratios \label{SEC-INDICES} }

Figure \ref{fig-stellarpops} shows the absorption line indices, H$\zeta _{\rm A}$, CN3883, D4000, and the M/L ratios 
versus velocity dispersion and versus each other. 
Table \ref{tab-indexrelations} lists the relations shown on panels (a)--(c).
On the other panels we overlay SSP models based on SEDs from Maraston \& Str\"{o}mb\"{a}ck (2011) and M/L information
for similar SSP models from Maraston (2005). All models use solar abundance ratios.  

The H$\zeta _{\rm A}$-velocity dispersion relation for the Lynx sample is offset from the relation for the 
low redshift reference sample, in agreement with expectations for stellar populations with ages of 1--3 Gyr.
Similarly as also found in Section \ref{SEC-FP}, the M/L ratios for the Lynx sample support ages of 1--3 Gyr.
In particular, the majority of the Lynx galaxies agree with the 1--3 Gyr model locations in the H$\zeta _{\rm A}$--M/L diagram
(Figure \ref{fig-stellarpops}f).
The RXJ0152.7--1357 and RXJ1226.9+3332 samples follow similar trends with slightly weaker H$\zeta _{\rm A}$ 
indices as expected as their redshifts are lower than the Lynx redshift.

\begin{deluxetable*}{lrrr rrr rrr}
\tablecaption{Scaling Relations for Line Indices\label{tab-indexrelations} }
\tablewidth{0pc}
\tabletypesize{\scriptsize}
\tablehead{
\colhead{Relation} & \multicolumn{3}{c}{Low redshift} & 
  \multicolumn{3}{c}{Lynx E} &\multicolumn{3}{c}{Lynx W}  \\
 & \colhead{$\gamma$} & \colhead{$N_{\rm gal}$} & \colhead{rms} 
 & \colhead{$\gamma$} & \colhead{$N_{\rm gal}$} & \colhead{rms} &  \colhead{$\gamma$} & \colhead{$N_{\rm gal}$} & \colhead{rms}  \\
\colhead{(1)} & \colhead{(2)} & \colhead{(3)} & \colhead{(4)} 
& \colhead{(5)} & \colhead{(6)} & \colhead{(7)} & \colhead{(8)} & \colhead{(9)} & \colhead{(10)} 
}
\startdata
$\log \rm{H}\zeta _{\rm A} = (-0.82 \pm 0.39) \log \sigma + \gamma$  &  1.919 & 213 & 0.16 &  2.379 & 13 & 0.21 &  2.338 &   9 & 0.19  \\    
$\rm{CN3883}    = (0.18 \pm 0.01) \log \sigma    + \gamma$           & -0.177 & 217 & 0.03 & -0.290 & 13 & 0.08 & -0.269 &   9 & 0.03  \\   
$\rm{D4000}     = (0.42 \pm 0.04) \log \sigma    + \gamma$           &  1.127 & 209 & 0.09 &  0.906 & 13 & 0.20 &  1.085 &  9 & 0.24  \\   
\enddata
\tablecomments{Column 1: Scaling relation. Column 2: Zero point for the low redshift sample (average of the zero points for Perseus and Coma). Column 3: Number of galaxies
included from the low redshift sample. Column 4: rms in the Y-direction of the scaling relation for the low redshift sample.
Columns 5, 6, and 7: Zero point, number of galaxies, rms in the Y-direction for the Lynx W sample.
Columns 8, 9, and 10: Zero point, number of galaxies, rms in the Y-direction for the Lynx E sample.
}
\end{deluxetable*}

\begin{figure*}
\begin{center}
\epsfxsize 17cm
\epsfbox{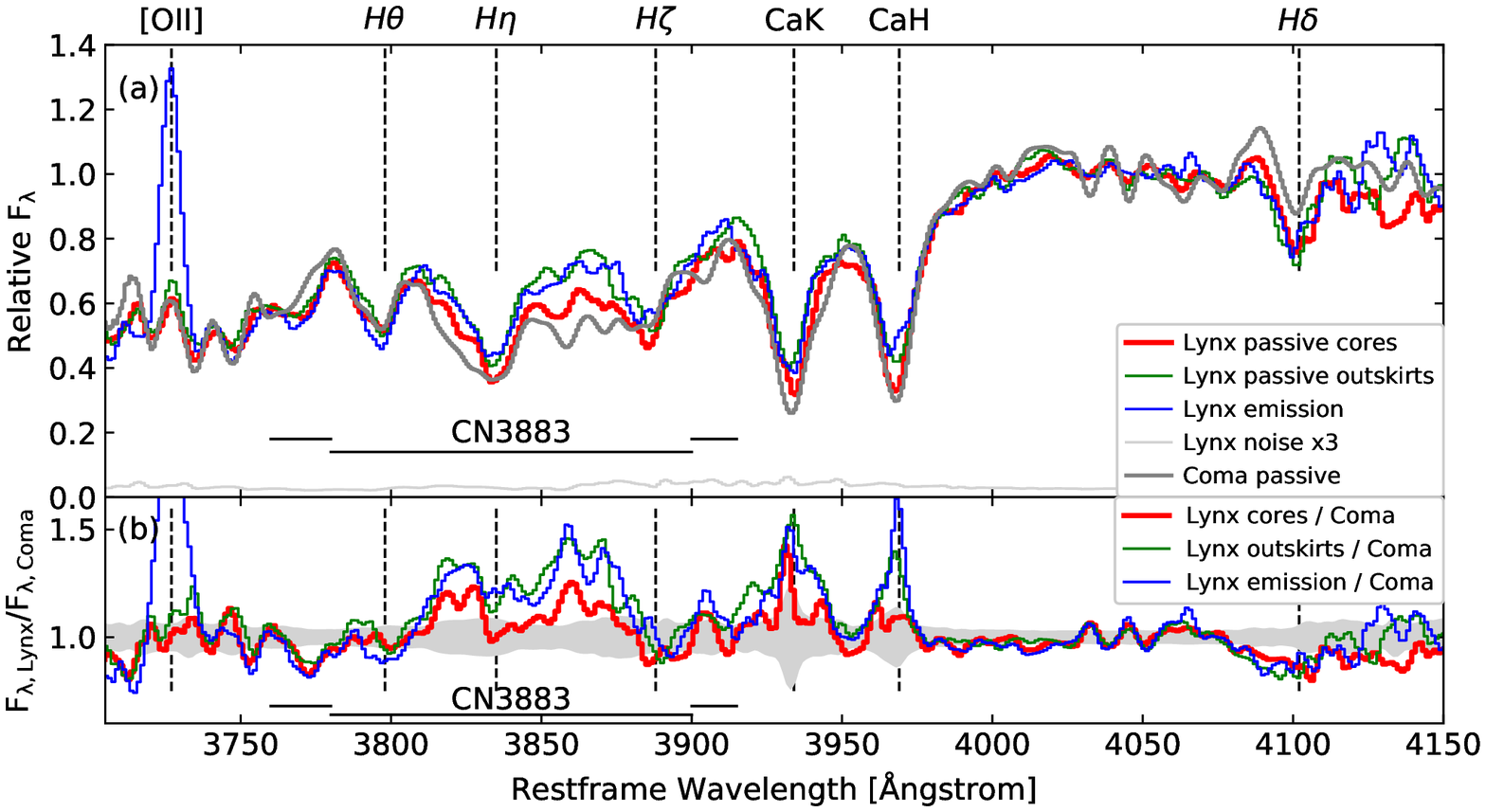}
\end{center}
\caption{
Panel (a): Composite spectra of bulge-dominated galaxies in the Lynx clusters. 
Red -- {\it cores-composite}, which includes the passive galaxies in cores of Lynx W and Lynx E, inside $R_{500}$.
Green -- {\it outskirts-composite}, which includes the passive galaxies outside the cores of Lynx E and W. 
Blue -- {\it emission-composite}, which includes the galaxies with significant [\ion{O}{2}] emission.
Light gray -- three times the representative noise spectrum for the Lynx composites, based on poisson noise and read out noise.  
At wavelengths longer than $\approx 4120${\AA} systematic errors from
the sky subtraction dominate the noise and the noise spectrum underestimates the real noise.
Dark gray -- for reference, the high S/N composite of passive galaxies in the Coma cluster made from SDSS spectra. 
Major spectral lines are marked with dashed black lines. 
The location of the CN3883 passband is marked with the long horizontal black line,
the shorter offset horizontal lines mark the continuum bands for the index.
Panel (b): The ratio between the Lynx composites and the Coma cluster composite.
Red -- Ratio of {\it cores-composite} to Coma cluster composite.
Green -- Ratio of {\it outskirts-composite} to Coma cluster composite.
Blue -- Ratio of {\it emission-composite} to Coma cluster composite.
Gray area mark three times the representative noise of the ratios. 
The {\it cores-composite} shows a stronger CN3883 feature than both
the {\it outskirts-composite} and the {\it emission-composite}, see text.
\label{fig-composite} }
\end{figure*}

The CN3883-velocity dispersion relation for the Lynx sample is significantly offset from the low redshift reference sample,
while the RXJ0152.7--1357 and RXJ1226.9+3332 samples show no significant offsets, see Figure \ref{fig-stellarpops}b.
Our results for RXJ0152.7--1357 and RXJ1226.9+3332 were originally published in J\o rgensen et al.\ (2005) and J\o rgensen \& Chiboucas (2013).
Additional results for the $z=0.2-0.9$ GCP clusters can be found in  J\o rgensen et al.\ (2017).
A small number of the Lynx galaxies, primarily the passive galaxies in the centers of the two clusters (shown with black
outlines on the points in Figure \ref{fig-stellarpops}) fall closer to the low redshift relation than the median
relation for the bulk of the Lynx galaxies. These galaxies still have strong H$\zeta _{\rm A}$ and low M/L ratios in 
agreement with ages of 1--3 Gyr (Figure \ref{fig-stellarpops}a and f).
Thus, their unusually strong CN3883 may reflect very high [CN/Fe] or more broadly high total metallicity [M/H].
Due to the short wavelength coverage of our spectra, we cannot evaluate whether only [CN/Fe] or also the total metallicity [M/H]
is high relative to the remainder of the sample. 

The D4000-velocity dispersion relation, Figure \ref{fig-stellarpops}c, for the Lynx sample has 
a very large scatter. Eight of the 22 Lynx galaxies have D4000
as strong or stronger as the bulk of the galaxies in the low redshift reference sample. 
The remainder of the Lynx galaxies have weaker D4000 for a given velocity dispersion.
While D4000 is usually used as an indicator of stellar population age (e.g., Gallazzi et al.\ 2014), 
we note that the blue passband for D4000 covers the CN3883 absorption feature. 
Thus, D4000 and CN3883 can be expected to be tightly correlated, as shown in Figure \ref{fig-stellarpops}e. 
The oldest Maraston \& Str\"{o}mb\"{a}ck models predict stronger D4000, for a given CN3883, than seen 
at low redshift, but nevertheless the models illustrate that variations in age and metallicity are completely degenerate 
in the two indices over the parameter ranges populated by the galaxies in our samples.
See also our discussion in J\o rgensen \& Chiboucas (2013) regarding the ability of the 
Maraston \& Str\"{o}mb\"{a}ck models to correctly model the CN features. 

Turning to Figure \ref{fig-stellarpops}g and h, it is clear that the strong CN3883 and D4000 values for 
part of the Lynx sample and the majority of the RXJ0152.7--1357 and RXJ1226.9+3332 galaxies are unassociated 
with a change in the M/L ratio, and that they are also not well-modeled with the Maraston \& Str\"{o}mb\"{a}ck models.
A possible explanation may be that these models assume solar abundance ratios, while
the quiescent massive galaxies are known to have above solar $\alpha$-element, carbon and nitrogen abundances, 
see Conroy et al.\ (2014) and J\o rgensen et al.\ (2017) and references therein.
Detailed stellar population models giving both M/L-ratios and SEDs, from which we can derive model 
CN3883 and D4000 values, are not yet available for such element enhanced models.

\subsection{Composite Spectra \label{SEC-COMPOSITES} }

To investigate further in particular the strong CN3883 indices, we derive composite spectra by stacking the 
available Lynx spectra according to their environment as well as the presence of significant [\ion{O}{2}] emission.
Based on the phase space diagram, Figure \ref{fig-phasespace} discussed in Section \ref{SEC-DIFFERENCES},
we establish three composite spectra as follows.
The {\it cores-composite} is made up of all passive galaxies inside $R_{500}$ in either Lynx E or Lynx W.
The {\it outskirts-composite} is made up of all passive galaxies outside the cluster cores.
Finally, the {\it emission-composite} is made up of all bulge-dominated galaxies with [\ion{O}{2}]$>$ 5 {\AA}, excluding
the two AGNs. As noted in Section \ref{SEC-DIFFERENCES} the galaxies included in both the {\it outskirts-composite} 
and the {\it emission-composite} populate the phase space diagram in areas that are either recently accreted by the clusters
or still infalling. 
We repeat the determination of the composite spectra and the following analysis also for 
assignments using the X-ray cluster centers. In this case, ID 545 is assigned to the {\it outskirts-composite}
instead of the {\it cores-composite}, and ID 148 is assigned to the {\it cores-composite} instead of the {\it outskirts-composite}.
As the effects of this alternative assignment are minimal, the figures only show the results for assignments
using the BCGs as cluster centers, while we comment on results from both in the text.
We also establish a high S/N composite spectrum of SDSS spectra of 60 galaxies in the Coma cluster 
spanning the same range in velocity dispersion ($\log \sigma =2.1-2.4$) as the Lynx sample.
Figure \ref{fig-composite} shows all of these composite spectra, as well as the ratios between the 
Lynx composites and the Coma cluster composite.

Except for the presence of the [\ion{O}{2}] emission, the {\it emission-composite} appears similar to 
the {\it outskirts-composite}. 
The equivalent width of [\ion{O}{2}] in the {\it emission-composite} is $14.4\pm 0.2$ {\AA}, which
we convert to a star formation rate (SFR) using the average luminosity of the galaxies in the 
composite ($L_B=10^{11.08} L_{B,\sun}$) and the calibrations from Kennicutt (1992) and Gallagher et al.\ (1989), 
see also J\o rgensen et al.\ (2014). We find ${\rm SFR}=3.39\pm 0.05 M_{\sun}\,{\rm yr^{-1}}$.
The {\it emission-composite} exclude the two galaxies with obvious neon emission presumably originating 
from AGNs in the galaxies, cf.\ Section \ref{SEC-SAMPLE}. 
However, we cannot rule out that lower level AGN emission contributes to the strength
of the [\ion{O}{2}] emission in the {\it emission-composite}. 
Therefore the SFR should be taken as an upper limit.
Given the limit on the SFR, the emission in the {\it emission-composite} 
may originate from weak residual star formation, 
which has little influence on bulk of the stellar population, and/or from weak AGNs.

The main difference between the {\it cores-composite} and the {\it outskirts-composite} is the strength of 
the absorption within the CN3883 passband. Relative to the Coma composite, representing
the low redshift reference, the {\it outskirts-composite} has very weak CN3883, as also seen from the measurements
of the individual line indices. The {\it cores-composite} on the other hand has stronger CN3883 approaching the 
absorption strength in the Coma composite. 
Thus, the composite spectra confirm our result from the individual line index measurements.

\begin{figure}
\begin{center}
\epsfxsize 8.5cm
\epsfbox{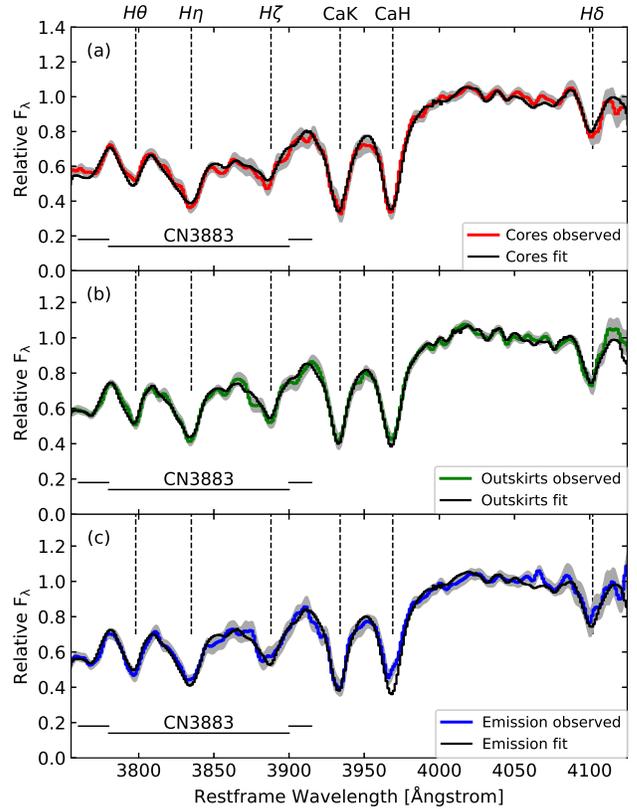}
\end{center}
\caption{
The composite spectra shown with the best fit models of SSP SEDs with ages of 1--3Gyr.
Panel (a): Red line shows the {\it cores-composite}.
Panel (b): Green line shows the {\it outskirts-composite}. 
Panel (c): Blue line shows the {\it emission-composite}. 
Black on all panels shows the best fit model SEDs, see Table \ref{tab-compositefit}.
Gray areas indicate three times the uncertainty on the observed spectra.
\label{fig-compositefit} }
\end{figure}

To quantify the origin of the difference and also explore any differences between the
{\it emission-composite} and the {\it outskirts-composite}, we fit all three composite spectra with linear combinations 
of a set of SSP models from Maraston \& Str\"{o}mb\"{a}ck (2011). The fitting was done with the same kinematics
fitting software used in our determinations of the velocity dispersions, see Gebhardt et al.\ (2000, 2003) for 
a description of the software.
We use SSP models with a Salpeter (1955) IMF. 
We use only models with ages between 1 Gyr and 3 Gyr. The absence of strong emission lines limit the ages to 1 Gyr or older
and is in agreement with our results based on the M/L ratios of the galaxies.
The upper limit on the ages is chosen to accommodate formation redshifts of up to $\approx$3.5 and is also in agreement with
our results from the strength of H$\zeta _A$ and the M/L ratios (see Figure \ref{fig-stellarpops}).
In reality the composite spectra are fully described with only four models. Figure \ref{fig-compositefit} shows the 
data and the best fits, while the linear compositions of the fits are summarized in Table \ref{tab-compositefit}.
The uncertainties are determined using Monte-Carlo simulations in which we add $1 \sigma$ additional noise to the composite spectra.
In each case, 100 realizations of such spectra are used, and the uncertainties are quoted as the rms scatter of 
the results from these realizations.

The {\it cores-composite} is dominated by stellar populations of super-solar metallicity ([M/H]=0.3), mostly with ages of 3 Gyr, combined 
with a $\approx 40$\% contribution by luminosity from a stellar population of sub-solar metallicity ([M/H]=--0.3) and an age of 1 Gyr.
The {\it outskirts-composite} can be modeled almost completely by 1 Gyr old stellar populations of both super-solar and sub-solar metallicity.
The {\it emission-composite} is modeled primarily by the models with sub-solar metallicity, with only about a quarter of the 
flux originating from a 1 Gyr old super-solar stellar population. In Section \ref{SEC-DISCUSSION} we will address to what
extent these mixes of stellar populations may evolve to match our low redshift reference sample spectra in the time available.
Table \ref{tab-compositefit} also lists the results for the {\it cores-composite} and {\it outskirts-composite}
if using the X-ray centers for assignments to the composites. In this case a larger fraction of the {\it outskirts-composite}
best fit has sub-solar metallicity. However, the main result is unchanged: The {\it outskirts-composite} is dominated by stellar populations
only 1 Gyr old, while the {\it cores-composite} contains a $\approx 50$\% contribution from 3 Gyr old stellar populations with super-solar metallicity.

\begin{deluxetable*}{llrrrrr}
\tablecaption{SED Fits to Composite Spectra\label{tab-compositefit} }
\tablewidth{0pc}
\tabletypesize{\scriptsize}
\tablehead{
                    & & \multicolumn{2}{c}{[M/H]=-0.3} & \multicolumn{2}{c}{[M/H]=0.3} & \\
\colhead{Composite} & \colhead{Cluster centers} & \colhead{1 Gyr} & \colhead{3 Gyr} & \colhead{1 Gyr} & \colhead{3 Gyr} & \colhead{$\chi ^2$} \\
\colhead{(1)} & \colhead{(2)} & \colhead{(3)} & \colhead{(4)} & \colhead{(5)} & \colhead{(6)} & \colhead{(7)}
}
\startdata
Cores     & BCGs    & $0.38\pm 0.05$ & $0.00\pm 0.00$ & $0.11\pm 0.08$ & $0.51\pm 0.03$ & 2.6  \\    
Outskirts & BCGs    & $0.33\pm 0.07$ & $0.05\pm 0.08$ & $0.62\pm 0.14$ & $0.00\pm 0.02$ & 2.6  \\  
Emission  & \nodata & $0.50\pm 0.06$ & $0.27\pm 0.05$ & $0.23\pm 0.11$ & $0.00\pm 0.04$ & 2.5  \\ 
Cores     & X-ray   & $0.28\pm 0.04$ & $0.06\pm 0.04$ & $0.12\pm 0.08$ & $0.54\pm 0.02$ & 3.1  \\    
Outskirts & X-ray   & $0.40\pm 0.06$ & $0.16\pm 0.13$ & $0.43\pm 0.13$ & $0.00\pm 0.02$ & 2.2  \\    
\enddata
\tablecomments{The fits are listed as fractions of luminosity originating from each SSP model.
Cluster centers based on BCGs or X-ray centers as noted for the {\it cores-composite} and {\it outskirts-composite}.}
\end{deluxetable*}

\section{Discussion \label{SEC-DISCUSSION} }

There are two main themes in our discussion of the results: 
(1) to what extent do the data support or allow structure evolution (sizes and velocity dispersions), and 
(2) can the stellar populations evolve passively from $z=1.27$ to the present and be consistent with
the observed stellar populations in the low redshift reference sample.

The cores of Lynx E and W are very different, Lynx E resembling the cores of very massive clusters at similar
redshifts, e.g. RDCS\,J1252.9--2927 at $z=1.24$ (Nantais et al.\ 2013), and 
XMMU\,J2235.2--2557 at $z=1.39$ (Gr\"{u}tzbauch et al.\ 2012), 
dominated by passive bulge-dominated galaxies, while Lynx W lacks a well-defined core of such galaxies. 
However, the passive bulge-dominated 
galaxies within $R_{500}$ of the Lynx W center appear similar to those in the Lynx E core.
Thus, in the following we focus on the evolution of the galaxy sample as a whole, rather than
the difference between the two clusters. 

\subsection{Structure Evolution \label{SEC-DISCSIZE}}

The bulk size growth of the bulge-dominated galaxies in the Lynx clusters from $z=1.27$ to the present
is limited to $\le 0.1$ dex at a fixed dynamical mass, whether we consider the full 
sample or only the passive galaxies.
This is consistent with our previous result from the smaller Lynx W sample (J\o rgensen et al.\ 2014).
Our low redshift reference sample contains 84 galaxies with ages $\le 9$Gyr (J\o rgensen et al.\ 2018b).
If we compare only to these galaxies and thereby correct for the progenitor bias due to the 
age of the stellar populations (van Dokkum \& Franx 2001), the allowed
bulk size growth of the bulge-dominated Lynx galaxies is reduced to about half.
This size growth is significantly less than found in other studies of passive bulge-dominated galaxies in 
clusters, e.g., Saglia et al.\ (2010) and Beifiori et al.\ (2017). 

The simple models for minor and major mergers (Bezanson et al.\ 2009) are overlaid in Figure \ref{fig-lrelmass}
as arrows. 
For clarity we show arrows corresponding to a change in size of $\Delta \log r_e =0.3$.
For minor mergers this corresponds to $\Delta \log {\rm Mass}=0.15$, while for major mergers
$\Delta \log {\rm Mass}=0.3$, see Section \ref{SEC-METHOD}.
If the bulk size growth of 0.1 dex allowed by the data is due to dry minor mergers, 
then under the assumption that the merged galaxies contain similar stellar populations then the 
luminosity (in the absence of evolution) will increase as the mass ($\Delta \log L = 0.05$), and
the M/L ratios will be unchanged.
The mean surface brightness will decrease by $\Delta \log \langle I \rangle _e = -0.15$.
As noted in Section \ref{SEC-FP}, even if we relax the assumption of the merging galaxies 
containing similar stellar populations, the very small size evolution and mass growth 
allowed by the data limit the changes in the M/L ratios to about $\Delta \log M/L < 0.1$
in the case of even large age differences.

The main issue from the previous paragraph is that
the small size growth that can be accommodated by our Lynx data and the presumably minor merging associated with such growth 
is insufficient to produce brightest cluster galaxies (BCG) in these clusters similar to those seen in low redshift clusters. 
The central galaxy in Lynx E (ID 4942) at a dynamical mass of $10^{11.1} M_{\sun}$ is a factor 6--10 less massive 
than the two central galaxies in the Coma cluster.
If it were to reach a mass similar to the Coma central galaxies by major merging with its nearest neighbors in the cluster core,
its size would grow by a similar factor, cf.\ Bezanson et al.\ (2009). 
With an effective radius of only $\log r_e {\rm [kpc]} =0.1$ ID 4942 is located well below the size-mass relation 
for the low redshift reference sample. 
A series of major mergers leading to a factor six size and mass growth would place it very close to the low 
redshift relation. 
To evaluate whether such a scenario is realistic given the galaxy content of the cluster core will require
more complete spectroscopic information for galaxies near the core.
We do not have spectroscopy of the triple galaxy in Lynx W that we suspect will form the central galaxy
in that cluster. If we use the total magnitudes of the three components
to estimate the final mass of a merger of all three cores,
the total mass would be of the order $10^{11.6} M_{\sun}$, or about half the mass of one of the 
central galaxies in the Coma cluster. 
Our brief discussion of major mergers as a path to building the BCGs is in agreement with
results from Lidman et al.\ (2013), who found that between $z\approx 1$ and the present 
major mergers may be the dominant path to growing the mass of the BCGs. 

Our sample in Lynx W also contains (newly accreted) bulge-dominated emission line galaxies 
of similar mass, specifically ID 209 at ${\rm Mass} = 10^{11.7} M_{\sun}$ and 
ID 293 at ${\rm Mass} = 10^{11.4} M_{\sun}$. 
These galaxies may represent galaxies that will enter the sample of passive bulge-dominated galaxies 
and by $z\approx 0$ affect the location of the size-mass relation. This type of progenitor bias
is discussed in detail by Belli et al.\ (2015) who estimate that half of the evolution of the 
size-mass relation from $z=2$ to 1.25 may be due to larger galaxies ceasing star formation
and entering the samples of passive galaxies. Presumably such an effect would still be of importance
at later epochs and the massive bulge-dominated emission line galaxies in Lynx W may be 
candidates to become the passive massive BCGs at low redshift.

In summary, the data for our sample of bulge-dominated galaxies set strict limits on the overall
size (and mass) growth of the galaxies. However, the data are consistent with either a limited
amount of major mergers and the associated structure evolution, or alternatively quenching of 
star formation in massive bulge-dominated emission line galaxies, both of which can be the 
source of the BCGs seen at low redshift.

\begin{figure}
\begin{center}
\epsfxsize 6.8cm
\epsfbox{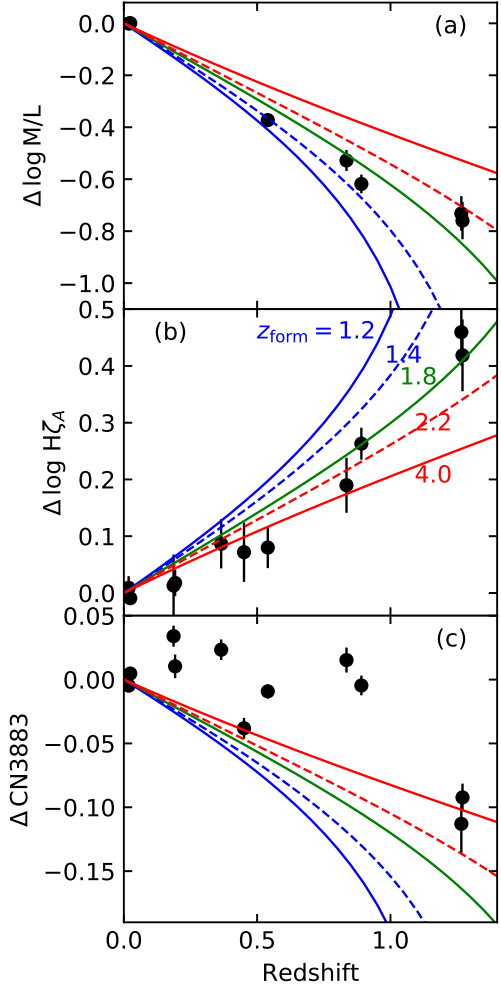}
\end{center}
\caption{
The zero point offsets of the scaling relations for the $z=0.2-1.3$ GCP cluster samples 
relative to the low redshift reference samples, shown as a function of redshift. 
The offsets are calculated from the zero points, $\gamma$, as 
$\Delta = \gamma_{\rm high-z} - \gamma_{\rm low-z}$.
Predictions from models for passive evolution based on models 
from Maraston (2005) and Maraston \& Str\"{o}mb\"{a}ck (2011)
are overplotted, labeled with the assumed formation redshift $z_{\rm form}$.
\label{fig-zp} }
\end{figure}

\begin{figure}
\begin{center}
\epsfxsize 8.5cm
\epsfbox{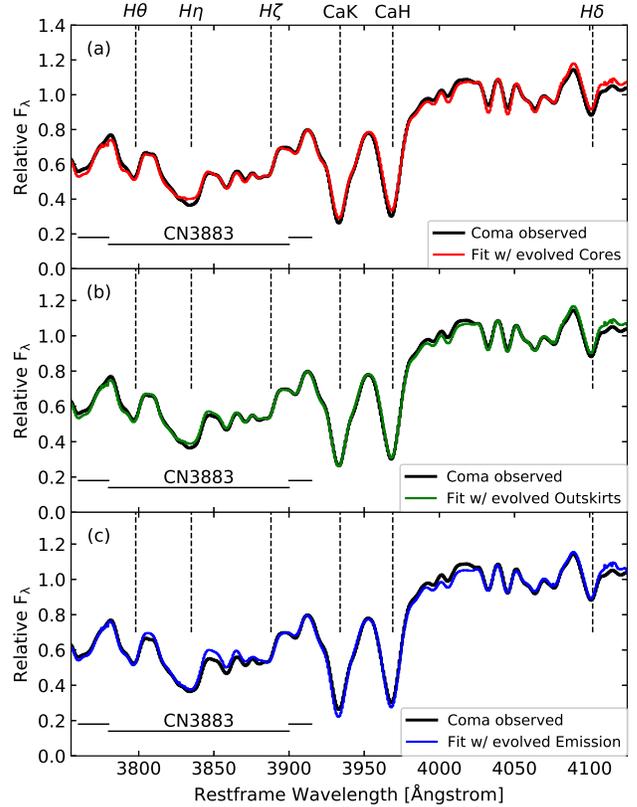}
\end{center}
\caption{
Fits to the Coma composite using templates made by evolving the Lynx best-fit multi-component fits by 8 Gyr.
Black line -- Coma composite spectrum. The uncertainty on the spectrum is within the width of the line.
Red line (panel a) -- best fit using the evolved Lynx {\it cores-composite} fit.
Green line (panel b) -- best fit using the evolved Lynx {\it outskirts-composite} fit.
Blue line (panel c) -- best fit using the evolved Lynx {\it emission-composite} fit.
\label{fig-comafit} }
\end{figure}

\subsection{Stellar Population Evolution Measured by the Scaling Relations \label{SEC-DISCPOPS}}

In Figure \ref{fig-zp} we show the zero point differences between the $z>0.2$ GCP clusters, including the two Lynx
clusters, and the low redshift reference sample, for the main scaling relations.
The zero point differences for the $z=0.2-0.9$ GCP clusters are
fully consistent with our results in J\o rgensen \& Chiboucas (2013) and J\o rgensen et al.\ (2014, 2017), but for 
consistency have been recalculated using the scaling relations listed in Tables \ref{tab-relations} and \ref{tab-indexrelations}
in the present paper. 
The zero point differences for the Lynx data are shown for Lynx E and W separately, though there is no significant 
difference between the results for the two clusters.
Predictions for passive evolution based on the models from Maraston (2005) and Maraston \& Str\"{o}mb\"{a}ck (2011)
are overlaid in Figure \ref{fig-zp}. 
The changes in M/L ratios and the H$\zeta _{\rm A}$ indices are both consistent with formation redshifts of 
$z_{\rm form}\approx 1.9$. 
When comparing our results to the literature, we focus on the highest mass galaxies $M_{\rm dyn}\ge 10^{11.5} M_{\sun}$,
for which the M/L ratios give ages of about 3 Gyr and $z_{\rm form}\approx 3.7$.
This is in agreement with recent results from Beifiori et al.\ (2017) who based on
galaxies in three $z=1.4-1.6$ clusters found ages of 1.2--2.3 Gyr and formation redshifts of $z_{\rm form}=2.3-3$.

Turning to the CN3883 index, Figure \ref{fig-zp}c, it is clear that the index is not well-modeled by the 
passive evolution model based on the Maraston \& Str\"{o}mb\"{a}ck (2011) SSP SEDs.
It is not clear if this is a shortcoming of the models, or due to more complex mixtures of the stellar populations
being present and not well-modeled by the SSPs. While in principle the Lynx cluster data on average are consistent with
a formation redshift of $z_{\rm form}=4$, as we saw in Section \ref{SEC-INDICES}, the sample contains
many galaxies with CN3883 indices as strong as in the low redshift reference sample.
There are no other studies of this index at $z>1$. 
Kriek et al.\ (2016) studied a high mass $z=2.1$ galaxy, COSMOS-11494, based on Keck/MOSFIRE spectroscopy and 
found a super-solar abundance ratio of [Mg/Fe]=0.45. 
A comparison of the spectrum from Kriek et al.\ to our composite spectra in the wavelength interval of the CN3883
index shows that COSMOS-11494 has a CN3883 feature stronger than the Lynx {\it outskirts-composite}
spectrum, but not as strong as the {\it cores-composite}.
Thus, the CN3883 feature supports the super-solar abundance ratio of this galaxy, and also
hints that strong CN3883 features may be present at very early epochs in some massive galaxies.  
Onodera et al.\ (2015) measured the CN$_2$ index from their stacked spectrum of passive $z\approx 1.6$ galaxies.
Converting the very weak CN$_2$ index measured by these authors to CN3883 using our empirical conversion 
based on $z=0.5-0.9$ galaxies (J\o rgensen \& Chiboucas 2013) gives CN3883$\approx 0.13$. 
This is similar to our results for the Lynx galaxies outside the cluster cores. 
The galaxies in the Onodera et al.\ sample may reside in groups or clusters,
evaluating from these authors' redshift information for the galaxies. However, it is not possible for
us to evaluate how massive these clusters or groups may be.

\subsection{Stellar Population Evolution Probed by the Composite Spectra}

To further evaluate to what extent passive evolution is sufficient to turn the stellar populations
of the bulge-dominated Lynx galaxies into stellar populations similar to those in the low redshift 
reference sample, we test whether the SED fits to the Lynx composite spectra can evolve into
models compatible with the Coma composite spectrum, in the time available between $z=1.27$ and the present.
We first convert the luminosity fractions from our best fits, Table \ref{tab-compositefit}, into mass fractions
using the M/L ratios for the models from Maraston (2005).
Then under the assumption of mass conservation we derive the model spectra of the best fits aged by 8 Gyr. 
Finally, we attempt to fit the Coma composite spectrum with each of the three such evolved
best fit spectra. The result is shown in Figure \ref{fig-comafit}.
From panels (a) and (b) it is clear that after 8 Gyr of passive evolution, we can no longer detect
any significant differences between what originated from the {\it cores-composite} and the {\it outskirts-composite}. 
This is a consequence of the stellar population models being quite insensitive to small
relative age differences once the ages are above about 8-10 Gyr.
On the other hand, the evolved {\it emission-composite} model does not fit the Coma composite well, showing significant
residuals in the vicinity of the CN3883 absorption feature, see Figure \ref{fig-comafit}c.
Thus, for the galaxies included in the {\it emission-composite} passive evolution is not 
sufficient for the galaxies to become similar to the Coma galaxies. The main problem is that the metallicity is too low. 
For this composite to result in a spectrum consistent with the Coma composite, the star formation needs to 
continue in order to recycle the metals into new stars such that the metallicity increases. 
However, such a scenario may be difficult to accommodate considering how low the SFR is 
already ($3.4 M_{\sun}\, {\rm yr^{-1}}$) and that the galaxies are in the process of falling into the clusters
and presumably completing the quenching process before they cross the cores of the clusters. 
For a galaxy at $R_{200}=1.52 R_{500}$ in these clusters to reach the cluster core takes typically 0.9 Gyr.
For simplicity we will assume that the SFR stays at $3.4 M_{\sun}\, {\rm yr^{-1}}$ producing another 
$3\cdot 10^9 M_{\sun}$ of new stars. However, this is only 1--10\% of the mass of the already 
existing stars and appears insufficient to change the luminosity weighted metallicity 
enough to reach those of the Coma cluster galaxies. 
A more complex star formation history involving an increased SFR followed by (rapid) quenching would 
be needed in order to reach the luminosity weighted metallicity of the passive Coma cluster galaxies
without the resulting luminosity weighted ages becoming too low.
Any prolonged star formation in these galaxies would also be in conflict with their possible evolution
into the largest passive galaxies in the clusters as proposed in Section \ref{SEC-DISCSIZE}.

\section{Conclusions \label{SEC-CONCLUSION}}

We have used deep ground-based optical spectroscopy from Gemini North and {\it HST}/ACS imaging to investigate the
structure and stellar populations of galaxies in Lynx E and W at redshift $z=1.27$, both part of the 
Lynx Supercluster.
Our main conclusions are as follows:

\begin{enumerate}

\item 
The Lynx E cluster core is similar to the cores of  RDCS\,J1252.9--2927 and  XMMU\,J2235.2--2557
dominated by bulge-dominated passive galaxies and void of star forming galaxies.
Lynx W lacks a well-defined core of bulge-dominated passive galaxies and appears to contain a
significant fraction of emission line galaxies.  
However, a closer inspection of the phase space diagram shows that the emission line galaxies
in general are either recently accreted or still infalling.

\item
At a given dynamical mass, the galaxies in the Lynx clusters show only
a very small difference in size and velocity dispersion when compared to
our low redshift reference sample. However, to produce BCGs similar to those seen in massive
low redshift clusters, the central galaxies will have to grow by at least a factor five,
possibly through major merging with available nearby galaxies of similar sizes and masses. 
Alternatively, large and massive galaxies with current star formation 
(e.g., emission line galaxies as present in the Lynx W sample) 
may become quenched and enter the passive population by a later epoch.
Such newly quenched galaxies may also merge with passive galaxies in the cluster
centers, though in some cases their masses and sizes are already similar to those of low redshift BCGs.

\item
The bulge-dominated galaxies in the Lynx clusters populate a Fundamental
Plane (FP) similar to that seen for lower redshift galaxies. However, the slope is steeper;
in the M/L-mass plane it is as steep as for the $z=0.9$ cluster galaxies. 
While the data are consistent with passive evolution with a median formation redshift
of $z_{\rm form} = 1.89 _{-0.10}^{+0.14}$, the steeper slope means that the low mass galaxies may be as 
young as 1 Gyr, while the highest mass galaxies have stellar population ages
of about 3 Gyr. The ages are also supported by the strong H$\zeta _{\rm A}$ absorption line indices.

\item
The passive galaxies in the cores of the two Lynx clusters have significantly stronger
CN3883 indices than found for the passive galaxies in the outskirts of the clusters.
Using composite spectra, the {\it cores-composite} is well modeled by approximately 
a 60:40 combination of 
a super-solar metallicity and 3 Gyr old stellar population and a sub-solar metallicity and 1 Gyr old stellar population.
The {\it outskirts-composite} on the other hand can be modeled almost completely by 1 Gyr old stellar populations
mixed 60:40 in super-solar metallicity and sub-solar metallicity. Thus, the galaxies must
have gone though different star formation histories.
Evolving the best-fit models passively by 8 Gyr shows that both models can then equally well
fit the spectra of the Coma cluster galaxies. 

\item 
The composite of bulge-dominated galaxies with emission lines, the {\it emission-composite},
is best fit with approximately a 75:25 combination sub-solar and super-solar metallicity models. 
Evolving such a model by 8 Gyr shows that it cannot satisfactorily fit the Coma cluster galaxies.
Thus, ongoing star formation is needed in these Lynx galaxies for them to evolve to consistency with
Coma cluster stellar populations. However, the current level of star formation as measured from the 
[\ion{O}{2}] emission seems to be insufficient to do so before the star formation may be quenched 
as the galaxies fall into the cluster core. In addition, weak AGN emission may
contribute to the [\ion{O}{2}] strength, making the contribution from current star formation even
lower.

\end{enumerate}

Further insight into the stellar populations of the Lynx bulge-dominated galaxies would 
benefit greatly from deep near-infrared spectroscopy enabling measurements of strengths of 
the iron and magnesium absorption lines. This would enable us to study the total metallicity
as well as abundance ratios in greater detail and possibly understand to what extent these 
galaxies can evolve passively into galaxies similar to those in our low redshift reference sample.

\vspace{0.5cm}

Acknowledgments:
Karl Gebhardt is thanked for making his kinematics software available.
Kristi Webb and Charity Woodrum are thanked for their contributions to the calibration
of the low redshift reference sample photometry. Ricardo Demarco is thanked for 
comments on a draft version of this paper.
The paper makes use of simulations software produced by Riley Peterson
(https://github.com/rileypeterson/Gemini-Python-Scripts).
The anonymous referee is thanked for comments and suggestions that helped improve this paper.

Based on observations obtained at the Gemini Observatory (processed using the Gemini IRAF package), 
which is operated by the Association of Universities for Research in Astronomy, Inc., under a 
cooperative agreement with the NSF on behalf of the Gemini partnership: the National Science 
Foundation (United States), the National Research Council (Canada), CONICYT (Chile), 
Ministerio de Ciencia, Tecnolog\'{i}a e Innovaci\'{o}n Productiva (Argentina), 
Minist\'{e}rio da Ci\^{e}ncia, Tecnologia e Inova\c{c}\~{a}o (Brazil), 
and Korea Astronomy and Space Science Institute (Republic of Korea).
The data presented in this paper originate from the Gemini programs GN-2014B-Q-22 and GN-2014B-DD-4.

In part, based on observations made with the NASA/ESA {\it Hubble Space Telescope}, 
obtained from the data archive at the Space Telescope Science Institute. 

The paper makes use of photometry and spectroscopy from Sloan Digital Sky Survey (SDSS).
Funding for SDSS-IV has been provided by the Alfred P. Sloan Foundation,
the U.S. Department of Energy Office of Science, and the Participating Institutions. SDSS-IV acknowledges
support and resources from the Center for High-Performance Computing at
the University of Utah. The SDSS web site is www.sdss.org.

Data from the {\it Chandra X-ray Observatory} archive were used to aid the interpretation
presented in this paper.

I.J., C.O., and R.C. acknowledge support from grant HST-AR-13255.01 from STScI.
STScI is operated by the Association of Universities for Research in Astronomy, 
Inc. under NASA contract NAS 5-26555.

The Cosmic Dawn Center (DAWN) is funded by the Danish National Research Foundation.
S.T. acknowledges support from the ERC Consolidator Grant funding scheme (project Context, grant No. 648179)


\appendix

\begin{deluxetable*}{rrrrr rrrr rrrr}
\tablecaption{Lynx Photometric Parameters from {\it HST}/ACS data \label{tab-phot} }
\tablewidth{0pc}
\tablehead{
\colhead{ID} & \colhead{ID$_{\rm J2014}$} & \colhead{RA (J2000)} & \colhead{DEC (J2000)} & \colhead{$m_{\rm tot,SEx}$} & \colhead{$(i_{775}-z_{850})$}&  
\colhead{$m_{\rm tot,dev}$} & \colhead{$\log r_{\rm e,dev}$} & \colhead{$m_{\rm tot,ser}$} & \colhead{$\log r_{\rm e,ser}$} 
& \colhead{$n_{\rm ser}$} & \colhead{PA} & \colhead{$\epsilon$} \\
\colhead{(1)} & \colhead{(2)} & \colhead{(3)} & \colhead{(4)} & \colhead{(5)} & \colhead{(6)} & \colhead{(7)} 
& \colhead{(8)} & \colhead{(9)} & \colhead{(10)} & \colhead{(11)} & \colhead{(12)} & \colhead{(13)}
}
\startdata
104 & 1698 & 132.15605 & 44.89076 & 24.45 & 0.552 & 24.36 & -0.906 & 24.36 & -0.906 & 1.5 & 79.9 & 0.58 \\
129 & 1533 & 132.17006 & 44.91985 & 24.64 & 0.679 & 23.98 & -0.401 & 24.51 & -0.753 & 0.6 & -88.4 & 0.63 \\
145 & 1644 & 132.15846 & 44.90064 & 22.36 & 0.704 & 21.30 & 0.536 & 22.24 & -0.130 & 0.4 & -36.8 & 0.34 \\
148 & 1748 & 132.15444 & 44.89276 & 23.07 & 0.933 & 22.76 & -0.556 & 22.87 & -0.643 & 3.0 & -86.4 & 0.39 \\
197 & 1809 & 132.15401 & 44.89896 & 24.35 & 0.931 & 23.96 & -0.256 & 24.33 & -0.562 & 1.0 & -2.9 & 0.50 \\
209 & 1763 & 132.14989 & 44.89334 & 21.68 & 0.971 & 21.12 & 0.200 & 21.36 & 0.021 & 2.9 & -13.9 & 0.26 \\
254 & 2015 & 132.14189 & 44.88400 & 24.42 & 0.886 & 24.20 & -1.121 & 24.14 & -1.086 & 4.8 & -52.5 & 0.13 \\
293 & 1888 & 132.15065 & 44.90477 & 22.14 & 0.919 & 21.98 & -0.333 & 21.63 & -0.063 & 5.8 & 44.2 & 0.24 \\
316 & 2063 & 132.14194 & 44.89228 & 23.18 & 0.781 & 22.77 & -0.393 & 23.04 & -0.593 & 2.1 & -70.4 & 0.28 \\
347 & 2138 & 132.13880 & 44.89084 & 24.12 & 0.494 & 23.56 & -0.284 & 24.12 & -0.670 & 0.8 & -24.0 & 0.52 \\
\enddata
\tablecomments{Column 1: Galaxy ID. Column 2: ID from J\o rgensen et al.\ (2014).
Columns 3 and 4: Positions consistent with USNO (Monet et al.\ 1998), with an rms scatter of $\approx 0.5$ arcsec.
Column 5: Total F850LP magnitude from SExtractor.
Column 6: Aperture color within an aperture with radius 0.5 arcsec.
Column 7: Total magnitude from fit with $r^{1/4}$ profile.
Column 8: Logarithm of the effective radius in arcsec from fit with $r^{1/4}$ profile. 
Columns 9 and 10: Total magnitude and logarithm of the effective radius, from fit with a S\'{e}rsic profile.
Column 11: S\'{e}rsic index. 
Column 12: Position angle of major axis measured from North through East; col.\ (13) ellipticity.
This table is available in its entirety as a machine readable table. A portion is shown here for guidance on its content.
}
\end{deluxetable*}

\begin{figure*}
\begin{center}
\epsfxsize 17cm
\epsfbox{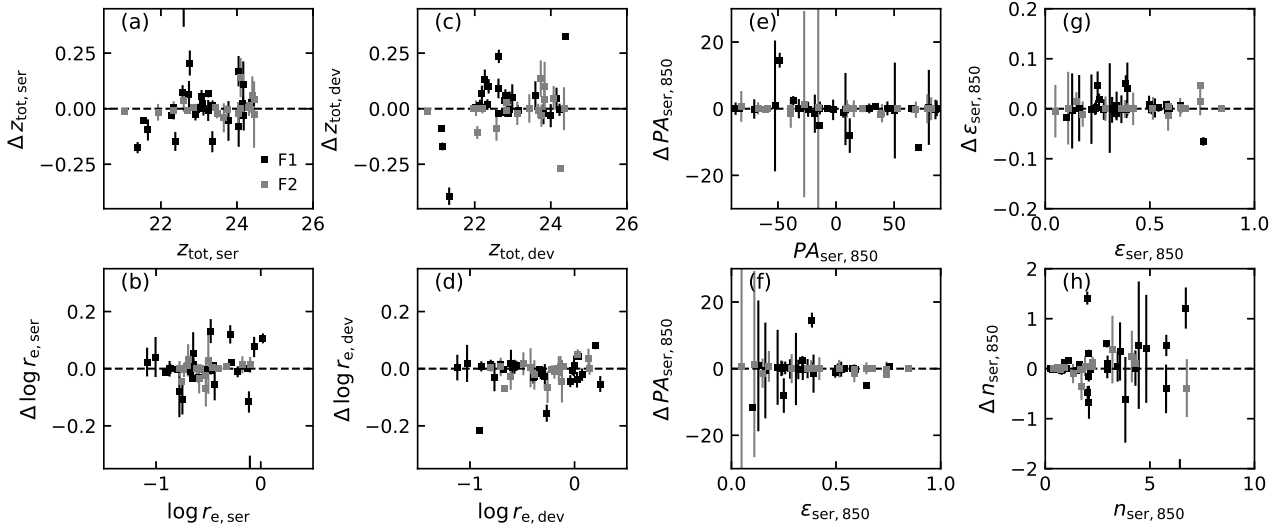}
\end{center}
\caption{
Comparison of our rederived {\it HST} photometry with the those used in J\o rgensen et al.\ (2014).
Black points -- galaxies in field 1 containing the center of Lynx W; 
grey points -- galaxies in field 2, which is between the cluster centers of Lynx E and W.
The comparisons are summarized in Table \ref{tab-phot2014comp_galfit}.
\label{fig-phot2014comp_galfit} }
\end{figure*}

\begin{deluxetable}{lrr}
\tablecaption{Comparison with J\o rgensen et al.\ (2014) \label{tab-phot2014comp_galfit} }
\tablehead{\colhead{Parameter} & \colhead{$\Delta$} & \colhead{rms}}
\startdata
$m_{\rm tot,ser}$ &  0.001 & 0.10 \\
$m_{\rm tot,dev}$ &  0.002 & 0.11 \\
$\log r_{\rm e,ser}$ & $-0.0003$ & 0.078 \\
$\log r_{\rm e,dev}$ & $-0.0001$ & 0.053 \\
PA                   & $-0.004$ & 3.2 \\
$\epsilon$           & 0.0001 & 0.018 \\       
$n_{\rm ser}$        & 0.005 & 0.54 \\
\enddata
\tablecomments{All comparisons are for the filter F850LP and include 47 galaxies. $\Delta$ list the median differences 
''this paper''--''J\o rgensen et al.\ (2014)''.}
\end{deluxetable}

\section{Photometry from {\it HST}/ACS \label{SEC-IMAGING}}

Table \ref{tab-phot} lists the photometric parameters for the spectroscopic sample as derived
from the {\it HST}/ACS observations in F850LP and F775W. The F850LP images were processed to derive
2-dimensional surface photometry using GALFIT (Peng et al.\ 2002). F775W was used for the color determinations, only.
The effective radii in Table \ref{tab-phot} are derived from the semi-major and -minor axes as 
$r_{\rm e} = (a_{\rm e}\,b_{\rm e})^{1/2}$. 
The following faint galaxies in the spectroscopic sample were not refit in the new processing of the {\it HST} imaging:
IDs 129, 197, 780, 1133, and 2911. 
Of these IDs 129 and 197 are faint disk galaxies in Lynx W, while none of the other galaxies are members of the Lynx clusters. 
In Table \ref{tab-phot}, we list the photometry from J\o rgensen et al.\ (2014) for these galaxies.
All magnitudes in Table \ref{tab-phot} have been corrected for galactic extinction.

In Figure \ref{fig-phot2014comp_galfit} and Table \ref{tab-phot2014comp_galfit}, we compare the new determinations for 
the Lynx W galaxies with those used in J\o rgensen et al.\ (2014).
There are no significant offsets between the two sets of parameters.

As in J\o rgensen et al.\ (2014), we derive rest-frame $B$-band from the observed $z_{\rm 850}$ magnitudes
and colors, using calibrations established based on
Bruzual \& Charlot (2003) stellar population models spanning the observed color range.
For the calibration, we adopt the $U$, $B$, and $V$ filter functions from Ma\'{i}z Apell\'{a}niz (2006), 
rather than the older filter functions distributed with the Bruzual \& Charlot models.
For the Lynx cluster redshift of $z=1.27$ the calibration using these new filter functions is
\begin{equation}
B_{\rm rest}=  z_{\rm 850} +1.416-1.038 (i_{775} - z_{850}) 
\end{equation}
The large color term is due to the fact that at the cluster redshift the F850LP filter spans the 
4000{\AA} break.
For Lynx member galaxies on the red sequence, this calibration will give $B_{\rm rest}$ values 0.1 mag 
fainter than the calibration used in J\o rgensen et al.\ (2014). 
As noted in J\o rgensen et al.\ (2018a), the effect of the new calibration for the $z<1$ GCP clusters is 
typically 0.05 mag, with the new calibration leading to fainter rest-frame B-band magnitudes. 
All GCP data have been recalibrated using the Ma\'{i}z Apell\'{a}niz (2006) filter functions.
The distance modulus for our adopted cosmology is $DM(z)$ = 44.74 at $z=1.27$.
The absolute B-band magnitude, $M_{\rm B}$, is then derived as
\begin{equation}
M_{\rm B} = B_{\rm rest} - DM(z).
\end{equation}
Techniques for how to calibrate to a ``fixed-frame'' photometric system are described in
detail by Blanton et al.\ (2003).

\begin{deluxetable}{lr}
\tablecaption{Spectroscopic Observations \label{tab-inst} }
\tablewidth{0pt}
\tabletypesize{\scriptsize}
\tablehead{}
\startdata
Instrument                  & GMOS-N       \\
CCDs                        & 3 $\times$ E2V DD 2048$\times$4608 \\
r.o.n.\tablenotemark{a}     & (3.17,3.22,3.46) e$^-$       \\
gain\tablenotemark{a}       & (2.31,2.27,2.17) e$^-$/ADU  \\
Pixel scale                 & $0\farcs 0727 {\rm pixel^{-1}}$ \\
Field of view               & $5\farcm5\times5\farcm5$ \\
Grating                     & R400\_G5305 \\
Spectroscopic filter        & OG515\_G0306 \\
Wavelength range\tablenotemark{b} & 5500-10500\AA  \\
Slit width x slit length    & $1\arcsec \times 2\farcs 75$ \\
Total exposure time         & 55,800 sec \\
Number of frames            & 31 \\
Image quality\tablenotemark{c} & $0\farcs 57$ \\
Instrumental resolution\tablenotemark{d} & 3.002 \AA, 100 $\rm km\,s^{-1}$ \\  
Aperture\tablenotemark{e}   & $1\arcsec \times 0\farcs 7$, $0\farcs 53$ \\
S/N\tablenotemark{f}        & 23.3 \\
\enddata
\tablenotetext{a}{Values for the three detectors in the array.}
\tablenotetext{b}{The exact wavelength range varies from slitlet to slitlet.}
\tablenotetext{c}{Image quality measured as the average FWHM at 8000{\AA} of the blue stars included in the masks.}
\tablenotetext{d}{Median instrumental resolution derived as sigma in Gaussian fits to the sky lines of the stacked 
spectra. The second entry is the equivalent resolution in $\rm km\,s^{-1}$ at 4000{\AA} in the rest frame of the cluster.}  
\tablenotetext{e}{The first entry is the rectangular extraction aperture 
(slit width $\times$ extraction length). The second entry is the radius in an equivalent 
circular aperture, $r_{\rm ap}= 1.025 (\rm {length} \times \rm{width} / \pi)^{1/2}$, cf.\ J\o rgensen et al.\ (1995).}
\tablenotetext{f}{Median S/N per {\AA}ngstrom for the 16 cluster members, in the rest frame of the cluster, measured in the wavelength interval 3750-4100{\AA}. }
\end{deluxetable}

\section{Spectroscopic Data \label{SEC-SPECTROSCOPY}}

Here we summarize processing of the spectroscopic data, and provide additional information
useful for evaluating the robustness of our results. 
The reader is also referred to the information in J\o rgensen et al.\ (2014), as we do
not repeat the tests described in that paper.

The spectroscopic observations were obtained in multi-object spectroscopic (MOS) mode with GMOS-N
in the period UT 2014 Nov 27 to 2015 Jan 12 under the Gemini program IDs GN-2014B-Q-22 and GN-2014B-DD-4.
All observations were carried out in queue mode.
Table \ref{tab-inst} summarizes the instrumentation, exposure times and other information
relevant for the observations.
The observations were processed using the same methods as used for Lynx W, and described in 
detail in J\o rgensen et al.\ (2014), see also J\o rgensen \& Chiboucas (2013). 
The processing includes standard steps for bias subtraction, flat fielding, and sky subtraction for nod-and-shuffle
observations. As for Lynx W, we use a spatially dependent correction for the charge diffusion effect.
All frames were then combined and 1-dimensional spectra extracted, and calibrated to a relative flux scale.

\begin{figure*}
\epsfxsize 17.5cm
\epsfbox{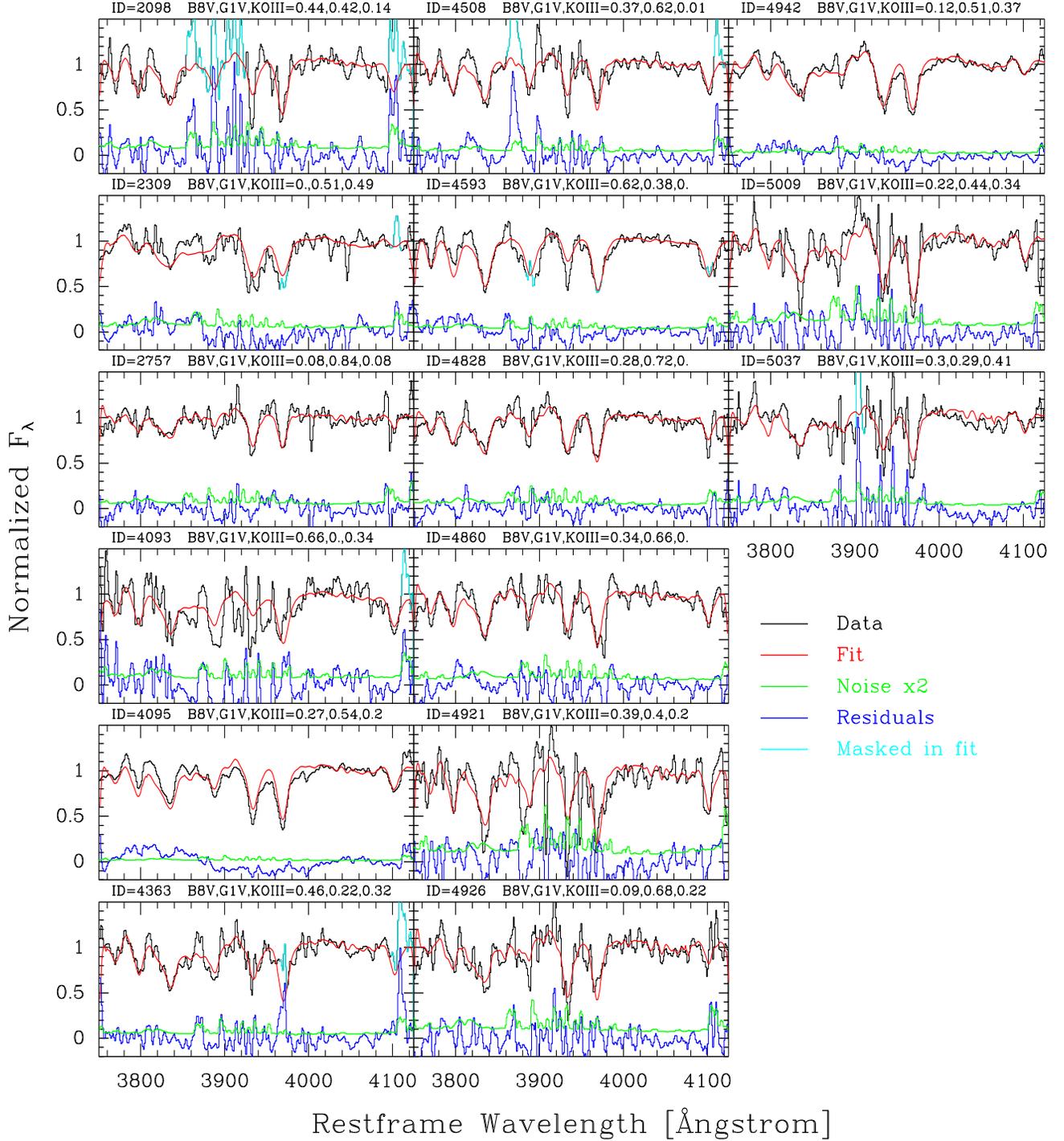}
\caption{
Summary of the kinematics fitting for the new spectra obtained primarily for members of Lynx E, and for which determination
of the velocity dispersion was possible.
Black -- normalized spectra; red -- best fit; green -- noise times two, normalized the same way as
the spectra; blue - residuals from the best fit; cyan -- wavelength regions excluded from the 
fits due to emission lines or strong sky subtraction residuals.
The figure shows the general quality of the fits and demonstrates that the use of the three
template stars adequately spans the stellar populations present in the sample.
\label{fig-kinfit} }
\end{figure*}

\begin{figure}
\begin{center}
\epsfxsize 8.5cm
\epsfbox{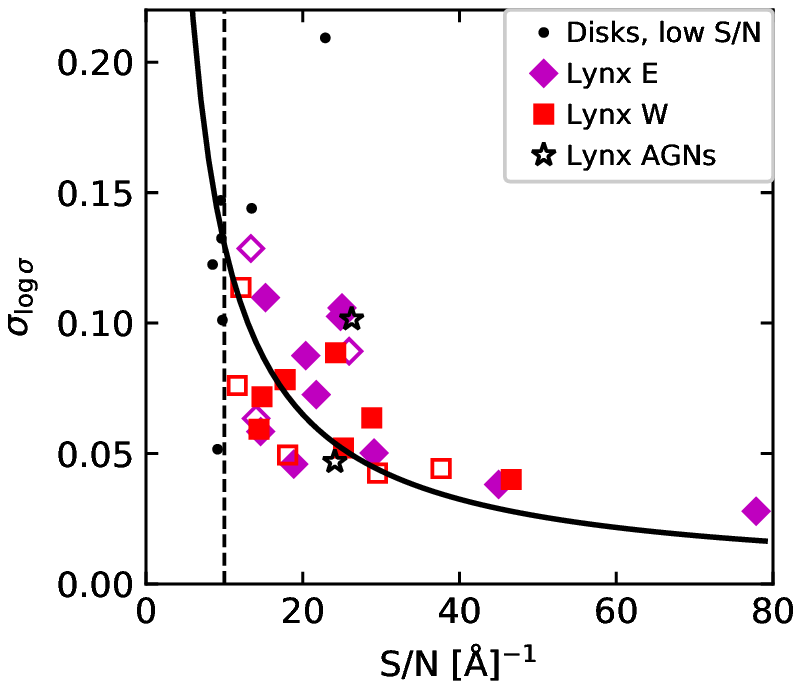}
\end{center}
\caption{
The uncertainty on the velocity dispersions as derived using a bootstrap method versus the 
S/N per {\AA} in the rest frame of the galaxies. Red boxes -- Lynx W. Magenta diamonds -- Lynx E.
Solid symbols -- bulge-dominated galaxies 
with EW[\ion{O}{2}] $\le 5${\AA}; open symbols --  bulge-dominated galaxies 
with EW[\ion{O}{2}] $> 5${\AA}; 
small black squares -- bulge-dominated galaxies with S/N $<10$ and disk-dominated galaxies. 
Solid line -- approximate relation between S/N and $\sigma _{\log \sigma}$: 
$\sigma _{\log \sigma} \approx 1.3 \times {\rm S/N^{-1}}$.
Dashed vertical line -- S/N=10. Galaxies below this S/N cutoff are excluded from the determination
of relations and zero points.
\label{fig-SNelsigma} }
\end{figure}

The calibrated spectra were fit with stellar templates as described in J\o rgensen et al.\ (2014).
This results in determination of the redshifts and the velocity dispersions.
The fits were performed with the kinematics fitting software made available by Karl Gebhardt, see
Gebhardt et al.\ (2000, 2003) for a description of the software.
Spectra of member galaxies were fit in the wavelength range 3750--4125{\AA}. 
Because the software determines the fits in pixel space, it is straightforward to mask wavelength ranges 
not to be included in the fit. We use this to flag areas of strong residuals from the sky subtraction
as well as emission lines. The instrumental resolution is determined 
from stacked sky spectra processed in the same way as the science spectra.
As in J\o rgensen \& Chiboucas (2013), we use three template stars with spectral types K0III, G1V and B8V.
Figure \ref{fig-kinfit} shows the normalized spectra, fits and residuals in the wavelength
region covered by the fits. The purpose of the set of template stars is to span the stellar 
populations in the galaxies. From Figure \ref{fig-kinfit} we conclude that the use of these three stars 
accomplishes this.  Further, we choose to use the same template stars as in our previous work 
to ensure consistency with our previous publications.
The software determines the line-of-sight-velocity-distribution (LOSVD) from the science spectra
and then the velocity dispersion is derived from the LOSVD through both a Gauss-Hermite polynomial fit
and a Gaussian fit. For the 15 member galaxies with velocity dispersion determined, the offset in 
log space between the two measurements is $-0.034$ with the Gauss-Hermite polynomial fits giving the 
smaller velocity dispersions.  
For consistency with our previous data, we use the velocity dispersions 
derived from the Gauss-Hermite polynomial fits to the LOSVD, rather than
the Gaussian fits. 
Aperture correction of the velocity dispersions were performed using the technique from 
J\o rgensen et al.\ (1995).

The uncertainties, $\sigma _{\log \sigma}$, on the velocity dispersions are derived by the fitting 
software using a bootstrap method. In Figure \ref{fig-SNelsigma} we show the resulting uncertainties 
versus the S/N of the spectra. The uncertainties scale approximately with S/N as 
$\sigma _{\log \sigma} \approx 1.3 \times {\rm S/N^{-1}}$. This relation is shown on the figure.
The relation is not used to determine uncertainties on our measurements, 
but may be useful for other researchers planning similar observations.

The spectra cover wavelengths to 4400{\AA} in the rest frame. However, the sky subtraction errors limit
the range for which the spectra can be used for determination of absorption line indices to 
$\approx 4100${\AA} in the rest frame (9250{\AA} observed).
Thus, we can derive the following indices CN3883, CaHK, D4000 and H$\zeta _{\rm A}$. 
The indices CN3883 and CaHK are defined in Davidge \& Clark (1994).
For the high order Balmer line index H$\zeta _{\rm A}$ we adopt the definition from Nantais et al.\ (2013). 
For D4000 we use a shorter red passband than used in the standard definition. The resulting
D4000 indices are calibrated to consistency with the usual definition from Gorgas et al.\ 1999), see J\o rgensen et al.\ (2014).

The line indices have been aperture corrected and corrected for velocity dispersion as described in 
J\o rgensen et al.\ (2005), see J\o rgensen \& Chiboucas (2013) for a discussion of the method
applied to intermediate redshift galaxies.
As for the lower order Balmer lines, we assume that H$\zeta _{\rm A}$ has no aperture correction.

For galaxies with detectable emission from  [\ion{O}{2}] we determined the equivalent width 
of the [\ion{O}{2}]$\lambda\lambda$3726,3729 doublet.
With an instrumental resolution of $\sigma \approx 3$\,{\AA} (FWHM $\approx 7$\,{\AA}), 
the doublet is not resolved in our spectra and we refer to it simply as the
``[\ion{O}{2}] line''. 

Table \ref{tab-spec} lists the results from the template fitting and the derived line strengths.

\section{Presentation of the imaging and spectra for the new data \label{SEC-SPECIM}}

\begin{figure*}
\epsfxsize 17.5cm
\epsfbox{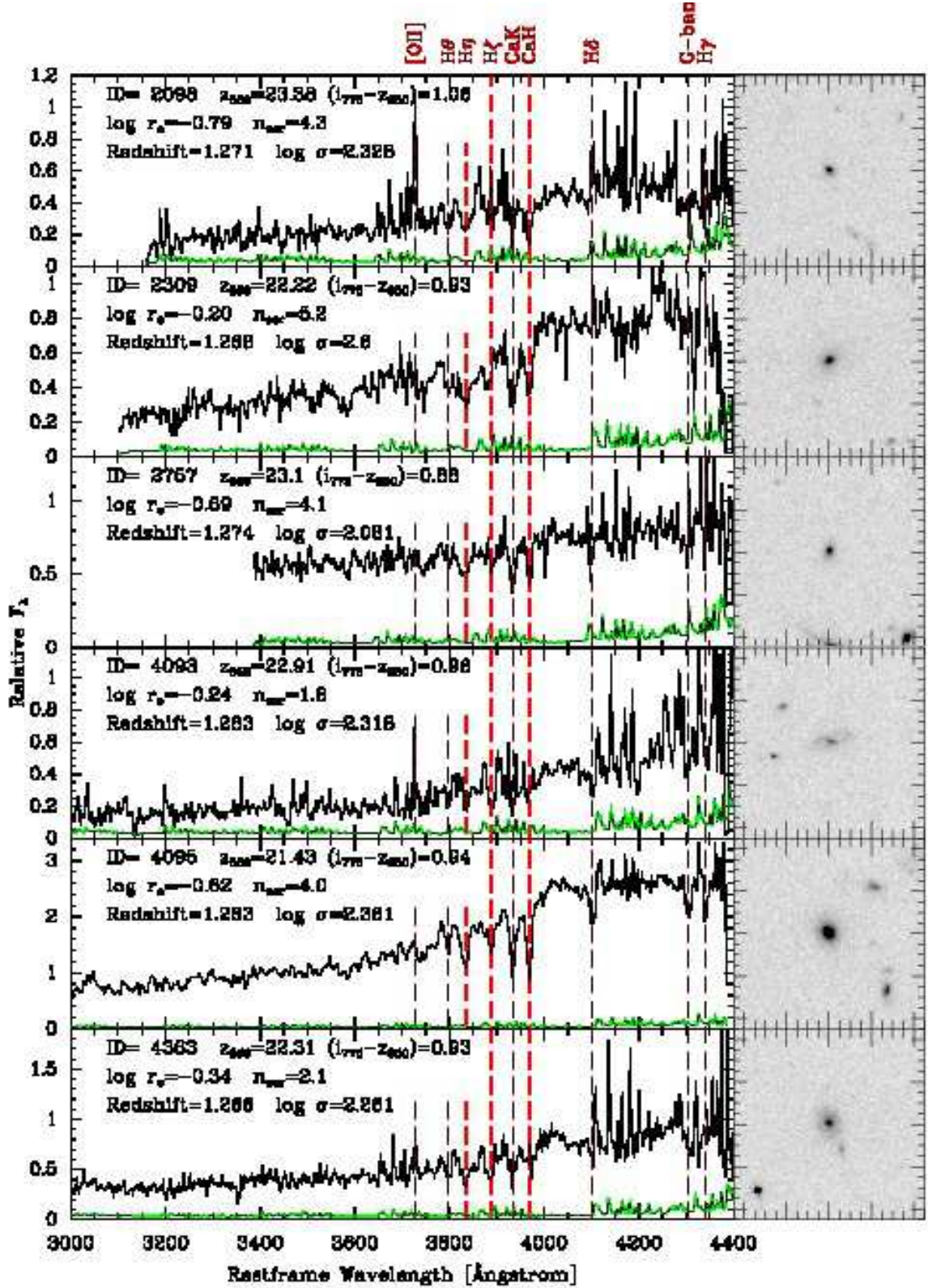}
\vspace{-0.5cm}
\caption{Spectra and gray scale images of the galaxies that are considered members of either Lynx E or W.
On the spectra, black lines show the galaxy spectra, green lines show the random noise multiplied by two.
At the strong sky lines, the random noise underestimates the real noise due to systematic
errors in the sky subtraction. 
Some of the absorption lines are marked. The location of the emission line [\ion{O}{2}],
is also marked, though emission is only present in some of the galaxies. 
The gray scale images are made from the {\it HST}/ACS images in the F850LP filter. 
Each image is 9 arcsec $\times $ 9 arcsec.
At the distance of Lynx E and W this corresponds to 75 kpc $\times$ 75 kpc for our adopted cosmology. 
North is up, East to the left.
The panels are labeled with key photometric and spectroscopic parameters. The effective radius, $r_e$,
is given as the logarithm of $r_e$ in arcsec from the S\'{e}rsic profile fit. The velocity dispersion, $\sigma$,
is given as the logarithm of $\sigma$ in $\rm km\,s^{-1}$.
The spectroscopic data shown in this figure are available in the online journal.
\label{fig-speclynx} }
\end{figure*}

The new spectra as well as stamp-sized images of the galaxies from the {\it HST}/ACS 
imaging of the cluster members are shown in Figure \ref{fig-speclynx}.
The stamps cover the equivalent of 75 kpc $\times$ 75 kpc at the distance of the cluster.
The spectra used to create Figure \ref{fig-speclynx} are available in the online journal.

\begin{deluxetable*}{rrrrrrrrrrr  rrr rrr rrr r}
\rotate
\tablecaption{Spectroscopic Parameters \label{tab-spec} }
\tablewidth{0pt}
\tabletypesize{\scriptsize}
\tablehead{
\colhead{ID} & \colhead{Redshift} & \colhead{Member\tablenotemark{a}} &  
\colhead{$\log \sigma$} & \colhead{$\log \sigma _{\rm cor}$\tablenotemark{b}} & 
\colhead{$\sigma _{\log \sigma}$} & \multicolumn{3}{c}{Template fractions} & \colhead{$\chi ^2$} & \colhead{S/N\tablenotemark{c}}
& \colhead{CN3883}& \colhead{$\sigma _{\rm CN3883}$} & \colhead{CaHK}& \colhead{$\sigma _{\rm CaHK}$} 
& \colhead{D4000} & \colhead{$\sigma _{\rm D4000}$}
& \colhead{H$\zeta _{\rm A}$}& \colhead{$\sigma _{\rm H\zeta _A}$} 
& \colhead{EW [\ion{O}{2}]} & \colhead{$\sigma _{\rm EW[OII]}$} \\
\colhead{}&\colhead{} &\colhead{} &\colhead{} &\colhead{} &\colhead{} & \colhead{B8V} & \colhead{G1V} & \colhead{K0III} &
}
\startdata
2076 & 1.1409 & 0 & \nodata & \nodata & \nodata & \nodata & \nodata & \nodata & \nodata & 5.8 & \nodata & \nodata & \nodata & \nodata & \nodata & \nodata & \nodata & \nodata & 56.3 & 28.3 \\
2098 & 1.2714 & 1 & 2.304 & 2.328 & 0.063 & 0.44 & 0.42 & 0.14 & 3.4 & 14.0 & 0.126 & 0.015 & 24.58 & 0.91 & 1.863 & 0.011 & 2.52 & 0.40 & 20.2 & 3.2 \\
2309 & 1.2682 & 1 & 2.576 & 2.600 & 0.073 & 0.00 & 0.51 & 0.49 & 3.4 & 21.7 & 0.297 & 0.012 & 19.89 & 0.61 & 2.098 & 0.009 & 2.10 & 0.32 & \nodata & \nodata \\
2333 & 1.0138 & 0 & \nodata & \nodata & \nodata & \nodata & \nodata & \nodata & \nodata & 24.1 & \nodata & \nodata & \nodata & \nodata & \nodata & \nodata & \nodata & \nodata & 3.2 & 0.3 \\
2478 & 1.0731 & 0 & 2.186 & 2.210 & 0.077 & 1.00 & 0.00 & 0.00 & 2.0 & 27.6 & \nodata & \nodata & 6.53 & 0.45 & 1.388 & 0.004 & 2.87 & 0.17 & 39.7 & 2.5 \\
2626 & 0.6777 & 0 & 2.410 & 2.432 & 0.226 & 0.74 & 0.26 & 0.00 & 1.6 & 10.8 & \nodata & \nodata & \nodata & \nodata & \nodata & \nodata & \nodata & \nodata & \nodata & \nodata \\
2688 & 1.2236 & 0 & 2.225 & 2.249 & 0.138 & 0.82 & 0.18 & 0.00 & 1.9 & 9.9 & 0.085 & 0.019 & 20.75 & 1.34 & 1.292 & 0.010 & 6.44 & 0.34 & 45.3 & 17.9 \\
2757 & 1.2740 & 1 & 2.057 & 2.081 & 0.052 & 0.08 & 0.84 & 0.08 & 2.3 & 25.2 & 0.116 & 0.009 & 11.12 & 0.59 & 1.545 & 0.006 & 1.54 & 0.28 & \nodata & \nodata \\
3047 & 1.0370 & 0 & 1.972 & 1.995 & 0.077 & 0.40 & 0.60 & 0.00 & 4.7 & 26.1 & \nodata & \nodata & 17.53 & 0.41 & \nodata & \nodata & 2.95 & 0.20 & \nodata & \nodata \\
4093 & 1.2635 & 1 & 2.294 & 2.318 & 0.129 & 0.66 & 0.00 & 0.34 & 3.8 & 13.4 & 0.139 & 0.017 & 16.89 & 1.12 & 1.889 & 0.013 & 6.31 & 0.35 & 18.1 & 1.9 \\
4095 & 1.2631 & 1 & 2.337 & 2.361 & 0.028 & 0.27 & 0.54 & 0.20 & 10.7 & 77.8 & 0.122 & 0.003 & 18.43 & 0.20 & 1.886 & 0.002 & 3.14 & 0.08 & \nodata & \nodata \\
4123 & 1.1371 & 0 & 2.203 & 2.227 & 0.052 & 0.81 & 0.19 & 0.00 & 2.3 & 17.2 & 0.080 & 0.012 & 17.01 & 0.83 & 1.511 & 0.008 & 4.61 & 0.28 & 42.9 & 9.4 \\
4128 & 1.1999 & 0 & 2.139 & 2.163 & 0.095 & 0.00 & 0.80 & 0.20 & 3.6 & 16.1 & 0.061 & 0.013 & 21.45 & 0.76 & 1.665 & 0.008 & 2.76 & 0.28 & 5.2 & 0.4 \\
4216 & 1.2293 & 0 & 2.408 & 2.432 & 0.086 & 1.00 & 0.00 & 0.00 & 2.1 & 16.8 & \nodata & \nodata & 6.29 & 1.00 & 1.230 & 0.006 & 3.47 & 0.28 & 35.9 & 3.6 \\
4298 & 1.1991 & 0 & 2.227 & 2.251 & 0.109 & 0.50 & 0.27 & 0.23 & 2.3 & 11.1 & 0.381 & 0.022 & 5.25 & 1.28 & 1.986 & 0.014 & 3.38 & 0.46 & 10.9 & 0.5 \\
4363 & 1.2662 & 1 & 2.237 & 2.261 & 0.106 & 0.46 & 0.22 & 0.32 & 3.5 & 25.0 & 0.140 & 0.010 & 13.47 & 0.56 & 1.874 & 0.007 & 3.39 & 0.25 & 2.9 & 0.2 \\
4508 & 1.2644 & 1 & 2.064 & 2.088 & 0.102 & 0.37 & 0.62 & 0.01 & 3.3 & 26.2 & 0.063 & 0.009 & 14.12 & 0.58 & 1.796 & 0.007 & 3.66 & 0.20 & 28.4 & 1.7 \\
4593 & 1.2692 & 1 & 2.391 & 2.415 & 0.050 & 0.62 & 0.38 & 0.00 & 2.7 & 29.1 & 0.079 & 0.008 & 16.69 & 0.47 & 1.864 & 0.006 & 6.21 & 0.16 & 4.5 & 0.5 \\
4628 & 1.1718 & 0 & \nodata & \nodata & \nodata & \nodata & \nodata & \nodata & \nodata & 6.6 & \nodata & \nodata & \nodata & \nodata & \nodata & \nodata & \nodata & \nodata & 35.9 & 1.7 \\
4638 & 1.5870 & 0 & \nodata & \nodata & \nodata & \nodata & \nodata & \nodata & \nodata & 2.6 & \nodata & \nodata & \nodata & \nodata & \nodata & \nodata & \nodata & \nodata & 162.1 & 77.8 \\
4722 & \nodata & 3 & \nodata & \nodata & \nodata & \nodata & \nodata & \nodata & \nodata & 1.6 & \nodata & \nodata & \nodata & \nodata & \nodata & \nodata & \nodata & \nodata & \nodata & \nodata \\
4828 & 1.2686 & 1 & 2.244 & 2.268 & 0.089 & 0.28 & 0.72 & 0.00 & 1.4 & 25.9 & 0.120 & 0.009 & 19.34 & 0.55 & 1.847 & 0.007 & 3.24 & 0.23 & 10.5 & 0.7 \\
4860 & 1.2587 & 1 & 2.166 & 2.190 & 0.088 & 0.34 & 0.66 & 0.00 & 1.6 & 20.4 & 0.102 & 0.012 & 15.63 & 0.81 & 2.003 & 0.009 & 2.50 & 0.29 & \nodata & \nodata \\
4906 & \nodata & 3 & \nodata & \nodata & \nodata & \nodata & \nodata & \nodata & \nodata & 2.2 & \nodata & \nodata & \nodata & \nodata & \nodata & \nodata & \nodata & \nodata & \nodata & \nodata \\
4921 & 1.2590 & 1 & 2.011 & 2.035 & 0.144 & 0.39 & 0.40 & 0.20 & 2.5 & 13.5 & \nodata & \nodata & 23.54 & 1.13 & 1.487 & 0.012 & 4.65 & 0.40 & \nodata & \nodata \\
4926 & 1.2681 & 1 & 2.053 & 2.077 & 0.110 & 0.09 & 0.68 & 0.22 & 1.7 & 15.2 & 0.224 & 0.016 & 18.89 & 0.90 & 1.770 & 0.011 & 3.06 & 0.42 & \nodata & \nodata \\
4942 & 1.2594 & 1 & 2.445 & 2.469 & 0.038 & 0.12 & 0.51 & 0.37 & 5.0 & 45.0 & 0.236 & 0.006 & 22.03 & 0.32 & 1.992 & 0.004 & 2.26 & 0.15 & \nodata & \nodata \\
5009 & 1.2622 & 1 & 2.206 & 2.230 & 0.058 & 0.22 & 0.44 & 0.34 & 1.8 & 14.6 & 0.323 & 0.017 & 22.13 & 0.94 & 2.079 & 0.013 & 3.09 & 0.44 & \nodata & \nodata \\
5037 & 1.2607 & 1 & 2.120 & 2.144 & 0.103 & 0.30 & 0.29 & 0.41 & 4.6 & 24.8 & 0.240 & 0.010 & 18.85 & 0.56 & 2.161 & 0.008 & 4.08 & 0.27 & \nodata & \nodata \\
\enddata
\tablenotetext{a}{Adopted membership: 1 -- galaxy is a member of Lynx E or W; 
0 -- galaxy is not a member of either Lynx E or W; 3 -- redshift cannot be determined.}
\tablenotetext{b}{Velocity dispersions corrected to a standard size aperture equivalent to a
circular aperture with diameter of 3.4 arcsec at the distance of the Coma cluster.}
\tablenotetext{c}{S/N per {\AA}ngstrom in the rest frame of the galaxy. The wavelength 
interval was chosen based on the redshift of the galaxy as follows: 
redshift 1.00-1.35 -- 3750-4100 {\AA}; 
redshift $<$1.00 -- 4100-4600 {\AA}.
For ID 4722 and 4906 a redshift of 1.27 was assumed for the S/N calculation.}
\tablecomments{The typical uncertainties on the redshifts are 0.0001. Uncertainties on the template fractions
are not derived by default the fitting software, but our independent Monte-Carlo simulations indicate uncertainties of 0.02-0.05.
The absorption line indices have been corrected for galaxy velocity dispersion and aperture corrected.
D4000 is measured as $\rm D4000_{short}$, see J\o rgensen et al.\ (2014). 
This table is also available as a machine readable table.}
\end{deluxetable*}

\section{Photometry for the Low Redshift Reference Sample \label{SEC-LOWZDETAIL} }

\subsection{Calibration of SDSS photometry}

In J\o rgensen et al.\ (2018b) we presented consistently calibrated spectroscopic parameters 
for a large low redshift reference sample of bulge-dominated galaxies in both the Perseus and Coma cluster.
The data were calibrated to consistency with our {\it Legacy Data} for the two clusters used in our previous 
GCP papers (Barr et al. 2005; J\o rgensen et al. 2005, 2014, 2017; J\o rgensen \& Chiboucas 2013).
In order to take advantage of this reference sample,
we establish the photometric parameters for the sample, and ensure that those 
are also consistent with the {\it Legacy Data}.
For the purpose of the use in the present paper we rely on the SDSS $g'$-band photometry of the galaxies.
The full detail and the calibrated data will be included in a future paper (I.\ J\o rgensen et al.\ 2019).
Here we summarize the main points and compare the calibrated data to the photometric
data from Simard et al.\ (2011), see Section \ref{SEC-SIMARD}.

The SDSS data release 14 (DR14) contains photometric parameters for the galaxies from $r^{1/4}$ profile fitting
({\tt devmag}, {\tt devrad}), exponential ({\tt expmag}, {\tt exprad}) profile fitting, and 
a total magnitude {\tt cmodelmag} from a best-fit linear combination of 
these two fits with information about the fraction {\tt fracdev} of the flux originating from the 
$r^{1/4}$ fit in the combination.
We use the SDSS photometric parameters to derive parameters that mimic S\'{e}rsic (1968) parameters.
These parameters are in the following referred to as {\it pseudo-S\'{e}rsic} parameters to not confuse them
with parameters from actual fits to S\'{e}rsic profiles. In the main body of the paper, we simply
refer to the parameters as S\'{e}rsic parameters.
We proceed as follows. We adopt {\tt cmodelmag} as the best-fit total magnitude, $m_{\rm tot,pseudo-Sersic}$. 
We then derive the matching half-light radius, $r_{\rm e,pseudo-Sersic}$, from the information 
about the linear combination of the $r^{1/4}$ profile and the exponential profile. 

Using the general equation the enclosed luminosity of a S\'{e}rsic profile within a radius (Graham \& Driver 2005) 
and the information about the linear combination of the $r^{1/4}$ and exponential profiles, we derive the following
\begin{eqnarray}
\label{eq-sersic}
\lefteqn{ f_{\rm dev} \left ( \frac{1}{\Gamma (8)} \gamma \left (8,b_4 (\frac{R}{r_{dev}})^{0.25} \right ) - 0.5 \right ) + } \\
& & (1-f_{\rm dev}) \frac{L_{exp}}{L_{dev}} \left ( \frac{1}{\Gamma (2)} \gamma \left (2,b_1 (\frac{R}{r_{exp}}) \right ) - 0.5 \right ) = 0 \nonumber
\end{eqnarray}
where $f_{\rm dev}$ is {\tt fracdev} from SDSS, the ratio of luminosities $L_{exp}\,L_{dev}^{-1}$ is calculated 
from the corresponding magnitudes {\tt devmag} and {\tt expmag}. 
The radii $r_{dev}$ and $r_{exp}$ are {\tt devrad} and {\tt exprad}, respectively, both circularized using the axis ratios from SDSS.
$\Gamma$ is the gamma function, $\gamma$ is the incomplete
gamma function, $b_1$ and $b_4$ are specific values of the parameter $b_n$ for which $\Gamma (2n) = 2\gamma (2n,b_n)$
and have values of $b_1=1.6783$ and $b_4=7.6692$.
Solving Equation \ref{eq-sersic} for $R$ gives $r_{\rm pseudo-Sersic}$ matching $m_{\rm tot,pseudo-Sersic}$.
From $r_{\rm pseudo-Sersic}$ and $m_{\rm tot,pseudo-Sersic}$ we can derive the mean surface brightness
within $r_{\rm pseudo-Sersic}$.

We correct the SDSS photometry for galactic extinction using the extinction info included in DR14.
The correction is based on the dust calibration from Schlafly \& Finkbeiner (2011). 
Finally we calibrate the SDSS $r^{1/4}$ and pseudo-S\'{e}rsic  parameters to consistency with the {\it Legacy Data} 
by comparing the SDSS $r^{1/4}$  parameters with to the {\it Legacy Data} $B$-band data for the Coma cluster.
The {\it Legacy Data} originally used the Burstein \& Heiles (1982) extinction correction. 
Here we instead adopt the calibration from
Schlafly \& Finkbeiner (2011), which gives 0.02 mag fainter extinction corrected magnitudes.

\begin{figure}
\begin{center}
\epsfxsize 8.5cm
\epsfbox{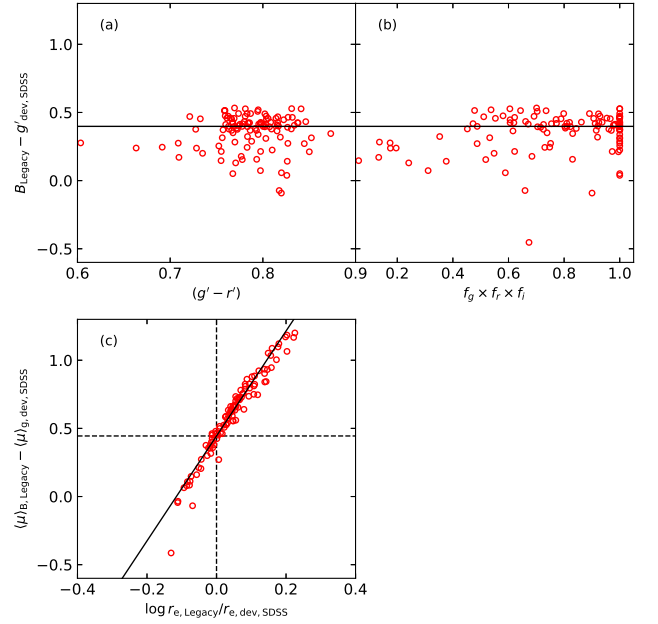}
\end{center}
\caption{
Comparison of the {\it Legacy Data} for the Coma cluster with the SDSS $r^{1/4}$ parameters.
Black solid lines on panels (a) and (b) show the median offset in magnitudes.
Black solid line on panel (c) is the best fit to the data, see text. 
\label{fig-legacy_sdss} }
\end{figure}

\begin{figure}
\begin{center}
\epsfxsize 8.5cm
\epsfbox{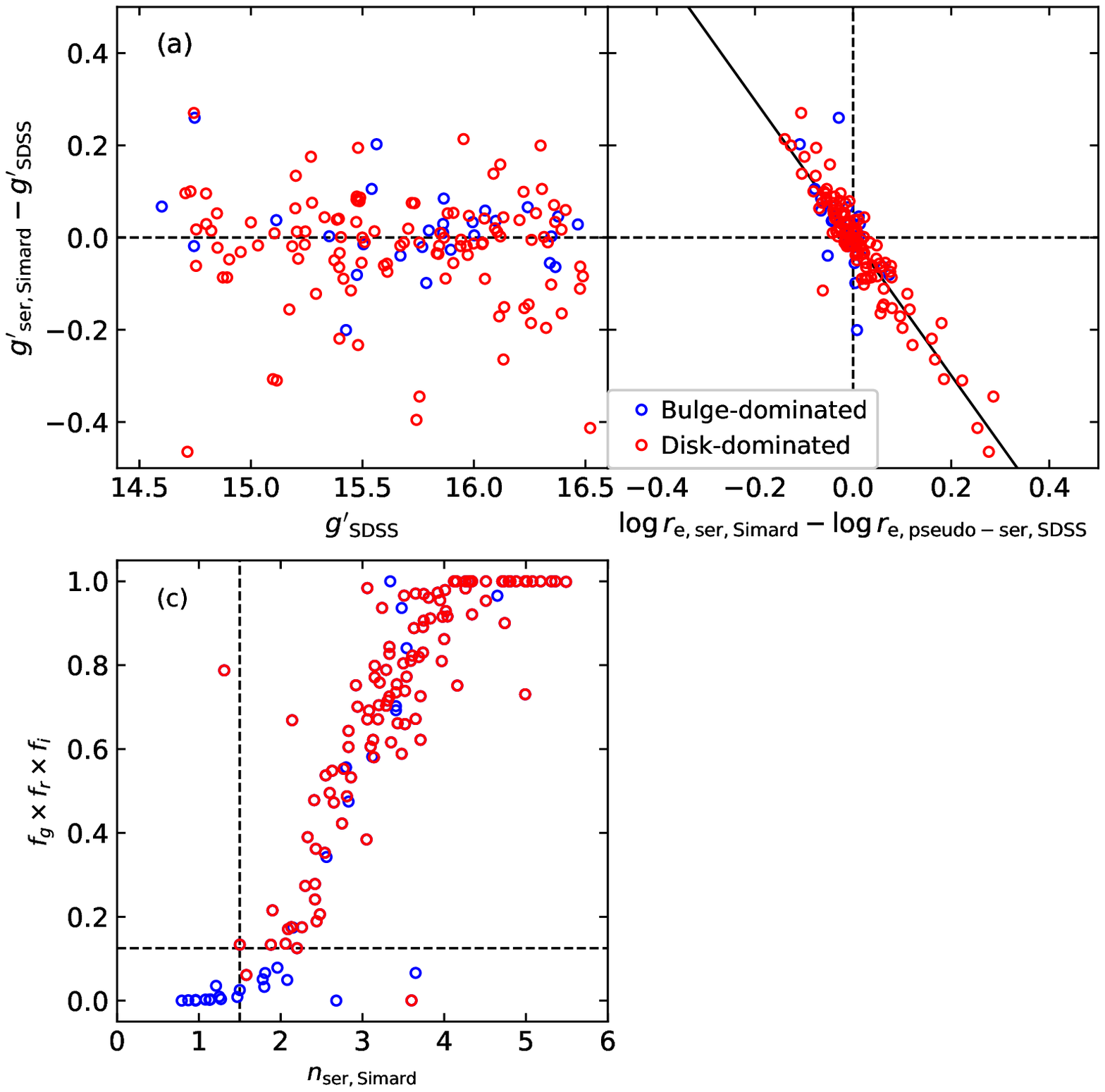}
\end{center}
\caption{
Panels (a) and (b): Comparison of the photometric parameters from Simard et al.\ (2011) for the Coma cluster 
with the SDSS pseudo-S\'{e}rsic parameters.
Red points -- bulge-dominated galaxies.
Blue points -- disk-dominated galaxies.
Black solid lines -- median offset on panel (b) and best fit on panel (b).
Panel (c) compares the product of the {\tt fracdev} parameters,
$f_g \times f_r \times f_i$, with the S\'{e}rsic index from Simard et al.\ (2011).
\label{fig-simard} }
\end{figure}

Figure \ref{fig-legacy_sdss} shows the comparison of the {\it Legacy Data} with the SDSS $r^{1/4}$ parameters.
On panels (a) and (b) the difference in total magnitudes, $B_{\rm Legacy}-g'_{\rm dev,SDSS}$, 
is shown versus color and the product of the three filters' {\tt fracdev}, $f_g \times f_r \times f_i$.
The magnitude offset does not depend on the color. About half of the tail of points scattering to low
values of $B_{\rm Legacy}-g'_{\rm dev,SDSS}$ are from galaxies with $f_g \times f_r \times f_i<0.4$. We take this as in indication
the growth curve method used for the {\it Legacy Data} (J\o rgensen et al.\ 1995) and the fitting implemented
in the SDSS pipeline handle galaxies with faint disks differently. 
The median magnitude difference is $0.398\pm 0.019$ mag.
In Figure \ref{fig-legacy_sdss}c, we show the difference in radii versus the difference in surface brightnesses.
The two photometric bands are close enough in wavelength space that we do not expect 
any significant difference in radii due to the color gradients of the galaxies.
As expected the differences are tightly correlated.
The best fit is shown as the solid line. The intercept gives the offset between the two magnitudes, $0.444\pm 0.005$ mag.
For both determinations of the magnitude difference, points deviating more than $3\sigma$ have been excluded.
To calibrate the SDSS $g'$-band photometry to consistency with the {\it Legacy Data} $B$-band photometry, we adopt the average 
of those two offsets, 0.421 mag.
Both $r^{1/4}$ SDSS and pseudo-S\'{e}rsic magnitudes are calibrated to the $B$-band using this offset. 
Due to the correlation between the errors on the surface brightnesses (or total magnitudes) and the 
effective radii, the intercept on Figure \ref{fig-legacy_sdss}c is less affected by the tail of points at 
lower $B_{\rm Legacy}-g'_{\rm dev,SDSS}$, than the median magnitude difference itself. 
Based on the difference between the adopted offset and the intercept value, we estimate that 
this tail in the distribution affects the calibration by no more than 0.02 mag, which is insignificant for our purpose.

The $B$-band photometry is calibrated to rest frame $B$-band using the $k$-correction info from Chilingarian et al.\ (2010). 
On average the correction for Coma cluster galaxies is 0.075 mag. In our use of the {\it Legacy data} we 
adopted the $k$-correction from Pence (1976), which is 0.111 mag for the Coma cluster.
Thus, together with the difference in the adopted galactic extinction the 
recalibrated new reference sample is 0.056 mag fainter than the {\it Legacy Data} used in previous GCP papers.

\subsection{Comparison with Literature Data \label{SEC-SIMARD} }

To ensure that our approach to converting the SDSS DR14 parameters to pseudo-S\'{e}rsic parameters
is valid, we compare the Coma cluster data with the S\'{e}rsic parameters from the 2-dimensional fitting of the 
SDSS imaging performed by Simard et al.\ (2011). 
The Perseus cluster is not included in the Simard et al.\ catalog. 
Figure \ref{fig-simard} summarizes the comparisons.
The agreement is very good with the total magnitudes as well as the effective radii being 
fully consistent with the 2-dimensional fitting parameters from Simard et al.\ (2011).
On  Figure \ref{fig-simard}c we take the opportunity to evaluate if our technique  
in J\o rgensen et al.\ (2018b) for selecting the bulge-dominated galaxies using the product 
of the SDSS {\tt fracdev} parameters, $f_g \times f_r \times f_i$, is consistent with using the S\'{e}rsic index, $n_{\rm ser}$.
The vertical and horizontal dashed lines on the figure show the limits in the two parameters
used in our previous sample selections, while the red points highlight the final selection of 
bulge-dominated Coma cluster galaxies (J\o rgensen et al.\ 2018b).
In general the two methods are equivalent. The few galaxies marked in blue at high $f_g \times f_r \times f_i$ and high
$n_{\rm ser}$ are galaxies that upon visual inspection were identified as spiral galaxies (cf.\ J\o rgensen et al.\ 2018b).

\section{Test of Sensitivity to Mass Determination Method \label{SEC-BETATEST} }

\begin{figure}
\epsfxsize 8.5cm
\epsfbox{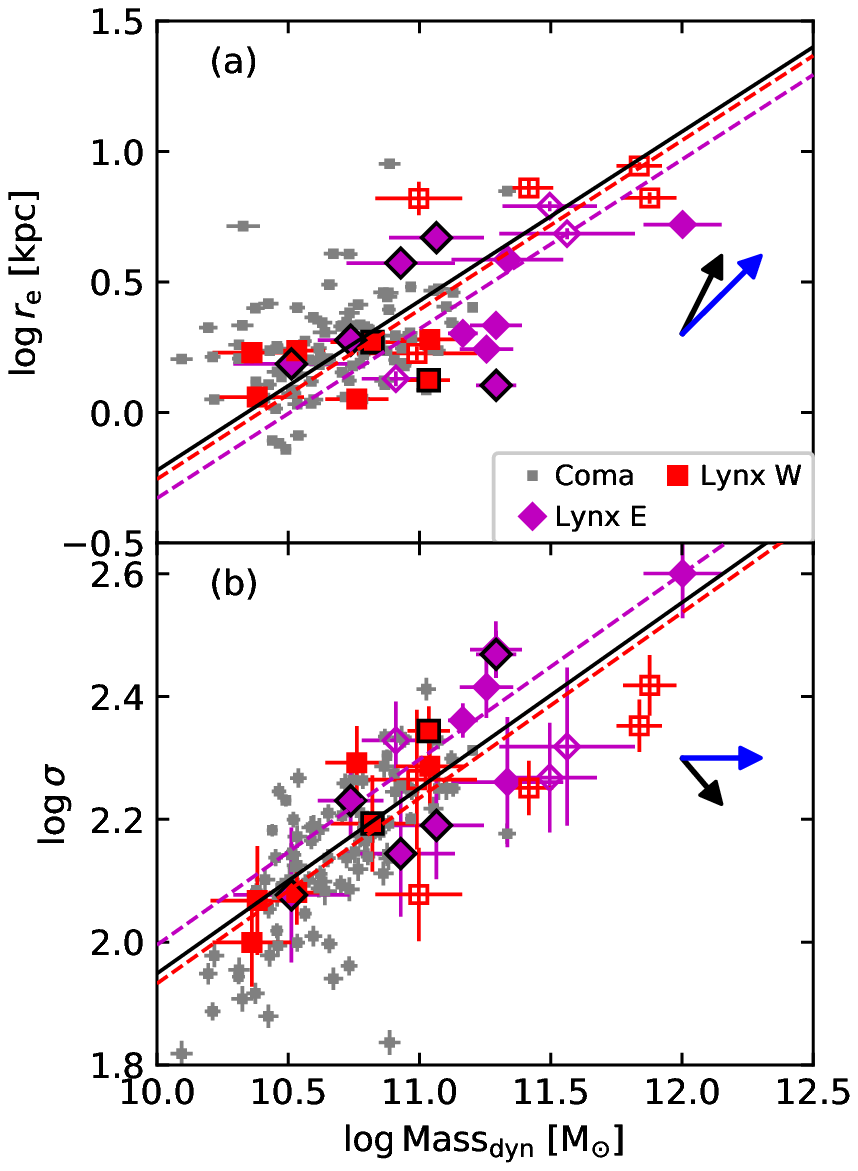}
\caption{
Effective radii and velocity dispersions versus dynamical masses derived as ${\rm Mass_{dyn}} = \beta r_e \sigma^2\,G^{-1}$, with
$\beta$ as a function of the S\'{e}rsic index. 
Dark gray -- Coma cluster members;  
magenta diamonds -- Lynx E; red boxes -- Lynx W. 
Open symbols for Lynx E and W show galaxies with EW[\ion{O}{2}]$>$5{\AA}.
Symbols with black edges show Lynx E and W passive galaxies within $R_{500}$ of the cluster centers.
Solid black lines -- best fit relations to the low redshift reference sample (Coma). 
Dashed colored lines -- the low redshift relations offset to the median zero points of the higher redshift 
clusters, color coded to match the symbols: magenta -- Lynx E; red -- Lynx W.
Arrows mark minor (black) and major (blue) merger tracks from Bezanson et al.\ (2009) equivalent to a factor two increase 
in size, $\Delta \log r_e = 0.3$. For clarity, the model tracks are shown offset from the data.
This figure can be compared with Figure \ref{fig-lrelmass} in the main text.
\label{fig-lrelmass_beta} }
\end{figure}

\begin{figure*}
\epsfxsize 14cm
\begin{center}
\epsfbox{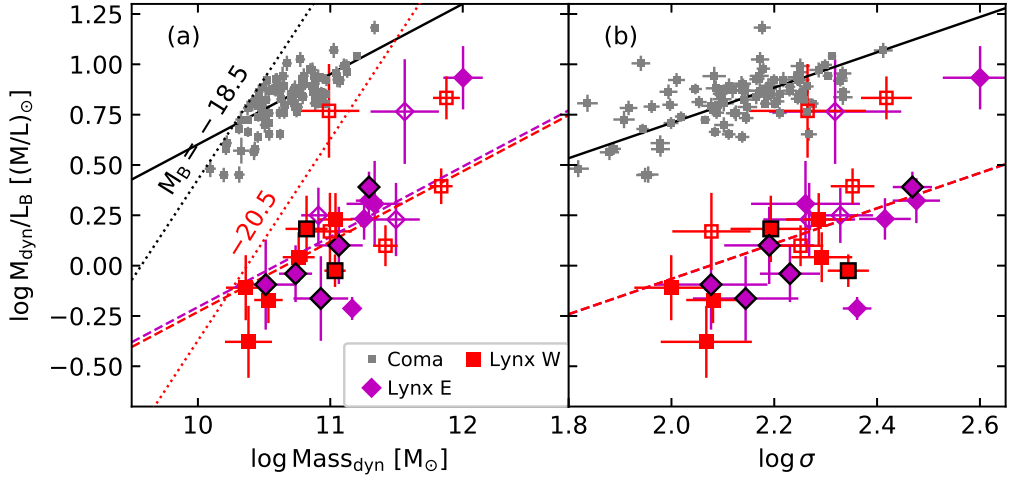}
\end{center}
\caption{ 
The dynamical M/L ratios versus the dynamical masses and versus the velocity dispersions. 
The masses are derived as ${\rm Mass_{dyn}} = \beta r_e \sigma^2\,G^{-1}$, with
$\beta$ as a function of the S\'{e}rsic index.
Symbols as in Figure \ref{fig-lrelmass_beta}.
Black lines -- best fit relations to the low redshift reference sample (Coma).
Dashed colored lines -- the low redshift relations offset to the median zero points of the higher redshift 
clusters, color coded to match the symbols: black -- low redshift sample; magenta -- Lynx E; red -- Lynx W. 
On panel (b) the magenta and red line are at identical locations, and only the red can be seen.
Dotted lines on panel (a) show the magnitude limits of the samples, $M_{\rm B,abs}=-18.5$ for the low 
redshift reference sample and typically $M_{\rm B,abs}=-20.5$ for the higher redshift samples.
This figure can be compared with Figure \ref{fig-MLonly} in the main text.
\label{fig-MLonly_beta} }
\end{figure*}

We here test to what extent our results are sensitive to our choice of deriving the dynamical masses as
${\rm Mass_{dyn}} = \beta r_e \sigma^2\,G^{-1}$ with $\beta =5$,
rather than using $\beta$ values dependent on the S\'{e}rsic (1968) index as proposed by Cappellari et al.\ (2006)
\begin{equation}
\label{eq-beta}
\beta (n_{\rm ser}) = 8.87 - 0.831\, n_{\rm ser} + 0.0241\, n_{\rm ser}^2.
\end{equation}
For this test we use the S\'{e}rsic indices from Simard et al.\ (2011) for the Coma cluster sample. 
This limits the low redshift sample to the 80 passive bulge-dominated galaxies with $\log {\rm Mass _{\rm dyn}}\ge 10.3$
and available S\'{e}rsic index. 
We first fit relevant relations to this smaller low redshift sample, using $\beta =5$. 
Then we repeat the fits using masses derived using $\beta$ from Equation \ref{eq-beta}. 
For the latter fits we also derive zero points for the Lynx E and W.
The fits and zero points are summarized in Table \ref{tab-betatest}, while Figures \ref{fig-lrelmass_beta} and \ref{fig-MLonly_beta} 
show the relations for masses using $\beta (n_{\rm ser})$.

Comparing the fits to the low redshift sample using fixed $\beta$ with the fits using $\beta (n_{\rm ser})$ shows that all slope differences
are well within the 1-$\sigma$ uncertainties, see Table \ref{tab-betatest}. Comparing the slopes from the fits using $\beta (n_{\rm ser})$ (Table \ref{tab-betatest})
to the relations derived using the full Perseus and Coma samples and $\beta =5$ (Table \ref{tab-relations} in the main text) shows slightly larger differences, 
but still well within the 1-$\sigma$ uncertainties on the slopes.

We then compare the zero point offsets for Lynx E and W as derived in the main text, using $\beta =5$, see Table \ref{tab-relations}, 
with the zero point offsets using $\beta (n_{\rm ser})$, see Table \ref{tab-betatest}. 
In both cases, there are no significant differences between Lynx E and W. 
The median zero point differences for the full Lynx sample using $\beta (n_{\rm ser})$ are 
$\Delta \log r_e = -0.07\pm 0.05$ and $\Delta \log \sigma = 0.016\pm 0.021$, to be compared with
$\Delta \log r_e = -0.02\pm 0.04$ and $\Delta \log \sigma = 0.003\pm 0.019$ for $\beta=5$, see Section \ref{SEC-SIZE}.
Similarly for the median offset in the M/L ratio at a fixed mass we find $\Delta \log {\rm M/L} = -0.82\pm 0.05$, 
to be compared with $\Delta \log {\rm M/L} = -0.75\pm 0.05$ for $\beta=5$.
None of the results for the two methods are different at more than the 1-$\sigma$ level. 

In summary, the results in this section confirm that our results as presented in the main text do not significantly depend
on our choice to derive the dynamical masses as ${\rm Mass_{dyn}} = \beta r_e \sigma^2\,G^{-1}$ with $\beta =5$.

\begin{deluxetable*}{lrrr rrr rrr}
\tablecaption{Structure Relations\label{tab-betatest} }
\tablewidth{0pc}
\tabletypesize{\scriptsize}
\tablehead{
\colhead{Relation} & \multicolumn{3}{c}{Low redshift} & 
  \multicolumn{3}{c}{Lynx E} &\multicolumn{3}{c}{Lynx W}  \\
 & \colhead{$\gamma$} & \colhead{$N_{\rm gal}$} & \colhead{rms} 
 & \colhead{$\gamma$} & \colhead{$N_{\rm gal}$} & \colhead{rms} &  \colhead{$\gamma$} & \colhead{$N_{\rm gal}$} & \colhead{rms}  \\
\colhead{(1)} & \colhead{(2)} & \colhead{(3)} & \colhead{(4)} 
& \colhead{(5)} & \colhead{(6)} & \colhead{(7)} & \colhead{(8)} & \colhead{(9)} & \colhead{(10)} 
}
\startdata
$\log r_e      = (0.72 \pm 0.18) \log {\rm Mass} + \gamma$\tablenotemark{a}  & -7.416 & 80 & 0.18 & \nodata & \nodata & \nodata & \nodata & \nodata & \nodata  \\   
$\log r_e      = (0.65 \pm 0.17) \log {\rm Mass} + \gamma$\tablenotemark{b}  & -6.712 & 80 & 0.16 & -6.819 & 13 & 0.22 & -6.745 & 12 & 0.20  \\   
$\log \sigma   = (0.33 \pm 0.05) \log {\rm Mass} + \gamma$\tablenotemark{a}  & -1.343 & 80 & 0.08 & \nodata & \nodata & \nodata & \nodata & \nodata & \nodata  \\ 
$\log \sigma   = (0.30 \pm 0.04) \log {\rm Mass} + \gamma$\tablenotemark{b}  & -1.071 & 80 & 0.09 & -1.024 & 13 & 0.09 & -1.087 & 12 & 0.09  \\ 
$\log \rm{M/L} = (0.32 \pm 0.06) \log {\rm Mass} + \gamma$\tablenotemark{a}  & -2.612 & 80 & 0.08 & \nodata & \nodata & \nodata & \nodata & \nodata & \nodata   \\ 
$\log \rm{M/L} = (0.35 \pm 0.06) \log {\rm Mass} + \gamma$\tablenotemark{b}  & -2.888 & 80 & 0.07 & -3.696 & 13 & 0.23 & -3.720 & 12 & 0.25   \\ 
$\log \rm{M/L} = (0.91 \pm 0.40) \log \sigma + \gamma$\tablenotemark{a}  & -1.193 & 80 & 0.10 & \nodata & \nodata & \nodata & \nodata & \nodata & \nodata   \\ 
$\log \rm{M/L} = (0.88 \pm 0.32) \log \sigma + \gamma$\tablenotemark{b}  & -1.045 & 80 & 0.11 & -1.820 & 13 & 0.26 & -1.820 & 12 & 0.28   \\ 
\enddata
\tablecomments{Column 1: Scaling relation, slopes from bet fit Coma sample with available S\'{e}rsic indices. 
Column 2: Zero point for the low redshift sample (Coma). 
Column 3: Number of galaxies included from the low redshift sample. 
Column 4: rms in the Y-direction of the scaling relation for the low redshift sample.
Columns 5, 6, and 7: Zero point, number of galaxies, rms in the Y-direction for the Lynx W sample.
Columns 8, 9, and 10: Zero point, number of galaxies, rms in the Y-direction for the Lynx E sample.
}
\tablenotetext{a}{Masses derived using $\beta =5$.}
\tablenotetext{b}{Masses derived using $\beta (n_{\rm ser})$ from Eq.\ \ref{eq-beta}.}
\end{deluxetable*}

\begin{figure*}
\epsfxsize 17.5cm
\epsfbox{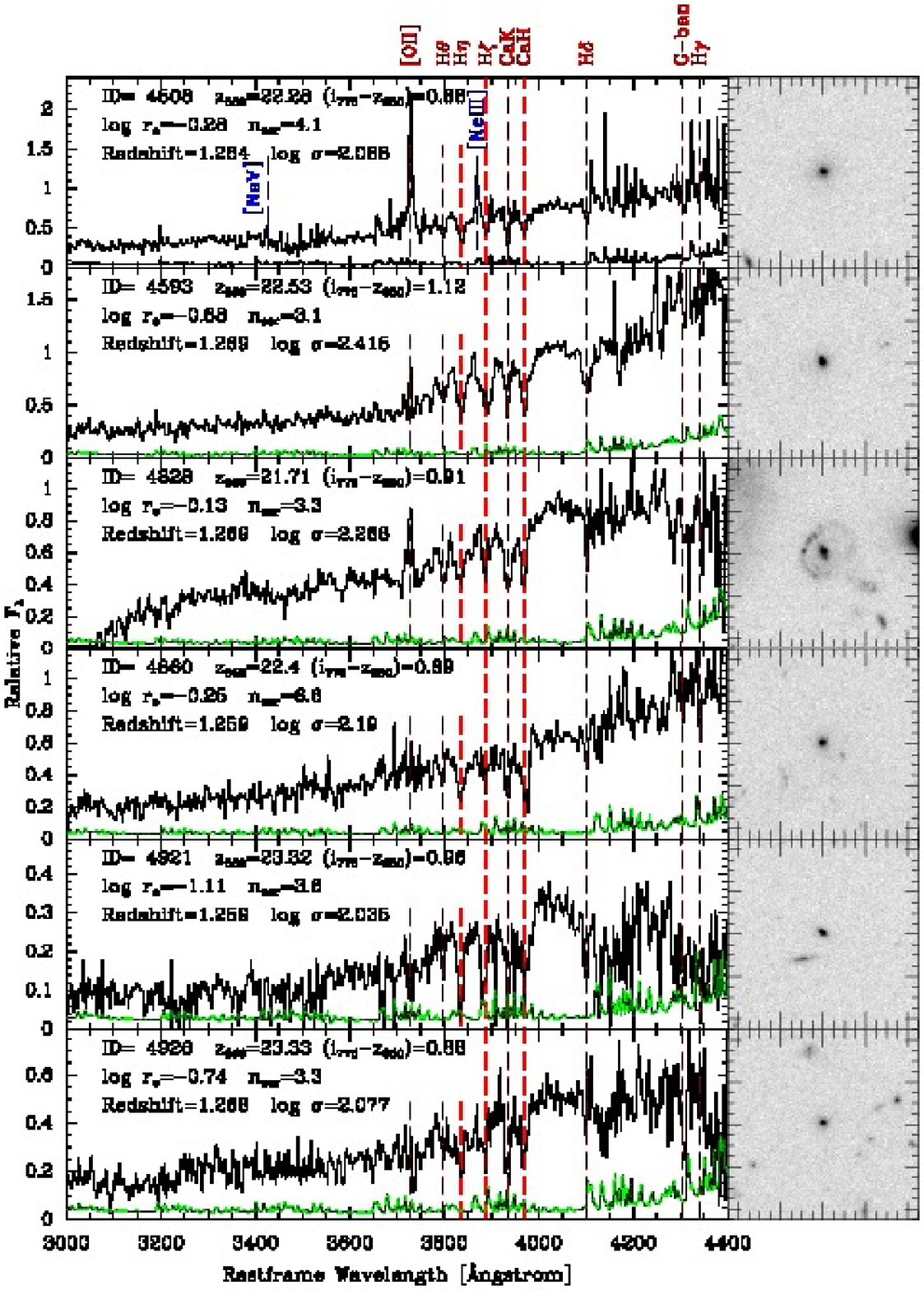}
\center{Fig.\ \ref{fig-speclynx} -- {\em Continued.}}
\end{figure*}
\begin{figure*}
\epsfxsize 17.5cm
\epsfbox{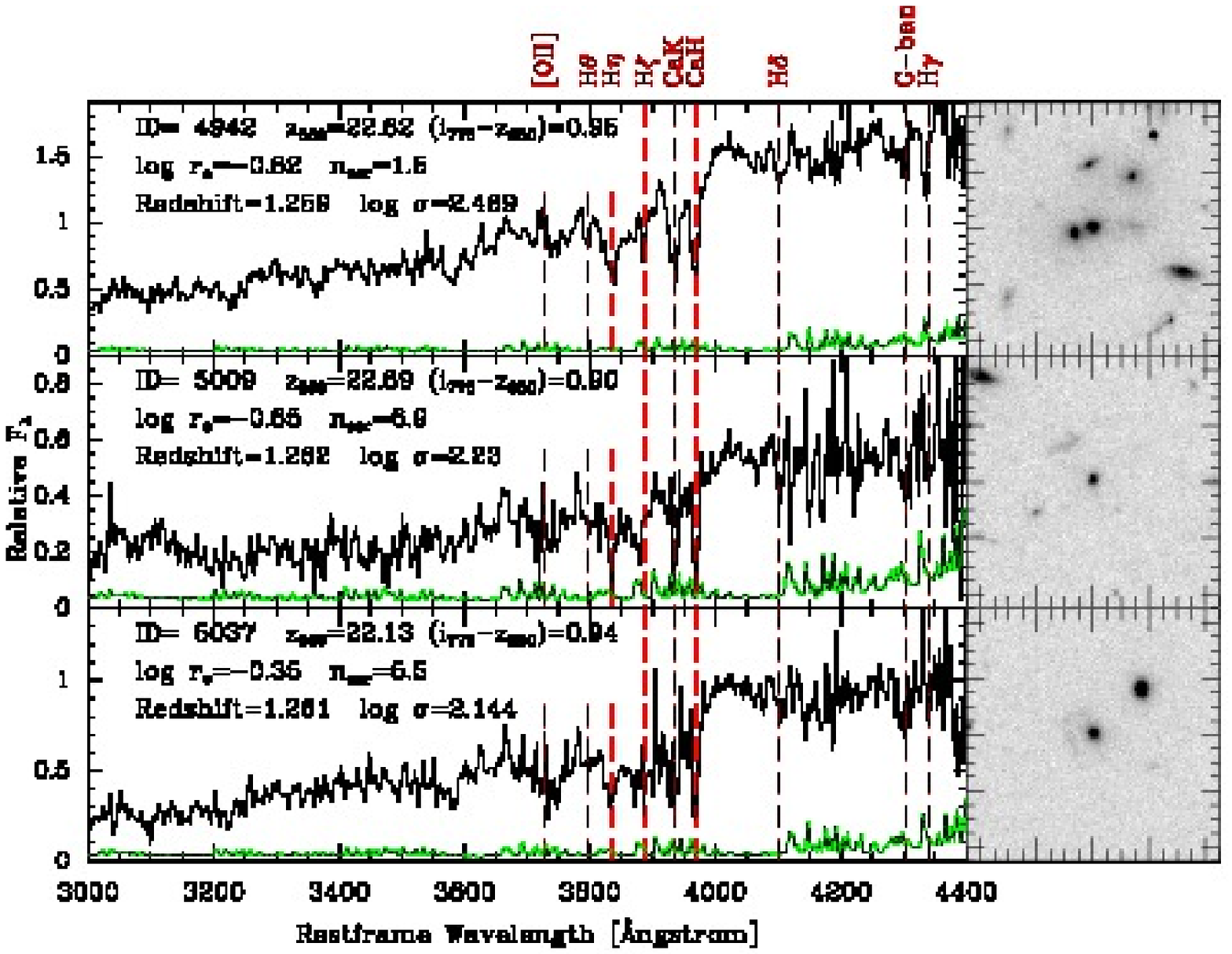}
\center{Fig.\ \ref{fig-speclynx} -- {\em Continued.}}
\end{figure*}

\end{document}